\documentclass[twocolumn]{aastex63}
\usepackage{graphics,graphicx}
\hypersetup{urlcolor=blue}
\usepackage{natbib}
\usepackage[outdir=./]{epstopdf}
\usepackage{textcomp,gensymb}
\usepackage{xfrac}
\usepackage{xcolor}
\usepackage{amsmath}
\usepackage{import}
\usepackage[figuresright]{rotating}
\usepackage{lineno}
\usepackage{color}
\usepackage{hyperref}
\hypersetup{colorlinks=true,allcolors=[rgb]{0,0,0.8}}

\newcommand{\toi}{1097}
\newcommand{\ticid}{360630575}

\newcommand\msun{\ensuremath{\text{M}_\odot}}

\newcommand{\mps}{m\,s$^{-1}$}

\newcommand{\vsini}{$v\sin{i_*}$}

\newcommand{\logg}{$log~g$ }

\newcommand{\fbol}{$F_{\mathrm{bol}}$}

\newcommand{\teff}{\ensuremath{T_{\text{eff}}}}
\newcommand\kms{km~s$^{-1}$}

\newcommand{\tess}{\textit{TESS}}
\newcommand{\ktwo}{{\textit K2}}

\newcommand{\gaia}{\textit{Gaia}}
\newcommand{\epscha}{$\epsilon$ Chamaeleontis}
\newcommand{\starname}{HD 109833}
\newcommand{\planetname}{HD 109833 \,b}
\newcommand{\planetnametwo}{HD 109833 \,c}
\newcommand{\assoc}{MELANGE-4}
\newcommand{\banyan}{\texttt{BANYAN $\Sigma$}}

\newcommand{\red}[1]{#1}

\usepackage[nolist]{acronym}
\newacro{sed}[SED]{spectral-energy distribution}
\newacro{cmd}[CMD]{color-magnitude diagram}
\accepted{Sept 28, 2022}
\submitjournal{AJ}

\shorttitle{New LCC Population and Planets}
\shortauthors{Wood et al.}
\graphicspath{{./}{}}

\begin{document}

\title{TESS Hunt for Young and Maturing Exoplanets (THYME) IX: a 27\,Myr extended population of Lower-Centaurus Crux with a transiting two-planet system}

\correspondingauthor{Mackenna Wood}
\email{woodml96@live.unc.edu}

%THYME, Lead, LDB, lithium, membership calcs, misttborn, writing, BANYAN models, etc
\author[0000-0001-7336-7725]{Mackenna L. Wood}%
\affiliation{Department of Physics and Astronomy, The University of North Carolina at Chapel Hill, Chapel Hill, NC 27599, USA} 

%% THYME, isochrone fit, supervision
\author[0000-0003-3654-1602]{Andrew W. Mann}%
\affiliation{Department of Physics and Astronomy, The University of North Carolina at Chapel Hill, Chapel Hill, NC 27599, USA} 

% found the 2nd planet, misttborn
\author[0000-0002-8399-472X]{Madyson G. Barber}
\altaffiliation{UNC Chancellor’s Science Scholar}
\affiliation{Department of Physics and Astronomy, The University of North Carolina at Chapel Hill, Chapel Hill, NC 27599, USA} 

%%Rotation analysis of target, friends, and BANYAN members
\author[0000-0002-9446-9250]{Jonathan L. Bush}%
\affiliation{Department of Physics and Astronomy, The University of North Carolina at Chapel Hill, Chapel Hill, NC 27599, USA}

% %%LCO data (along with Ben)
%% also did the mass calculation notebook
\author[0000-0001-9811-568X]{Adam L. Kraus}%
\affiliation{Department of Astronomy, The University of Texas at Austin, Austin, TX 78712, USA}

% LCO data
\author[0000-0003-2053-0749]{Benjamin M. Tofflemire}
\altaffiliation{51 Pegasi b Fellow}
\affiliation{Department of Astronomy, The University of Texas at Austin, Austin, TX 78712, USA}

% %%THYME, LC, etc.
\author[0000-0001-7246-5438]{Andrew Vanderburg}%
\affiliation{Department of Physics and Kavli Institute for Astrophysics and Space Research, Massachusetts Institute of Technology, Cambridge, MA 02139, USA}

% %THYME, 
\author[0000-0003-4150-841X]{Elisabeth R. Newton}%
\affiliation{Department of Physics and Astronomy, Dartmouth College, Hanover, NH 03755, USA}

% M dwarf models
\author[0000-0002-2012-7215]{Gregory A. Feiden}%
\affiliation{Department of Physics \& Astronomy, University of North Georgia, Dahlonega, GA 30597, USA}

%% CHIRON data and resulting RVs
\author[0000-0002-4891-3517]{George Zhou}
\affiliation{Centre for Astrophysics, University of Southern Queensland, West Street, Toowoomba, QLD 4350 Australia }

%% CHIRON data and resulting RVs
\author[0000-0002-0514-5538]{Luke G. Bouma}
\altaffiliation{51 Pegasi b Fellow}
\affiliation{Cahill Center for Astrophysics, California Institute of Technology, Pasadena, CA 91125, USA}

%% CHIRON data and resulting RVs
\author[0000-0002-8964-8377]{Samuel~N.~Quinn}
\affiliation{Center for Astrophysics $\vert$ Harvard \& Smithsonian, 60 Garden St, Cambridge, MA, 02138, USA}

%%HARPS DATA and resulting RVs
\author[0000-0002-5080-4117]{David J. Armstrong}
\affiliation{Department of Physics, University of Warwick, Gibbet Hill Road, Coventry, CV4 7AL, UK}
\affiliation{Centre for Exoplanets and Habitability, University of Warwick, Gibbet Hill Road, Coventry, CV4 7AL, UK}

\author[0000-0002-5899-7750]{Ares Osborn}
\affiliation{Department of Physics, University of Warwick, Gibbet Hill Road, Coventry, CV4 7AL, UK}
\affiliation{Centre for Exoplanets and Habitability, University of Warwick, Gibbet Hill Road, Coventry, CV4 7AL, UK}

% HARPS abundances:
\author{Vardan Adibekyan}
\affiliation{Instituto de Astrof\'isica e Ci\^encias do Espa\c{c}o, Universidade do Porto, CAUP, Rua das Estrelas, 4150-762 Porto, Portugal}
\affiliation{Departamento de F\'{\i}sica e Astronomia, Faculdade de Ci\^encias, Universidade do Porto, Rua do Campo  Alegre, 4169-007 Porto, Portugal}
  
\author{Elisa~Delgado Mena}
\affiliation{Instituto de Astrof\'isica e Ci\^encias do Espa\c{c}o, Universidade do Porto, CAUP, Rua das Estrelas, 4150-762 Porto, Portugal}
  
\author[0000-0001-9047-2965]{Sergio~G.~Sousa}
\affiliation{Instituto de Astrof\'isica e Ci\^encias do Espa\c{c}o, Universidade do Porto, CAUP, Rua das Estrelas, 4150-762 Porto, Portugal}

% BANYAN Help
\author[0000-0002-2592-9612]{Jonathan Gagné}
\affiliation{Plan\'etarium Rio Tinto Alcan, Espace pour la Vie, 4801 av. Pierre-de Coubertin, Montr\'eal, Qu\'ebec, Canada}
\affiliation{Institute for Research on Exoplanets, Universit\'e de Montr\'eal, Department de Physique, C.P. 6128 Succ. Centre-ville, Montr\'eal, Qu\'ebec, Canada}

% inclination and his stellar parameter code
\author[0000-0002-9641-3138]{Matthew J. Fields}
\affiliation{Department of Physics and Astronomy, The University of North Carolina at Chapel Hill, Chapel Hill, NC 27599, USA} 

%% SOAR Obs
\author[0000-0002-1312-3590]{Reilly P. Milburn}%
\affiliation{Department of Physics and Astronomy, The University of North Carolina at Chapel Hill, Chapel Hill, NC 27599, USA} 

\author[0000-0001-5729-6576]{Pa Chia Thao}%
\affiliation{Department of Physics and Astronomy, The University of North Carolina at Chapel Hill, Chapel Hill, NC 27599, USA} 

\author[0000-0003-3654-1602]{Stephen P. Schmidt}%
\affiliation{Department of Physics and Astronomy, The University of North Carolina at Chapel Hill, Chapel Hill, NC 27599, USA} 

\author[0000-0003-2519-6161]{Crystal~L.~Gnilka}
\affiliation{NASA Ames Research Center, Moffett Field, CA 94035, USA}
\affiliation{NASA Exoplanet Science Institute, Caltech/IPAC, Mail Code 100-22, 1200 E. California Blvd.,
Pasadena, CA 91125, USA}

\author[0000-0002-2532-2853]{Steve~B.~Howell}
\affil{NASA Ames Research Center, Moffett Field, CA 94035, USA}

%SOAR Speckle Obs:
\author[0000-0001-9380-6457]{Nicholas M. Law}
\affiliation{Department of Physics and Astronomy, The University of North Carolina at Chapel Hill, Chapel Hill, NC 27599, USA}

\author[0000-0002-0619-7639]{Carl Ziegler}
%carl.ziegler@dunlap.utoronto.ca
\affiliation{Dunlap Institute for Astronomy and Astrophysics, University of Toronto, 50 St. George Street, Toronto, Ontario M5S 3H4, Canada}

\author[0000-0001-7124-4094]{C\'esar Brice\~no}
%cbriceno@ctio.noao.edu
\affiliation{Cerro Tololo Inter-American Observatory, Casilla 603, La Serena, Chile}

%%%% TESS architects %%%%%
\author[0000-0003-2058-6662]{George~R.~Ricker}%
\affiliation{Department of Physics and Kavli Institute for Astrophysics and Space Research, Massachusetts Institute of Technology, Cambridge, MA 02139, USA}

\author[0000-0001-6763-6562]{Roland~Vanderspek}%
\affiliation{Department of Physics and Kavli Institute for Astrophysics and Space Research, Massachusetts Institute of Technology, Cambridge, MA 02139, USA}

\author[0000-0001-9911-7388]{David~W.~Latham}%
\affiliation{Center for Astrophysics $\vert$ Harvard \& Smithsonian, 60 Garden St, Cambridge, MA, 02138, USA}

\author[0000-0002-6892-6948]{Sara~Seager}%
\affiliation{Department of Physics and Kavli Institute for Astrophysics and Space Research, Massachusetts Institute of Technology, Cambridge, MA 02139, USA}
\affiliation{Department of Earth, Atmospheric and Planetary Sciences, Massachusetts Institute of Technology, Cambridge, MA 02139, USA}
\affiliation{Department of Aeronautics and Astronautics, MIT, 77 Massachusetts Avenue, Cambridge, MA 02139, USA}

\author[0000-0002-4265-047X]{Joshua~N.~Winn}%
\affiliation{Department of Astrophysical Sciences, Princeton University, 4 Ivy Lane, Princeton, NJ 08544, USA}

\author[0000-0002-4715-9460]{Jon M. Jenkins}%
\affiliation{NASA Ames Research Center, Moffett Field, CA, 94035, USA}
%% end tess architechts

%% POC/GI/MAST authors:
\author[0000-0001-5347-7062]{Joshua~E.~Schlieder}
\affiliation{NASA Goddard Space Flight Center, 8800 Greenbelt Rd, Greenbelt, MD 20771, USA}

%% POC/GI/MAST authors:
\author[0000-0002-4047-4724]{Hugh~P.~Osborn}
\affiliation{Physikalisches Institut, University of Bern, Gesellsschaftstrasse 6, 3012 Bern, Switzerland}
\affiliation{Department of Physics and Kavli Institute for Astrophysics and Space Research, Massachusetts Institute of Technology, Cambridge, MA 02139, USA}

%% SPOC author:
\author[0000-0002-6778-7552]{Joseph~D.~Twicken}
\affiliation{SETI Institute, Mountain View, CA  94043, USA}
\affiliation{NASA Ames Research Center, Moffett Field, CA  94035, USA}

%% TSO/EXOFOP authors:
\author[0000-0002-5741-3047]{David~R.~Ciardi}
\affiliation{Caltech/IPAC, NASA Exoplanet Science Institute, 770 S. Wilson Avenue, Pasadena, CA 91106, USA}

%% TSO/EXOFOP authors:
\author[0000-0003-0918-7484]{Chelsea~ X.~Huang}		
\affiliation{University of Southern Queensland, Centre for Astrophysics, West Street, Toowoomba, QLD 4350, Australia}

% We report the discovery and characterization of a nearby (~ 85 pc), older (27 +/- 3 Myr), distributed stellar population near Lower-Centaurus-Crux (LCC), initially identified by searching for stars co-moving with a candidate transiting planet from TESS (HD 109833; TOI 1097). We determine the association membership using Gaia kinematics, color-magnitude information, and rotation periods of candidate members. We measure it's age using isochrones, gyrochronology, and Li depletion. While the association is near known populations of LCC, we find that it is older than any previously found LCC sub-group (10-16 Myr), and distinct in both position and velocity. In addition to the candidate planets around HD 109833 the association contains four directly-imaged planetary-mass companions around 3 stars, YSES-1, YSES-2, and HD 95086, all of which were previously assigned membership in the younger LCC. Using the Notch pipeline, we identify a second candidate transiting planet around HD 109833. We use a suite of ground-based follow-up observations to validate the two transit signals as planetary in nature. HD 109833 b and c join the small but growing population of <100 Myr transiting planets from TESS. HD 109833 has a rotation period and Li abundance indicative of a young age (< 100 Myr), but a position and velocity on the outskirts of the new population, lower Li levels than similar members, and a CMD position below model predictions for 27 Myr. So, we cannot reject the possibility that HD 109833 is a young field star coincidentally nearby the population.

\begin{abstract}
We report the discovery and characterization of a nearby ($\sim 85\,pc$), older (27$\pm$3\,Myr), distributed stellar population near Lower-Centaurus-Crux (LCC), initially identified by searching for stars co-moving with a candidate transiting planet from \tess{} (\starname; TOI \toi). We determine the association membership using \gaia{} kinematics, color-magnitude information, and rotation periods of candidate members. We measure its age using isochrones, gyrochronology, and Li depletion. While the association is near known populations of LCC, we find that it is older than any previously found LCC sub-group (10--16\,Myr), and distinct in both position and velocity. In addition to the candidate planets around \starname{} the association contains four directly-imaged planetary-mass companions around 3 stars, YSES-1, YSES-2, and HD 95086, all of which were previously assigned membership in the younger LCC. Using the Notch pipeline, we identify a second candidate transiting planet around \starname{}. We use a suite of ground-based follow-up observations to validate the two transit signals as planetary in nature. \starname{} b and c join the small but growing population of $<100$\,Myr transiting planets from \tess. \starname{} has a rotation period and Li abundance indicative of a young age ($\lesssim100$ \,Myr), but a position and velocity on the outskirts of the new population, lower Li levels than similar members, and a CMD position below model predictions for 27\,Myr. So, we cannot reject the possibility that \starname{} is a young field star coincidentally nearby the population.

\end{abstract}

\keywords{exoplanets, exoplanet evolution, young star clusters- moving clusters, planets and satellites: individual (TOI1097)}

\section{Introduction}\label{sec:intro}

% Young associations are cool
% - - - - - - - - - - - - - - - - - - - - - - - - - - - - - -
Young associations --- coeval stellar populations thought to share a common age, metallicity, and kinematics --- are crucial for studies of stellar and planetary evolution, including circumstellar disk lifetimes \citep{haisch_disk_2001}, planet formation and migration \citep{mann_k2-33_2016, david_neptune-sized_2016, donati_hot_2016}, pre-main-sequence evolution \citep{stassun_empirical_2014, kraus_mass-radius_2015}, and planetary mass loss \citep{rockcliffe_lyman_2021, zhang_detection_2022}. However, the utility of a young association for such studies depends on a robust understanding of the association's structure and properties. In the last decades, detailed surveys of nearby associations have shown a much more complicated picture of the properties, structure, and formation history of stellar associations than previously known. 

The advent of the high-precision position, velocity, and color measurements of \gaia{} \citep{gaia_collaboration_gaia_2016} combined with new techniques for locating associations \citep[e.g., HDBSCAN;][]{mcinnes_hierarchical_2017} have significantly altered our view of young associations near the Sun. One revelation has been the level of sub-grouping within well-known associations. For example, \citet{wright_kinematics_2018} and \citet{krolikowski_gaia_2021} found that the distribution of stellar ages and kinematics within large associations is likely the result of many localized star-formation events occurring over periods of $5-10$ Myr as opposed to the collapse of a single molecular cloud. Some studies have revealed entirely new young populations \citep{oh_comoving_2017, kounkel_close_2019}, including some that are far more diffuse than those previously known \citep{meingast_extended_2019}. Other studies have found that groups thought to be distinct may be fragments of a single star-formation event \citep{gagne_number_2021}. The new discoveries and changes to known populations create a more complex picture of young associations and change our understanding of the origin of nearby stellar associations \citep[e.g.,][]{zucker_star_2022}.
%This simultaneous fragmenting and merging of known groups, combined with the discovery of new young stellar associations has created a complex picture of young associations and has direct implications on our understanding of the origin of nearby stellar associations \citep[e.g.,][]{zucker_star_2022}.

% Sco-Cen is an important and interesting young association
% - - - - - - - - - - - - - - - - - - - - - - - - - - - - - -
The Scorpius-Centaurus Association (Sco-Cen) is the nearest OB association to the Sun, harboring $\simeq$150 B-type stars and tens of thousands of lower-mass members \citep{de_zeeuw_hipparcos_1999}. The association is classically divided into three populations; Upper Scorpius (US), Upper Centaurus Lupus (UCL), and Lower Centaurus Crux (LCC), with ages varying from 11 to 17\,Myr \citep{pecaut_revised_2012}. The Sco-Cen complex (sometimes called Greater Sco-Cen) includes many other molecular clouds and star-forming groups (e.g., Ophiuchus and Lupus), but none of the main three groups show evidence of ongoing star formation \citep{preibisch_nearest_2008}. This combination of factors makes Sco-Cen an exceptional laboratory to test models of early stellar evolution, and has motivated more than a century of intense research \citep[e.g.,][]{kapteyn_individual_1914, pecaut_revised_2012, feiden_magnetic_2016, wright_kinematics_2018}.

% But we don't know everything about it, and some members and substructure can be missed.
% - - - - - - - - - - - - - - - - - - - - - - - - - - - - - -
However, there is still much to be understood about the structure and formation of Sco-Cen, particularly with the arrival of astrometry from \gaia{} \citep{gaia_collaboration_gaia_2016}. \cite{pecaut_star_2016} found a strong age gradient across Sco-Cen in general and LCC in particular, noting that the southern region of LCC is younger than the northern. \citet{goldman_large_2018} found a separate moving group within the southern portion of LCC, dividing it into multiple sub-populations with ages ranging from $7-10$ Myr. This result was confirmed by \citet{kerr_stars_2021}, who recovered those four groups and included the younger \epscha{} as a fifth. These discoveries raise the possibility that the census of Sco-Cen is still incomplete. % These discoveries paint a more complicated picture of Sco-Cen than the classical three single-aged populations, and raise the possibility that the census of the Sco-Cen region is still incomplete.

% Young associations are also important samples for planet discovery, and Sco-Cen in particular because of direct imaging
% - - - - - - - - - - - - - - - - - - - - - - - - - - - - - -
Mapping out Sco-Cen is particularly important because of its outsized role in the study of young planets. It is nearby and young enough for direct imaging of young planets on wide orbits \citep[e.g.,][]{hinkley_companions_2015, bohn_two_2020}. The association is also diffuse enough to separate out individual stars even with large \ktwo{} and \tess{} pixels, enabling the discovery of young transiting planets. The youngest known transiting planet, \citep[K2-33 b;][]{david_neptune-sized_2016, mann_k2-33_2016} the youngest transiting hot Jupiter, \citep[HIP 67522 b;][]{rizzuto_tess_2020}, and the largest planet known to orbit a mid-M dwarf \citep[TOI 1227 b;][]{mann_tess_2022} are all in Sco-Cen. Such systems are critical to our understanding of the early evolution of planetary systems, but the sample is still small. New sub-structures in Sco-Cen would provide new regions to search for such planets, potentially with a wider range of (young) ages. 

% Paper Summary
% - - - - - - - - - - - - - - - - - - - - - - - - - - - - - -
In this paper, we report the discovery of a new, older (27$\pm$3\,Myr) population outside LCC. Our discovery of the young association was prompted by the detection of \planetname{} by the \tess{} survey. The host, \starname, was included as a member of the Theia 64 moving group by \citet{kounkel_close_2019} and a member of the Sco-Cen region by \citet{kerr_stars_2021}. After confirming indicators of youth in \starname{} (see Section \ref{sec:stellar_properties}), we searched its nearest kinematic and spatial neighbors to see if they also appeared to be in a young association. The resulting color-magnitude diagram (CMD) showed a significant population of pre-main-sequence M dwarfs, indicating a young ($<50$\,Myr) association. We investigate the age and membership of this association and its relationship with existing Sco-Cen populations and statistically validate the planet candidates \planetname{} and c.

In Section \ref{sec:assoc_members}, we describe our iterative process for locating members and removing field and LCC interlopers. We discuss our observational program in Section \ref{sec:obs}, with observations both of the planet host and the association members. We describe the properties of the association, including its age, in Section \ref{sec:assoc_properties}, and some notable directly-imaged planet-hosting candidate members in Section \ref{sec:direct_imaging}. Section \ref{sec:stellar_properties} discusses the properties of the planet host \starname{}, and Section \ref{sec:planet_params} the properties of the planets. We conclude in Section \ref{sec:conclusion} with a summary of our work and brief discussion of the implications of this new population.
 
% - - - - - - - - - - - - - - - - - - - - - - - - - - - - - - -
% - - - - - - - - - - - - - - - - - - - - - - - - - - - - - - -

\section{The Membership of the \assoc{} association}\label{sec:assoc_members}

In this Section, we attempt to confirm that the population of stars spatially and kinematically near \starname{} is part of a real, co-eval association and to separate out interlopers from nearby LCC populations and the field. As we show later, this does not appear to be part of a known group, so we refer to the association as \assoc{} following the naming convention from \citet{tofflemire_tess_2021}.

To select the membership of \assoc{} we use an iterative four-step process, using a mix of kinematic and age-based indicators of membership. During the first two steps, our goal is to produce a clean membership list, preferring to exclude member stars than to include non-member stars. We then use that clean list to define the properties of the group. \red{This initial list is used to define the group's kinematics for a more expansive search for members in the next two steps.}  This process is outlined below, and detailed in Sections \ref{sec:mem_initial} - \ref{sec:mem_final}.

\begin{enumerate}
    \item Initial Selection - we select nearby co-moving stars using the \texttt{Comove} algorithm.
    \item Remove Interlopers - we apply cuts to the initial candidate list using color, magnitude, and rotation to remove interlopers from LCC and the field.
    \item \banyan{} - we use the refined candidate list to define the group kinematics, then use \banyan{} and full 6D kinematics (or 5D for those lacking a radial velocity measurement) of each candidate to determine kinematic membership probabilities and search for additional candidate members.
    \item Reapply Cuts - We reapply the color and magnitude cuts to the kinematic candidates. This produces the final candidate list, comprised of stars that are clustered in color, magnitude, kinematics, and rotation.
\end{enumerate}
We note that these steps inevitably create some biases in the list. For example, the use of colors and magnitudes to select members may bias the age (depending on how the cuts are applied). For this reason, we provide the candidate membership lists from each step so that readers can apply their own cuts on the data based on the specific scientific case. 

% - - - - - - - - - - - - - - - - - - - - - - - - - - - - - - -
\subsection{Initial Selection}\label{sec:mem_initial}

We initially select the co-moving neighbors of \starname{} using \texttt{Comove}\footnote{\url{https://github.com/adamkraus/Comove}}. 
Details of the algorithm are given in \citet{tofflemire_tess_2021}. 
To summarize, \texttt{Comove} uses astrometry from the \gaia{} Data Release 3 \citep[DR3,][]{lindegren_gaia_2021, riello_gaia_2021, gaia_collaboration_gaia_2022} and a user-provided velocity of \starname{} to compute its $XYZ$ position, and $UVW$ velocity \footnote{This is using a galactic coordinate system in which the sun is at $<0,0,0>$, X points towards the galactic center, Y is in the direction of galactic rotation, and Z is out of the galactic plane. U, V, and W are the velocities in the X, Y, and Z directions, respectively.}.
\texttt{Comove} then selects every \gaia{} star within a user-defined threshold of \starname{} in three-dimensional distance and expected tangential velocity ($V_{t,exp}$) assuming a $UVW$ matching \starname. We opted to use thresholds of 30\,pc and 2\,\kms. This tight limit likely removes many real members (particularly fainter stars with tangential velocity uncertainties larger than 2\,\kms), but larger search radii led to significant contamination from younger LCC stars. From this sample we took those which had valid \gaia{} $B_P$ and $R_P$ magnitudes, yielding an initial selection of $207$ stars. 

In addition to \gaia{} astrometry, velocities, and photometry, we retrieve the \gaia{} renormalized unit weight error (RUWE) for each candidate member star. The RUWE value is related to the goodness-of-fit from the \gaia{} astrometry, normalized to correct for color and brightness dependent  effects\footnote{\url{https://gea.esac.esa.int/archive/documentation/GDR2/Gaia_archive/chap_datamodel/sec_dm_main_tables/ssec_dm_ruwe.html}}. The RUWE should be around 1 for well-behaved sources, and higher values suggest the presence of a stellar companion \citep{ziegler_measuring_2018, belokurov_unresolved_2020, wood_characterizing_2021}. 

% Figure: Friend Finder Results; Updated 2022-08-01
\begin{figure}[tb]
    \centering
    \includegraphics[width=0.49\textwidth]{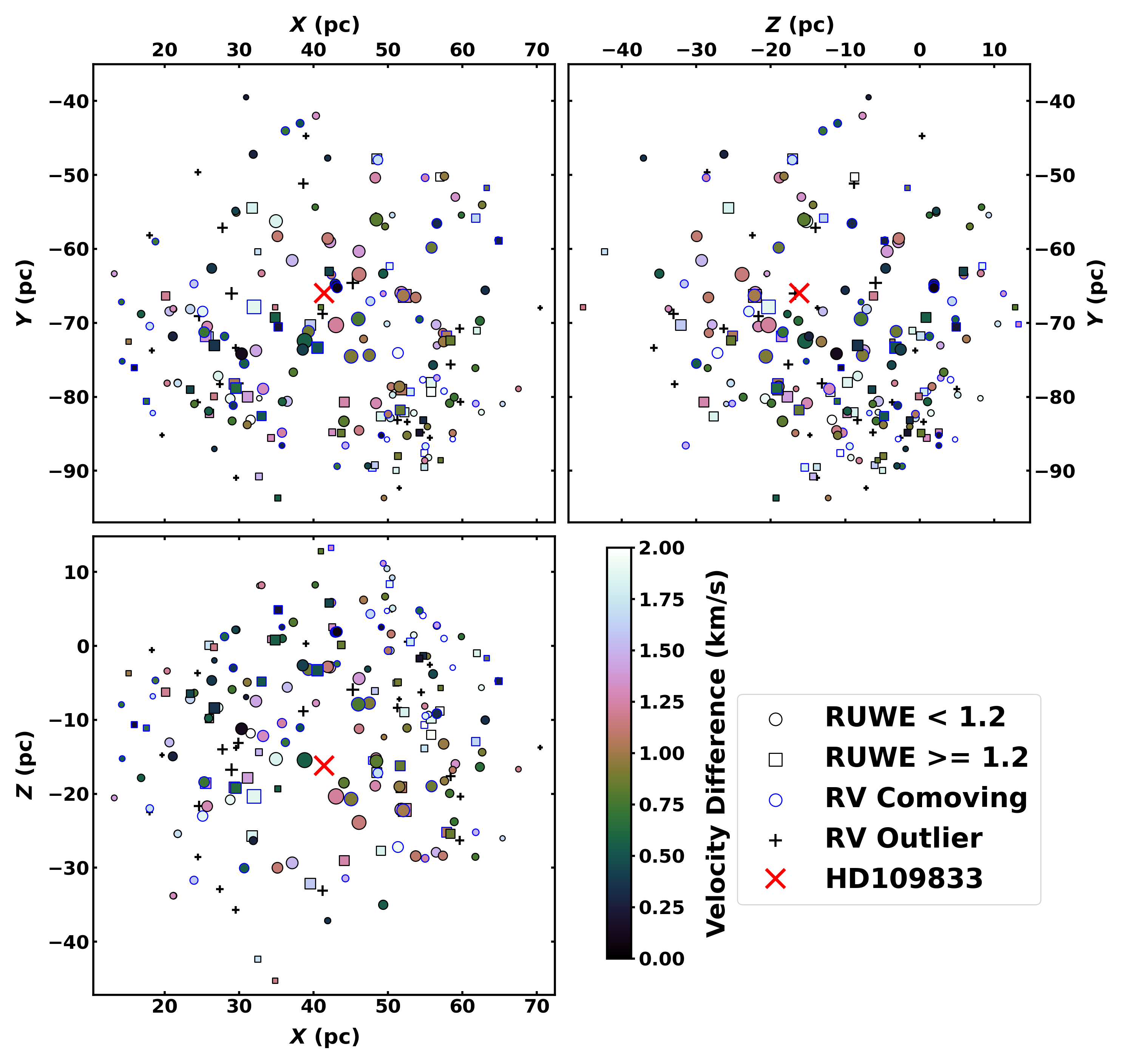}
    \caption{The results of \texttt{Comove}. \starname{} and neighbors within 30\,pc and 2\kms, are shown in X, Y, and Z coordinates. The size of the points corresponds inversely to their distance from \starname, and the color with the velocity difference, so that the largest points are closest to \starname{} and the darkest points most similar in velocity. Stars with \gaia{} $RUWE \geq 1.2$ and are represented with squares. The velocity differences of binary stars are poor indicators of membership, due to velocity from binary motion. Circles are used to represent stars with \gaia{} $RUWE < 1.2$.}
    \label{fig:friend_finder}
\end{figure} 

The selection reveals a population of pre-MS stars close to \starname{} in both position and velocity (Figure~\ref{fig:cmd}), indicating a young co-moving population. \red{We use this population as our initial membership list.} However, it includes many stars which do not appear to be a part of the main population (e.g., main-sequence field interlopers). To remove these probable interlopers, we make a series of cuts on the initial list.

% - - - - - - - - - - - - - - - - - - - - - - - - - - - - - - - -
\subsection{Removing LCC and Field interlopers}\label{sec:interlopers}

We make four cuts on the initial sample of candidate members, divided into those meant to remove LCC interlopers and those meant to remove field interlopers. Nearby members of LCC are young, so they cannot be distinguished from this population by CMD position or other age-based qualities, but are kinematically distinct. In contrast, nearby field stars may have similar position and motion to population members, but are unlikely to be young and thus can be identified through differences in age indicators. 

\begin{figure}[tb]
    \centering
    \includegraphics[width=0.49\textwidth]{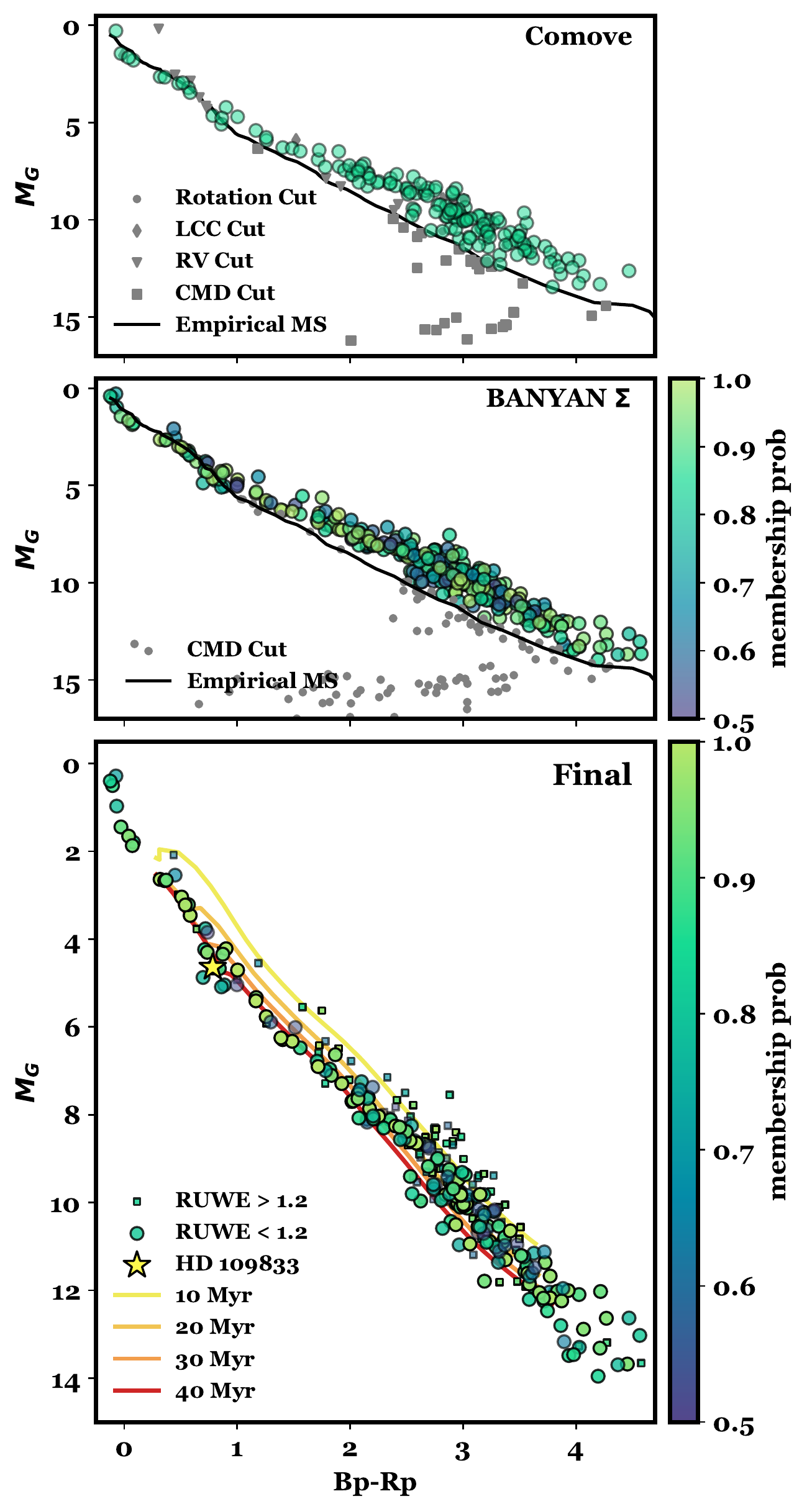}
    \caption{A CMD showing each step of the membership search. The top panel shows the results of \texttt{Comove}, and the cuts made on those results to make a tight core membership. The second panel shows the results of \banyan{} and the subsequent cut on CMD position to remove old interlopers and stars with poor color measurements. Candidates which survived this cut are colored by their kinematic membership probability. The bottom panel shows the final membership list, colored by membership probability. Stars with $\gaia{}RUWE > 1.2$, possible binaries, are marked as small squares, and form a binary sequence above the main association. Four DSEP magnetic isochrones \citep{feiden_magnetic_2016}, showing $10$, $20$, $30$, and $40$ Myrs are plotted alongside the sequence. In the top two panels, the ZAMS is shown as a black line, derived empirically by \citet{pecaut_intrinsic_2013}.}
    \label{fig:cmd}
\end{figure}

For our first cut, we use the empirical main-sequence (MS) defined in \citet{pecaut_intrinsic_2013}. We remove all candidates which have $B_P-R_P > 1$, $G-R_P > 0.5$ and are fainter than the interpolated empirical MS (Figure \ref{fig:cmd}). This removes 28 candidates. The anomalous CMD positions of some of these stars are likely caused by poor $B_P$ magnitude and parallax measurements from \gaia, but those are still `contaminants' from the perspective of our analysis as including them would bias estimates of the group's age.

The rotation period of a star can also be used as an indicator of age since stars are known to `spin down' with increasing age. To this end, we measure the rotation periods of $166$ candidate members using the Lomb-Scargle periodogram of \tess{} light curves (see Section \ref{sec:rotation} for more details). We remove any stars which have high-quality \tess{} data, but do not show a reliable rotation period. Stars with \tess{} magnitude $T > 15$ are not removed, since they are too faint to have adequate signal-to-noise ratio (SNR) for rotation measurement, \red{and thus we cannot reject them as members, as our goal here is to remove only those which we are confident are not members.} A total of eight candidates show no reliable rotation, of which four also had a low CMD position (and hence are removed by the cut above).

We remove 18 candidates with \gaia{} DR3 radial velocities $>5$\kms{} from the values predicted by \texttt{Comove}, and radial velocity errors $<3$\kms{}. These stars may be binaries that are genuine members, but we err on the side of a clean rather than a complete sample.

Lastly, to remove interlopers from the nearby LCC populations, we cross-match our candidate list against the membership list from \citet{goldman_large_2018}, and remove all candidates which are considered LCC members in that paper. We remove five stars for this reason, with an additional Goldman member removed by the earlier velocity cut. It is possible that these stars, which were believed to be members of one of the LCC subpopulations, are actually members of \assoc{}, but at this stage the goal is to get a clean list even at the cost of removing some true members.

These steps are outlined in the top panel of Figure~\ref{fig:cmd}. The cuts give a final sample of $152$ candidate members. 

%207(initial) - 28(cmd) - 18 (velocity) - 4 (rotation) - 5 (goldman) = 152

%Figure: CMD showing cuts; Updated 2022-08-04

% - - - - - - - - - - - - - - - - - - - - - - - - - - - - - - - -
\subsection{BANYAN}\label{sec:BANYAN}

The \texttt{Comove} selection has a sharp radius cutoff, which misses more distant (spatially and kinematically) stars and gives us only general information about the relative probability that a given star is a member. Once a general sense of the spatial distribution is known, a better approach is to use Bayesian membership probabilities that weight the relative likelihood that a star is within \assoc{} compared to the field or a nearby association \citep{rizzuto_multidimensional_2011, malo_bayesian_2012}. 

To this end, we use the \banyan{} tool. \banyan{} is a Bayesian probability tool to determine membership probabilities of stars in young moving groups \citep{gagne_banyan_2018}\footnote{\url{https://github.com/jgagneastro/banyan.sigma}}.
For each star, \banyan{} computes the membership probability using kinematic models of 27 nearby, young moving groups defined in \cite{gagne_banyan_2018} and the field population.

Significant substructure has been discovered within the LCC association since the publication of \banyan{}, and is not accounted for there, so to correctly select members of \assoc, we use updated parameters for the nearby LCC subpopulations in \banyan{}. This update is detailed in Appendix \ref{app:lcc}.

We add \assoc{} to \banyan{} following \cite{gagne_banyan_2018} by calculating the covariance matrix and center vector of the candidate members which survived all cuts from Section~\ref{sec:interlopers}, and had \gaia{} DR2 radial velocity measurements. These are, in units of \,pc and \kms:

\small
\begin{align} % Updated 2022-08-04
    \bar{\bar\Sigma} &= \begin{bmatrix}
        152.32 & 30.929 & 12.03 & 2.813 & 8.634 & 1.424 \\
        30.929 & 143.941 & -22.231 & -0.124 & 4.127 & -1.1\\
        12.03 & -22.231 & 86.511 & 3.7 & -1.383 & -1.119 \\
        2.813 & -0.124 & 3.7 & 2.732 & -3.434 & -0.535 \\
        8.634 & 4.127 & -1.383 & -3.434 & 7.304 & 1.027 \\
        1.424 & -1.1 & -1.119 & -0.535 & 1.027 & 1.048 \\
    \end{bmatrix}\notag
\end{align}
\normalsize

\begin{align}
    \bar{x}_{0}&=\begin{bmatrix} 
    42.238 & -69.627 & -8.228 & -9.386 & -20.475 & -5.596 \\
    \end{bmatrix}\notag
\end{align}

Using the updated $UVWXYZ$ matrix for \assoc{} and LCC sub-populations, we run \banyan{} on all stars within 100pc of \starname{} with a \gaia{} DR3 parallax and a parallax over error of $>20$. There may be stars beyond this range, but those more distant stars may require a more sophisticated model than the multivariate Gaussians used by \banyan{}, and we expect this to be a negligible fraction of the overall population at such a young age. 
 
\banyan{} yielded $424$ candidate members with membership probability greater than $50\%$. \red{Of the original $152$ stars found by Comove, $122$ are included in the \banyan{} results.} There are $168$ candidate members with membership probability greater than $90\%$, which we consider high-probability members. These kinematic candidate members are shown in the middle panel of Figure~\ref{fig:cmd}.

% 424 Objects w/ PROB > 50%, 167 Objects w/ PROB > 90% <---- DR3 results

% - - - - - - - - - - - - - - - - - - - - - - - - - - - - - - -
\subsection{Final Membership}\label{sec:mem_final}

Since \banyan{} only considers kinematic information when determining membership probabilities, it is possible that old interlopers, which match the association in $UVWXYZ$ but not in age, are included in the membership list. As with our initial selection from \texttt{Comove}, the \banyan{} selection includes a large number of stars that are below the main sequence. Therefore, we reapply the CMD cut discussed in Section~\ref{sec:interlopers}. We also remove $20$ stars without \gaia{} Bp or Rp measurements. In total, this step removes 118 stars, yielding a final membership list of $306$ stars. Many of these targets may be real members with poor magnitudes or parallaxes. The final membership is shown in the last panel of Figure~\ref{fig:cmd}. Their membership probabilities and stellar properties are listed in Table \ref{table:members}.

% 424(P>50%) - 20(Bp missing) - 98(cmd) = 306

% - - - - - - - - - - - - - - - - - - - - - - - - - - - - - -
% - - - - - - - - - - - - - - - - - - - - - - - - - - - - - -
\section{Observations and Reduction}\label{sec:obs}

Below we describe observations of both \starname{} and candidate members of the parent association. The goal of the former was to characterize the host star and the candidate transiting planets. The latter set of observations focused on confirming membership and measuring the age and kinematic properties of the association. For details of how the sample of \red{$306$} candidate association members are identified, see Section~\ref{sec:assoc_members}. 

% - - - - - - - - - - - - - - - - - - - - - - - - - - - - - -
\subsection{TESS Photometry}\label{sec:tess}

We use \tess{} photometry to measure the transit properties and to measure rotation periods for the planet host and association members. We use different light curve extractions for these two purposes, as described below. 

\subsubsection{Observations of the planet host}
The planet host, \starname{} (TIC \ticid, TOI \toi), was observed by the {\it TESS} mission \citep{ricker_transiting_2015} from 2019 Apr 22 through 2019 Jun 18 (Sectors 11 and 12), then again from 2021 Apr 29 through 2021 Jun 24 (Sectors 38 and 39). The first two sectors have 30\,m cadence data, and the Sectors 38 and 39 have 20\,s, 2\,m, and 10\,m cadence data. We employed the 30\,m data from the first two sectors and the 20\,s data from the last two sectors in this work.

We retrieve light curves for the planet host and candidate members of the parent association from the Mikulski Archive for Space Telescopes (MAST\footnote{\url{https://mast.stsci.edu/portal/Mashup/Clients/Mast/Portal.html}}). For analysis of the planet host, we use the Presearch Data Conditioning Simple Aperture Photometry \citep[PDCSAP; ][]{smith_kepler_2012, stump_kepler_2012, stump_multiscale_2014} \tess{} light curve produced by the Science Processing Operations Center \citep[SPOC; ][]{jenkins_tess_2016}.

\subsubsection{Observations of association members}
The PDCSAP reduction can weaken long-term trends in the light curve, which favors exoplanet discovery, but impedes measuring rotation periods beyond $\simeq$10\,days. So, to measure the rotation periods of candidate association members, we extract light curves from \tess{} full-frame images using Causal Pixel Models \citep[CPM;][]{wang_causal_2016} with the \texttt{unpopular} package \citep{hattori_unpopular_2021}. Parameters for running \texttt{unpopular} are identical to those described in \citet{Barber_thyme_2022}. We do not extract light curves for stars that were too faint ($T>15$) or too contaminated by nearby stars (\textit{contratio} $>1.6$). In total, we are able to extract usable light curves for $203$ of the $306$ candidate members. All the \tess{} light curves used in this paper can be found in MAST: \dataset[10.17909/4cwh-0n56]{http://dx.doi.org/10.17909/4cwh-0n56}.

We use the resulting \tess{} CPM light curves to measure single-sector rotation periods for stars. For each individual star, we search for periods from $0.1$-$30$ days using a Lomb-Scargle periodogram \citep{lomb_least_1976, scargle_studies_1982, press_fast_1989}, \red{While we search up to $30$ days, only stars with  $ P_{rot} < 12$ days are considered, as the narrow observing window of \tess{} (27 days), makes longer periods unreliable.} We perform an eye check following \citet{rampalli_three_2021}, assigning Quality 0 for clear spot-modulated light curves, Quality 1 to clearly young stars with some ambiguity in their periodogram peaks, Quality 2 to spurious measurements, and Quality 3 to complete non-detections. We visually remove eclipsing binary signals and targets with blended light curves (not all get captured by our contamination ratio requirements). If multiple sectors are available, we report the average of the single-sector measurements after clipping ones that disagree with the clearest signal (strongest Lomb-Scargle power) by more than 25\% to eliminate double or half harmonics. Out of \red{$185$} candidate members with usable light curves, we assign Quality 0 or 1 to \red{$173$} stars. 

% - - - - - - - - - - - - - - - - - - - - - - - - - - - - - -                     
\subsection{High Contrast Imaging}\label{sec:phot}

We use High Contrast Imaging to search for close companion stars to \starname{}. The data we use can be found on ExoFOP-TESS (\url{https://exofop.ipac.caltech.edu/tess/target.php?id=360630575}).

\subsubsection{ZORRO/Gemini}

To search for close companion stars that might dilute the transit signal, we observed \starname{} on 2020 Mar 13 UT with the Gemini South speckle imager, Zorro \citep{scott_twin_2021}. We used the standard speckle imaging mode with narrowband $562$ nm and $832$ nm filters. The 'Alopeke-Zorro instrument team took all data as part of their program queue operations and reduced the data with their standard pipeline \citep{howell_speckle_2011}. 

No close companions are detected in either band. The $832$ nm filter sets stronger contrast limits, ruling out equal-mass companions at separations $\rho$ $>0.05\arcsec$, additional companions with $\delta 832 < 4.6 $ magnitudes at $0.1\arcsec$, and with increasing contrast sensitivity from there out to $\rho = 1.2\arcsec$.

\subsubsection{HRCam/SOAR}

We also search for previously unknown companions to \starname{} using data from the SOAR speckle imaging camera \citep[HRCam;][]{tokovinin_ten_2018} taken on 2020 Feb 10 UT. Observations were taken using the $I$ band. As with the Zorro data, we detect no companions in the HRCam data out to $\rho = 3\arcsec$. Companions with \red{$\delta I < 1$} are ruled out for $\rho > 0.1\arcsec$, and those with \red{$\delta I < 4.5 $} ruled out at $\rho > 0.25 \arcsec$.

% - - - - - - - - - - - - - - - - - - - - - - - - - - - - - -
\subsection{Spectroscopy}\label{sec:spec}

To confirm membership in the association and measure the association's age we gather spectra of candidate association members. Our goals are to estimate the equivalent width of the Li I 670.8 nm line. The extracted equivalent widths are given in Table~\ref{table:members}. 

We also obtain spectra of \starname{} over three years with the goal of checking for signs of binarity. \red{Our analysis of the resulting velocities, detailed in Section \ref{sec:false_pos} , shows no evidence of binarity.} All radial velocities, organized by instrument, are given in Table \ref{table:rvs}.

% HD 109833 RV Table
\begin{table}
    \begin{center}
    \caption{RV Measurements of \starname{}.}
    \begin{tabular}{|ccc|}
    \hline
    Telescope & BJD & $RV$ \\
    & UT & \kms \\
    \hline
    HARPS/La Silla   & 2458858.8137  & $10.563\pm0.002$ \\
    HARPS/La Silla   & 2458859.8226  & $10.598\pm0.003$ \\
    HARPS/La Silla   & 2458861.8281  & $10.571\pm0.002$ \\
    HARPS/La Silla   & 2458862.8580  & $10.614\pm0.003$ \\
    HARPS/La Silla   & 2458863.8108  & $10.548\pm0.002$ \\
    HARPS/La Silla   & 2458864.8653  & $10.577\pm0.002$ \\
    HARPS/La Silla   & 2458865.8660  & $10.603\pm0.002$ \\
    CHIRON/SMARTS    & 2459240.8846 & $10.577\pm0.052$ \\
    CHIRON/SMARTS    & 2459306.7098 & $10.568\pm0.03$  \\
    CHIRON/SMARTS    & 2459308.7235 & $10.689\pm0.064$ \\
    CHIRON/SMARTS    & 2459312.6771 & $10.621\pm0.04$  \\
    CHIRON/SMARTS    & 2459314.7458 & $10.576\pm0.042$ \\
    CHIRON/SMARTS    & 2459322.7661 & $10.528\pm0.058$ \\
    CHIRON/SMARTS    & 2459337.6306 & $10.646\pm0.083$ \\
    CHIRON/SMARTS    & 2459338.6142 & $10.558\pm0.115$ \\
    CHIRON/SMARTS    & 2459339.6936 & $10.573\pm0.053$ \\
    CHIRON/SMARTS    & 2459346.5901 & $10.571\pm0.074$ \\
    CHIRON/SMARTS    & 2459742.5434 & $10.599\pm0.037$ \\
    NRES/LCO         & 2459672.5589  & $10.757\pm0.138$ \\
    NRES/LCO         & 2459673.5576  & $10.809\pm0.203$ \\
    NRES/LCO         & 2459675.6047  & $10.917\pm0.162$ \\
    \hline
    \end{tabular}
    \tablecomments{Measured radial velocities of \starname{}. A $1.4 $ \kms{} offset was added to the measurements from CHIRON/SMARTS in accordance to the zero-point of that instrument.}
    \label{table:rvs}
    \end{center}
\end{table}

\subsubsection{Goodman/SOAR Spectroscopy}\label{sec:goodmanspec}
We observe a total of $26$ candidate association members using the Goodman High Throughput Spectrograph \citep{clemens_goodman_2004}. Goodman is part of the Southern Astrophysical Research Telescope (SOAR) atop Cerro Pachon, Chile. Observations were taken over 8 nights between 2021 Mar 29 and 2021 Sept 20, under mostly photometric conditions. 

From the list of candidate association members, these 26 are selected for observation with the goal of mapping out the lithium depletion boundary (LDB). To choose stars to observe we estimate the age of the association from an isochrone, and use that age to predict the magnitude of the LDB in $M_{Ks}$. We then select stars with $\Delta K<1$ from the predicted boundary. This estimate was updated as we took more data and revised the age of the group, so there was no single observing list. We choose stars for observing based on their magnitude in \gaia{} $R_P$ (prioritizing brighter stars that need shorter exposures), \gaia{} RUWE (omitting stars with $RUWE > 1.2$ as they are more likely to be binaries), and location on sky (prioritizing short slews between targets and middle elevations).

Observations were designed to measure the EW of the Li I $670.8$nm line; we use the red camera, the 1200 l/mm grating, and the M5 mode, which provides a wavelength coverage of $630-740$nm. We use either the 0.45" slit, or the 0.6" slit, depending on the magnitude of the target and atmospheric seeing. This setup should give a resolution of $R=4500-5800$, although in practice the true resolution is lower and varies with exposure time (see below). For each target, we take five spectra with exposure times varying from 10\,s to 300\,s each.

For reduction, we perform standard bias subtraction, flat-fielding, and optimal extraction of the target spectrum. The spectra show large wavelength shifts while observing, likely due to issues with the mount model and flexure compensation system. In extreme cases, this shifts the spectrum by 5-10\,pixels between exposures of the same target, which corresponds to several resolving elements (depending on the slit). To mitigate the effect, we take Ne arcs prior to each target and use simultaneous skyline spectra to calibrate the wavelength solution of individual spectra. The combination of a pre-target arc and skylines performs better than bracketing the data with arcs. We make an initial map of pixels to wavelengths using a fourth-order polynomial derived from the nearest Ne arc, then apply a linear correction to each spectrum based on the sky lines. We stack the extracted and wavelength calibrated spectra using a robust weighted mean. The stacked spectra have mean $SNR > 40$ for all targets.

We correct each star to its rest wavelength using radial-velocity standards taken with the same setup. Although the resulting spectra were sufficient for spectral typing and measuring relevant equivalent widths (e.g., EW[Li]), we find the radial velocities to be poor ($\sigma_{\rm{rv}}\simeq5-10$\kms\ based on stars with known velocity). This is likely due to non-linearity in the wavelength shifts impacting the edges of the spectrum and regions with fewer sky lines and non-uniform shifts during an exposure. As a result, we do not report velocities based on these spectra.

% - - - - - - - - - - - - - - - - - -
\subsubsection{NRES/LCO}\label{sec:lco}

To increase the baseline of our RV characterization of \starname{}, we obtain three spectra of \starname{} using the Network of Robotic Echelle Spectrographs (NRES)  \citep{siverd_nres_2018} at the Las Cumbres Observatory. Observations were taken the nights of 2022 Apr 3, 4, and 5. 

NRES spectra cover $380-860nm$ at high resolution ($R\sim 53,000$). The data are reduced using the LCO NRES pipeline \texttt{BANZAI-NRES}\footnote{\url{https://github.com/LCOGT/banzai-nres}}. This includes extraction of radial velocities by cross-correlating observed spectra with PHOENIX model atmospheres \citep{husser_new_2013}.

% - - - - - - - - - - - - - - - - - - - - - - - - - - - - - -
\subsubsection{HARPS}

For RV characterization of the planet host \starname, we also obtain seven spectra of that star taken with the High Accuracy Radial velocity Planet Searcher (HARPS) fiber-fed Echelle Spectrograph on the ESO 3.6m telescope at La Silla Observatory under the NCORES large programme (ID 1102.C-0249, PI: Armstrong). The spectra are high-resolution ($R\sim 115000$), and cover a spectral range of $378$nm$-691$nm. Observations were taken on the nights of 2020 Jan 10-11 and 13-17 in high-accuracy mode (HAM), with an exposure time of 1500-1800s, depending on observing conditions, and a typical SNR per pixel of 100. The standard online HARPS data reduction pipeline reduces the data, using a G2 template to form the weighted cross-correlation function (CCF) to determine the radial velocities (RVs). We find a typical error on the RVs of 2-3\mps. 

% - - - - -- - - - - - - - - - - - - - - - - - - - - - - - -
\subsubsection{CHIRON/SMARTS} \label{subsec:chiron}

We acquire twelve spectra of \starname{} using CHIRON at the SMARTS 1.5\,m telescope at Cerro Tololo Inter-American Observatory \citep{tokovinin_chironfiber_2013}. These observations were acquired between 2021 Jan 26 and 2022 June 12. We use CHIRON in its image slicer mode, which gives a resolution of $\approx 79{,}000$ across $415--880$nm.  

We also acquire spectra of two \assoc{} candidate members on 2021 Apr 30, and 2021 May 4 using the same setup and reduction. 

To derive the radial velocities and stellar parameters for the eleven spectra that met our signal-to-noise requirements, we follow the methods described in \citet{zhou_warm_2018}. We perform a least-squares deconvolution of the spectra using non-rotating synthetic spectral templates \citep{donati_spectropolarimetric_1997}. These templates are constructed using the ATLAS9 atmosphere models \citep{castelli_new_2004} and the \texttt{SPECTRUM} script \citep{gray_calibration_1994}.  The resulting line profiles were fit using a broadening kernel that included terms for the rotational, macroturbulent, and instrumental broadening. We then fit the line profile from each observation independently, yielding the radial velocities listed in Table~\ref{table:rvs}, as well as a mean rotational broadening velocity of $v\sin i_\star = 10.5 \pm 0.2 {\rm km\,s}^{-1}$.

% - - - - - - - - - - - - - - - - - - - - - - - - - - - - - - -
% - - - - - - - - - - - - - - - - - - - - - - - - - - - - - - -
\section{Properties of the \assoc{} Association}\label{sec:assoc_properties}

\subsection{\assoc{} in the context of nearby associations}\label{sec:kinematics}

\begin{figure*}[tb]
    \centering
    \includegraphics[width=0.99\textwidth]{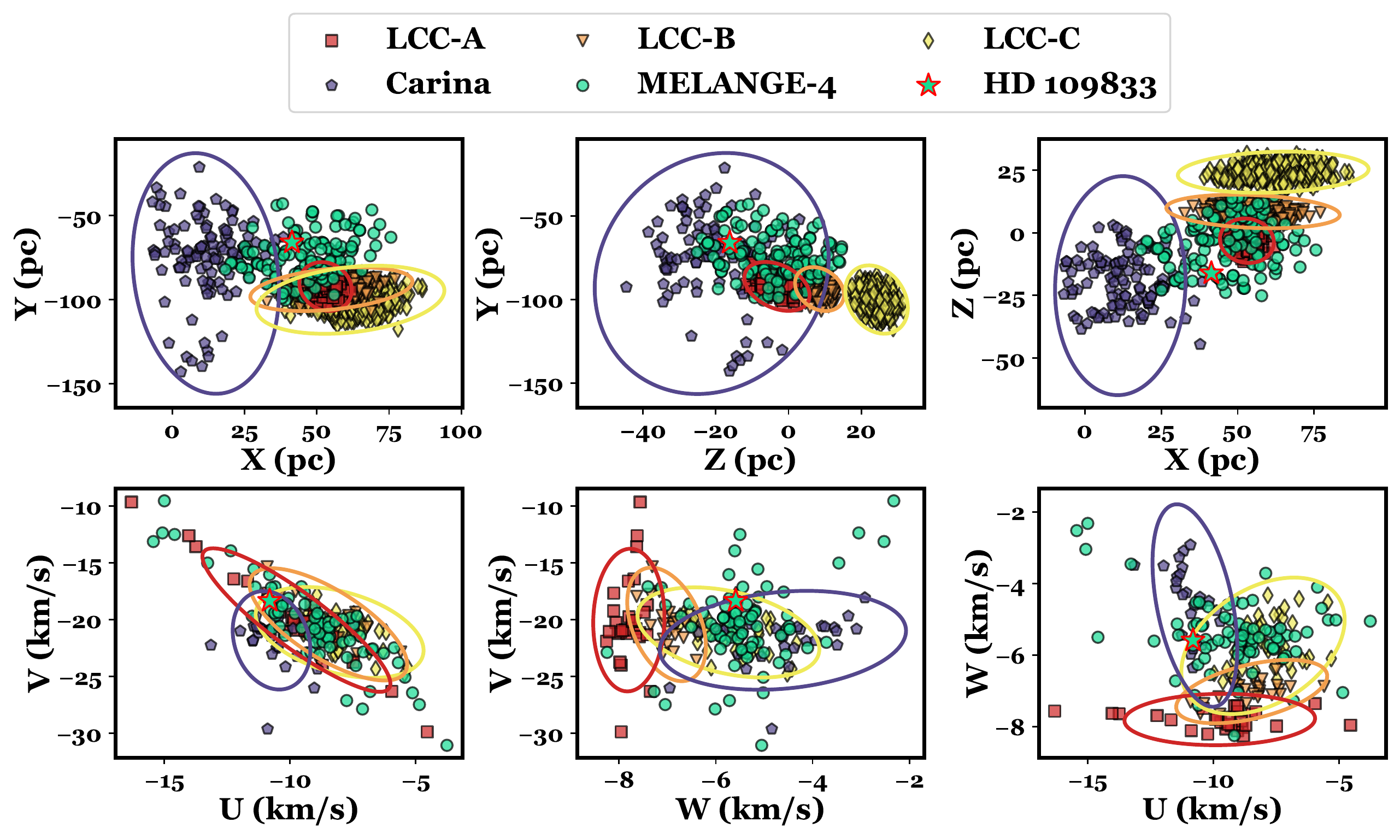}
    \caption{The spatial and velocity spreads of \assoc{}, the nearby LCC sub-populations, and Carina. The top row shows the galactic position (X, Y, and Z) for each population, and the bottom row the velocities (U, V, and W). The membership candidates for \assoc{} are shown in turquoise, and high probability ($>99\%$) members of LCC-A,  LCC-B, LCC-C, and Carina are shown in red, orange, yellow, and purple respectively. The position of \starname{} is marked with a turquoise star.}
    \label{fig:xyz}
\end{figure*}

The central position and velocity of \assoc{} is near the LCC population on the southern part of Sco-Cen and is on the western edge of the Carina association. Despite its proximity to these populations, the positions and velocities of members make it clear that \assoc{} is not part of any known associations. This is demonstrated in Figure~\ref{fig:xyz} and described in further detail below. 

% Figure: XYZ and UVW of the association and nearby populations, showing ellipses and scatterplot members; Updated 20220804

Recent research has split LCC into 4 sub-populations \citep{goldman_large_2018, kerr_stars_2021}, which we consider separately. \citet{goldman_large_2018} (G18) found four sub-populations: LCC-A0, LCC-A, LCC-B, and LCC-C, from youngest to oldest. The same sub-populations were found by \citet{kerr_stars_2021} (K21), where they were named LCC-B, LCC-C, LCC-E, and LCC-D, respectively. We use the names from G18, and focus on LCC-A, LCC-B, and LCC-C, as these are the most well-defined. LCC-A0, which is renamed as the Musca association by \citet{mann_tess_2022}, is the youngest of the sub-populations ($11\pm2$Myr), and the furthest from \assoc{}. See Appendix \ref{app:lcc} for details on these sub-populations.

\assoc{} shows significant spatial overlap with LCC-A and partial spatial overlap with LCC-B, but is completely separate from both sub-populations in $W$. In terms of velocity, it overlaps partially with LCC-C, from which it is furthest in $XYZ$. Only 3-4 stars in our list of candidate members of \assoc{} overlap with a single LCC group in both $XYZ$ and $UVW$ (the sources with the most negative $W$ in Figure~\ref{fig:xyz}). As expected, these have the lowest membership probabilities (50-60\%). As we show later in this Section, the age of \assoc{} is also significantly older than the closest LCC sub-groups. 

In addition to the LCC subgroups, \assoc{} has kinematic overlap with the Carina association. However, the mean $X$ of \assoc{} is $\simeq 40 \,{\rm pc}$ from the center of Carina. Similarly, most age estimates for Carina find an age of $\simeq40\,{\rm Myr}$ \citep[e.g.,][]{torres_young_2008, bell_self-consistent_2015, wood_carina_prep} which are inconsistent with the age we find for \assoc{}. \citet{booth_age_2021} find a younger age for Carina ($\simeq$15\,Myr), but this is still inconsistent with our age of \assoc. 

One thing that stands out is how diffuse the group is in $XYZ$ space compared to known LCC populations. This makes it look like a group transitioning from the more tightly packed population within large complexes towards the more diffuse moving groups \citep[e.g.,][]{kraus_greater_2017}. The broad distribution can explain why the group was not noticed prior to the arrival of \gaia{} data, as well as why many members were previously thought to be part of LCC or Carina.

% - - - - - - - - - - - - - - - - - - - - - - - - - - - - - - -
\subsection{Age from rotation} \label{sec:rotation}

The rotation sequence can be used to constrain the age of a population \citep[e.g.][]{tofflemire_tess_2021, andrews_young_2022, Newton_tess_2022, Messina_gyro_2022}. For $<100$\,Myr associations, this is more challenging because many late-type stars are still spinning up as they contract onto the main sequence and Sun-like or warmer stars have not yet moved onto the slow-rotating sequence \citep{rebull_rotation_2018}. However, the {\it spread} in rotation periods within a group is still a useful proxy for age. For example, the period spread in 10\,Myr Upper Sco is much greater than for 40-60\,Myr associations like Tucana-Horologium and $\mu$\,Tau \citep{gagne_mu_2020} because the rotation spread at 10\,Myr is driven mostly by initial rotation rather than sculpting effects. 

\begin{figure}[tb] % Updated with Gaia DR3 results
    \centering
    \includegraphics[width=0.46\textwidth]{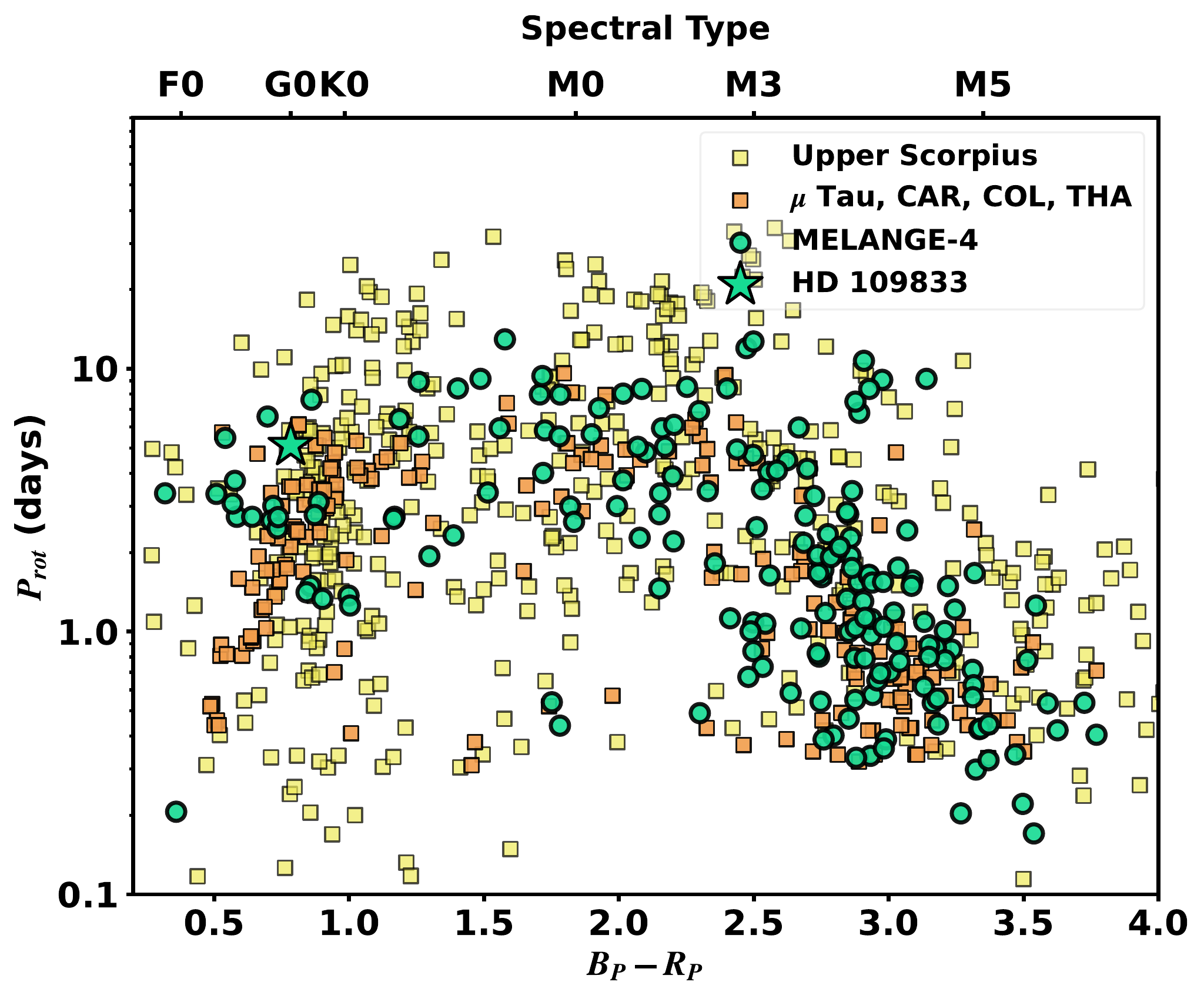}
    \caption{Rotation periods of candidate members of \assoc{} as a function of \gaia{} $B_P-R_P$ color. We compare the color-period sequence of candidate members to the younger ($\sim 10$ Myr) population of Upper Scorpius \citep{rebull_rotation_2018} and older ($45$-$60$ Myr) populations of Carina (CAR), Columba (COL), Tucana-Horologium (THA), and $\mu$ Tau \citep{gagne_mu_2020}. As expected, the \assoc{} sequence shows more coherence than the sequence of the younger Upper Scorpius. The low-mass candidates rotate slower than those in the $45$-$60$ Myr populations, indicating a younger population that will spin up as they contract onto the main sequence. The high-mass candidates are a mix of still rapidly-rotating stars and stars that have begun to spin down.}
    \label{fig:rotation}
\end{figure}

We show the rotation distribution of candidate members of \assoc{} with Quality 0 or 1 periods alongside other young populations in Figure~\ref{fig:rotation}. As expected, low-mass members are rotating slower than slightly older stars as they are still spinning up while contracting towards the main sequence. Similarly, the higher-mass members contain a mix of rapidly-rotating stars and those that have started to move to the slow-rotating sequence. The overall rotation distribution is consistent with a $20$-$40$\,Myr population. 

% - - - - - - - - - - - - - - - - - - - - - - - - - - - - - - -
\subsection{Age from Lithium Depletion}\label{sec:lithium}

Li is rapidly burned in the core of most stars, and for fully convective stars on the main sequence (early M and later) there is an age-dependent cut-off in $T_{eff}$ above which all Li has been consumed and below which the stars retain nearly all initial lithium known as the lithium depletion boundary (LDB). The LDB is sharp, generally providing a more precise age measurement than isochrones or rotation \citep{burke_theoretical_2004}. The location of the LDB is less sensitive to model assumptions, which makes the LDB a ``semi-fundamental'' age estimator \citep{soderblom_ages_2014}. 

\begin{figure}[tb]
    \centering
    \includegraphics[width=0.48\textwidth]{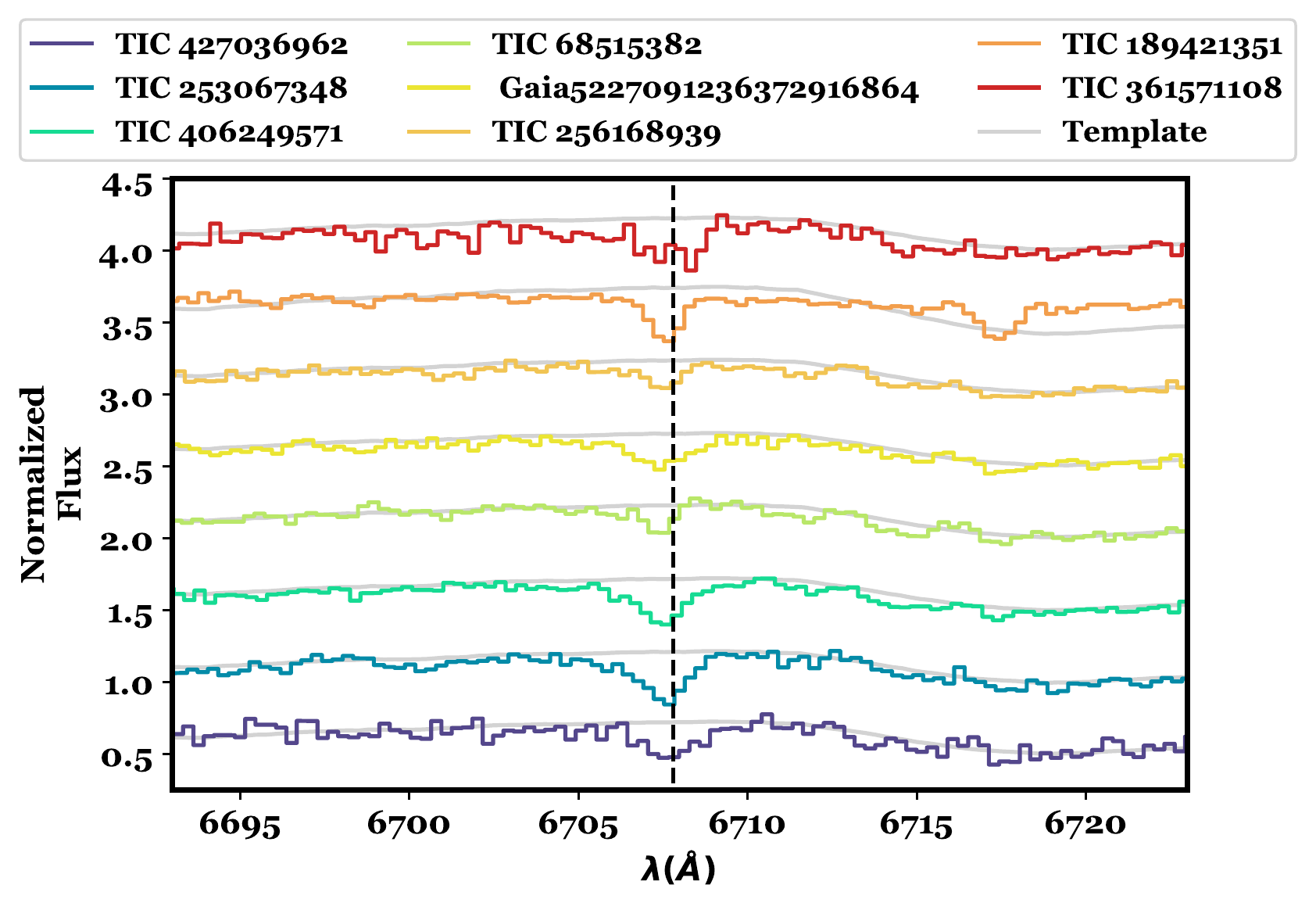}
    \caption{Spectra of all association members with measured Li absorption, overplotted with a Li free M-dwarf template \citep[gray;][]{bochanski_low-mass_2007}. The location of the Li line, $6707.8$\,\AA, is marked with a dashed black line. Two of these stars, TIC 68515382 (light green) and TIC 256168939 (light orange), have $EW(Li) < 300\,$m\AA, and thus are not fully Li-depleted.}
    \label{fig:spectra}
\end{figure}

\begin{figure*}[tb]
    \centering
    \includegraphics[width=0.45\linewidth]{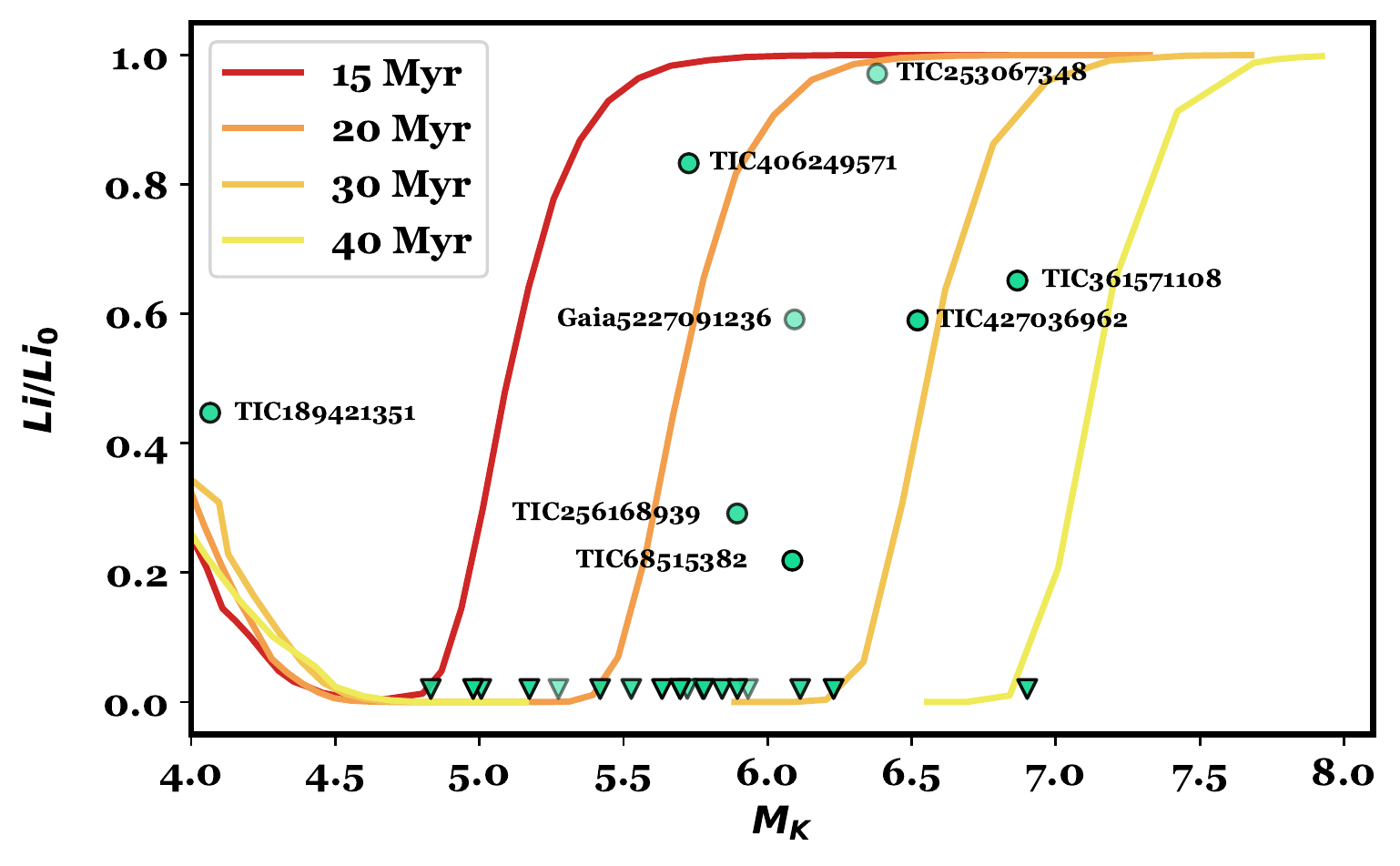}
    \includegraphics[width=0.45\linewidth]{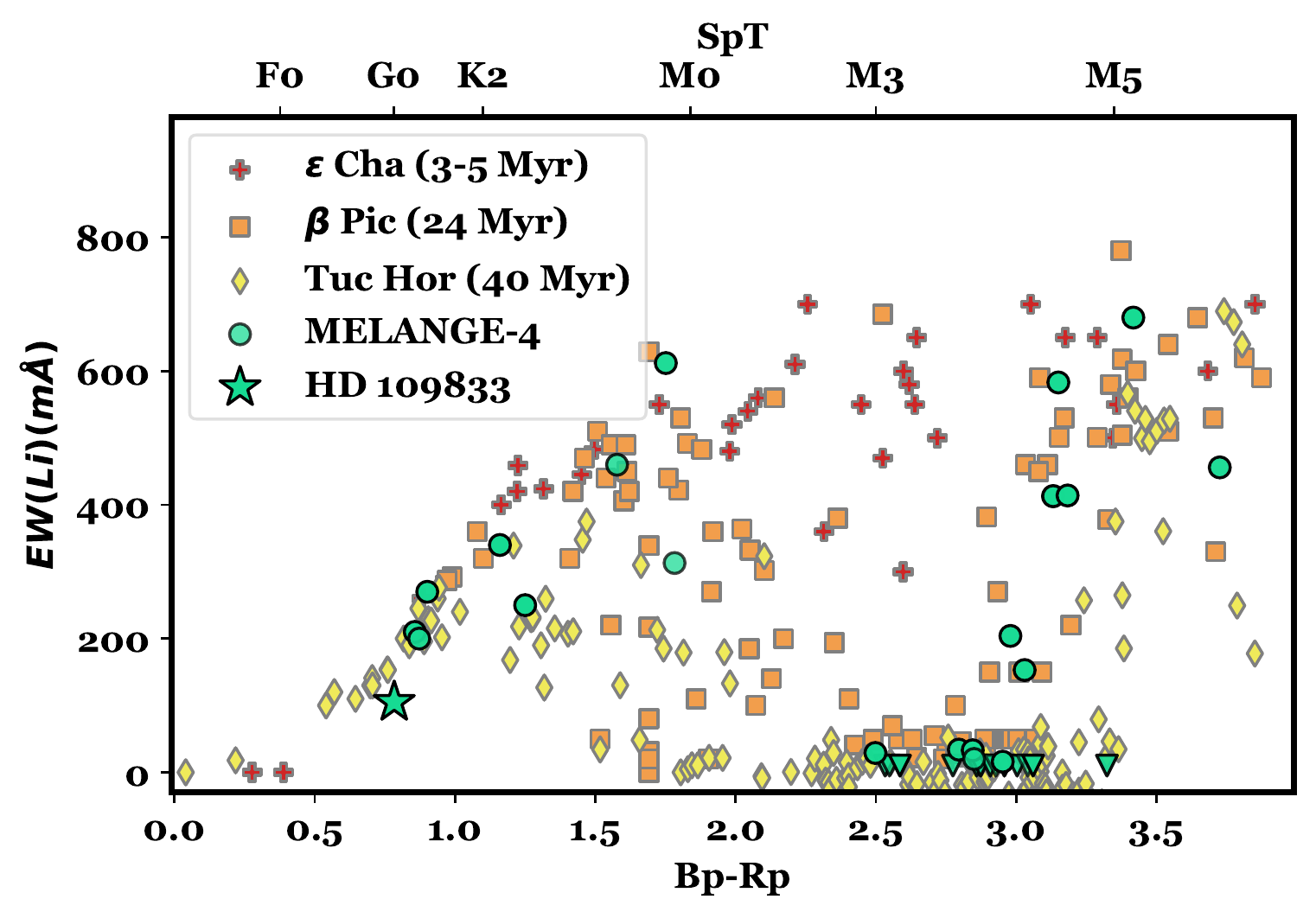} 
    \caption{Lithium measurements for \assoc{}.
    Left) Fraction of initial Li abundance as a function of absolute K magnitude, overplotted with DSEP magnetic isochrones \citep{feiden_magnetic_2016}. Equivalent widths of association members were converted to $Li/Li_0$ using the curve of growth from \cite{zapatero_osorio_lithium_2002}. The transparency of the points corresponds to their kinetic membership probability from \banyan{}, such that the most likely interlopers are the faintest. 
    Right) Comparison of \assoc{} Li sequence and Li sequences of known associations. For members of each association, Li equivalent widths are shown as a function of \gaia{} Bp-Rp. Equivalent widths of \assoc{} members marked with downward triangles are upper limits. The transparency of the points corresponds to their kinetic membership probability from \banyan{}, such that the most likely interlopers are the faintest. Li measurements of \assoc{} members are supplemented with additional five measurements from \citet{mamajek_post_2002, torres_vizier_2006}, and \citet{pecaut_star_2016}. The Li measurements of $\beta$ Pic members are taken from \citet{shkolnik_all-sky_2017}, measurements of \epscha{} members from \citet{murphy_re-examining_2013}, and measurements for Tuc-Hor from \citet{da_silva_search_2009} and \citet{kraus_stellar_2014}}.
    \label{fig:li_sequence}
\end{figure*}

We measure the LDB age of the association using the SOAR/Goodman spectra described in Section \ref{sec:goodmanspec}. We estimate the equivalent width of the Li I 670.8nm line (EW[Li]), using a pseudo-continuum estimate from a linear fit to the region on either side of the line. To account for variations in the line width (\vsini{} and resolution differences between spectra) we manually adjust the width of Li region to include both line edges. \red{Spectra of all association members with measured Li absorption are shown in Figure \ref{fig:spectra}.} Our analysis does not account for contamination from the nearby Fe line ($6707.4$\,\AA) in the FGK stars nor molecular contamination of the continuum in the cooler M dwarfs. As a result, the uncertainties are likely no better than 10\% independent of SNR. The EWs are listed in Table \ref{table:obs} and plotted in Figure \ref{fig:li_sequence}.

\begin{figure}[tb]
    \centering
    \includegraphics[width=0.48\textwidth]{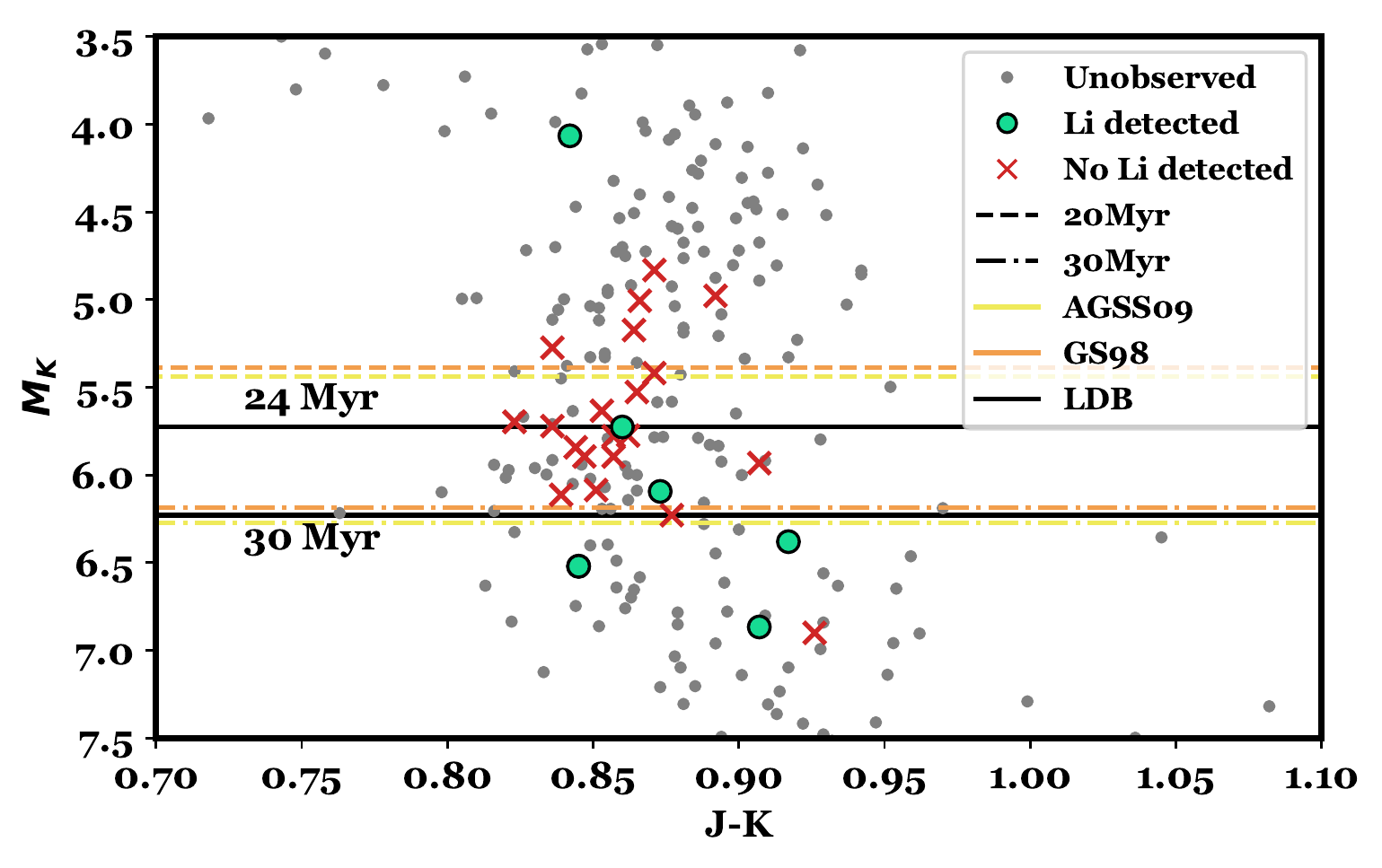}
    \caption{The Lithium Depletion Boundary of \assoc. Shown is the  color (J-K) and absolute magnitude ($K_s$) of members of \assoc{}. Observed stars with $EW(Li) > 300$ \,m\AA\ are shown as colored dots, and stars with $EW(Li) < 300$\,m\AA\ as red X's. Candidate association members which were not observed are shown as gray dots. Orange and yellow lines show the predicted $20$ and $30$\,Myr $99\%$ lithium depletion boundaries using the DSEP magnetic models with two different solar abundances. The $30$ \,Myr GS98 line has been moved up by 0.03 magnitudes to increase visibility. The solid black lines through the brightest M-dwarf with Li absorption, and the faintest likely member without Li absorption show the edges of the boundary region.}
    \label{fig:ldb}
\end{figure} 

To locate the LDB we must define a threshold between Li-rich and Li-poor stars. We have chosen a threshold of EW[Li]$\ge300 m\AA$, following the reasoning of \citet{binks_gaia-eso_2021}, and using the curve of growth from \citet{zapatero_osorio_lithium_2002}. As a proxy for stellar mass or \teff, we use the absolute $K_S$-band magnitudes from 2MASS with parallaxes from \gaia{} DR3. $K_S$ is less sensitive to metallicity \citep{mann_how_2019}, reddening, and spots \citep{somers_spots_2020} than \gaia{} colors, and is broadly available for our targets. 

As expected, we identify a region above which no Li is present and below which all high-probability members have Li \red{(see Figure \ref{fig:ldb})}. The single star with a Li detection at $M_K=4$ (TIC 189421351) is more massive than the other stars and is expected to have undepleted Li (see Figure \ref{fig:li_sequence}). The star with no detected Li and $M_K=6.9$ (TIC 443273186) has a low CMD position and is a likely field interloper, so we exclude it from this analysis. The upper boundary of this region is defined as the magnitude of the brightest star which we have observed to have $pEW(Li) > 300$\,m\AA, and the lower boundary by the magnitude of the faintest star with $pEW(Li) < 300$\,m\AA. Using that definition, we find that the LDB region spans $ 5.7 < M_K < 6.1$. Using a more generous definition of 200\,m\AA{} for Li-rich, \citep[e.g.,][]{binks_lithium_2014} adds one additional Li-rich star at $M_K = 5.9$. This did not affect our LDB region or measured age. 

We determine the age of \assoc{} by comparing this LDB location to several stellar evolution models with varying assumptions (e.g., magnetic field strength and spot coverage). For each model, we define the LDB as the magnitude at which Li has been depleted from the initial amount by $99\%$. We use isochrones taken from \citet{baraffe_new_2015} (BHAC15), and the Dartmouth Stellar Evolution Program \citep[DSEP;][]{dotter_dartmouth_2008} as our baseline models. For the treatment of magnetic fields, we use isochrones from \cite{feiden_magnetic_2016}, which are built on top of the DSEP models. The magnetic DSEP models include grids built on two different sets of solar abundances, from \citet{grevesse_standard_1998} (GS98), and \citet{asplund_chemical_2009} (AGSS09). These different abundance scales produce slightly different predictions for Li depletion, leading to different LDB ages, as shown in Table \ref{table:ldb_ages}. For spots, we used the spot models from \citet{somers_garrett_2019, somers_spots_2020}.

We list the resulting age bounds from each model in Table \ref{table:ldb_ages}. The age ranges from each model are broadly consistent, with lower age bounds ranging from 23 Myr to 26 Myr, and upper age bounds between 28 Myr and 32 Myr. The largest age discrepancies come from using the SPOT models with spot fractions $>$50\%. While individual stars may have large spot fractions \citep{Gully-Santiago_spots_2017}, we expect the bulk of the stars here to have spot fractions $\lesssim30\%$ \citep{cao_age_2022, savanov_activity_2018, klein_one_2022}. Thus, we include the ages from the SPOT model of only $17\%$ and $34\%$ in the table, both of which are consistent with the ages from the other models tested.

\begin{table}[ht!]
    \centering
    \begin{tabular}{|c|c|c|}
     \hline
     Model & Lower Bound & Upper Bound\\ 
     \hline
    BHAC15            & 24 Myr & 28 Myr \\
    DSEP (GS98)       & 23 Myr & 29 Myr \\
    DSEP Mag (GS98)   & 24 Myr & 30 Myr \\
    DSEP Mag (AGSS09) & 26 Myr & 32 Myr \\
    SPOT (17\%)       & 23 Myr & 31 Myr \\
    SPOT (34\%)       & 24 Myr & 31 Myr \\
     \hline
    \end{tabular}
    \caption{Upper and lower age bounds given by each of the models used. The upper bound corresponds to the age given an LDB at the magnitude of the faintest observed star without Li, and the lower bound corresponds to the age given an LDB at the magnitude of the brightest observed stars with Li.}.
    \label{table:ldb_ages}
\end{table}

We take $27\pm3$ Myr as the age of the association to encompass all these estimates.

A number of effects could cause Li poor stars to appear above the LDB or vice versa. Unresolved binary stars which are Li-rich could appear to be as much as $0.75$ magnitudes brighter than the individual components, raising them on the CMD to look younger. 
\citet{baraffe_effect_2010} suggested that cold, episodic accretion onto young low-mass stars could cause early Li depletion in individual stars, leading a star to look older, but \citet{sergison_no_2013} find no evidence for this in two young associations, and it should not impact the age estimates here because we have multiple reliable detections. Poor parallaxes (e.g. on binaries), and stellar variability may also affect the CMD position at the $\lesssim0.1$~mag level, which can explain some of the spread.

Likely the largest cause of anomalous stars is non-member interlopers, either younger interlopers from the nearby LCC populations or older interlopers from the field. A possible example of this is the star TIC 443273186, which is the lowest-luminosity star we observed at $ M_K = 6.9$, but has no significant Li detection. This star has a low CMD position compared to other association candidate members, and is likely a field interloper rather than an association member.

While the LDB in low-mass stars is the most accurate method of using Li measurements to determine association age, it is also possible to estimate an association's age by examining the full sequence of Li abundance as a function of color \citep[see ][for a review]{soderblom_ages_2014}. Because this method uses Li abundance, rather than the simple threshold used by the LDB, and requires conversion between modeled A(Li) and measured EW(Li), it is more model-dependent than the LDB method, but serves here as an additional check on the association age.

First, we compare the magnetic DSEP stellar evolution models \citep{feiden_magnetic_2016} against the Li measurements of \assoc{} M-dwarf members, shown in the left panel of Figure \ref{fig:li_sequence}. We convert EW(Li) to $Li/Li_{0}$ by dividing each by the predicted initial EW(Li) for M-dwarfs from \citet[][; $700$\,m\AA]{zapatero_osorio_lithium_2002}. The measured values lie between the models for a $20$ and $30$ Myr association.
Next, we compare the Li sequence of \assoc{} against that of three benchmark associations with ages ranging from $3 - 40$ Myr. We supplement our EW(Li) measurements of \assoc{} members (see Section \ref{sec:spec} and Table \ref{table:obs}) with literature measurements for five higher-mass stars, taken from \citet{mamajek_post_2002, torres_vizier_2006}, and \citet{pecaut_star_2016}. The Li sequence of \assoc{} lies on top of that of the 24 Myr old $\beta$ Pic association, shown in orange in the right panel of Figure \ref{fig:li_sequence}. The association has more Li at $B_P-R_P \simeq 3.0$ than the older Tuc-Hor association (\red{$40\,Myr$}), and lower Li (more depletion) at $ 1.7 < B_P-R_P < 3.0 $ when compared to the younger 3-5 Myr \epscha{} association.
Both of these tests support our measured LDB age of $27\pm3$ Myr.

% - - - - - - - - - - - - - - - - - - - - - - - - - - - - - - -
\subsection{Age from Isochrones}\label{sec:isochrone}

We independently estimate the age of \assoc{} by comparing the CMD to solar-metallicity isochrones using a Gaussian mixture model. For this analysis, we use the solar-metallicity PARSEC (v1.2S) models \citep{bressan_parsec_2012} rather than one of the models used in the lithium analysis (Section \ref{sec:lithium}), as those models do not reach the highest mass association members critical for differentiating between ages. Following \citet{mann_tess_2022}, we use a mixture model \footnote{\url{https://github.com/awmann/mixtureages}}, based on the method outlined in \citet{hogg_data_2010} and a Monte-Carlo Markov-Chain framework with \texttt{emcee} \citep{foreman-mackey_fast_2017}. The basic method is to fit the population with the combination of two models, one describing the single-star sequence of members, and one describing everything else (outliers). The second population may itself contain multiple populations, such as binaries, field interlopers, and young stars in Sco-Cen but not part of \assoc.

The fit has six free parameters (units in brackets): the association age ($\tau$ [myr]), the average reddening across the association ($E(B-V)$ [mags]), the amplitude of the outlier population ($P_B$), the offset of the outlier population from the main population CMD ($Y_B$ [mags]), the variance of the outliers around the mean ($V_B$ [mags]), and a term to capture missing uncertainties or differential reddening across the association ($f$ [mags]). All parameters evolve under uniform priors bounded by physical barriers, although E(B-V) is allowed to go negative to avoid Lucy-Sweeney bias. We re-sample the isochrone grid to ensure uniform sampling in age. We run the MCMC with 50 walkers until it passed at least 50 times the autocorrelation time after a burn-in of 5,000 steps (a total of 30,000 steps). 

We perform the comparison using \gaia{} photometry and parallaxes for the final membership list described in Section~\ref{sec:mem_final}. While the mixture model can handle outliers, it can be sensitive to multiple kinds of outliers as we expect here (binaries, LCC members, field stars, targets with poor parallaxes or photometry). So we remove stars with RUWE$>1.4$ \citep[likely to be binaries;][]{ziegler_soar_2019, wood_characterizing_2021}, stars with SNR$<$30 in their parallax or any photometry, and any target outside the range of our model grid. This reduces the list of stars to 219. 

\begin{figure}[tb]
    \centering
    \includegraphics[width=0.49\textwidth]{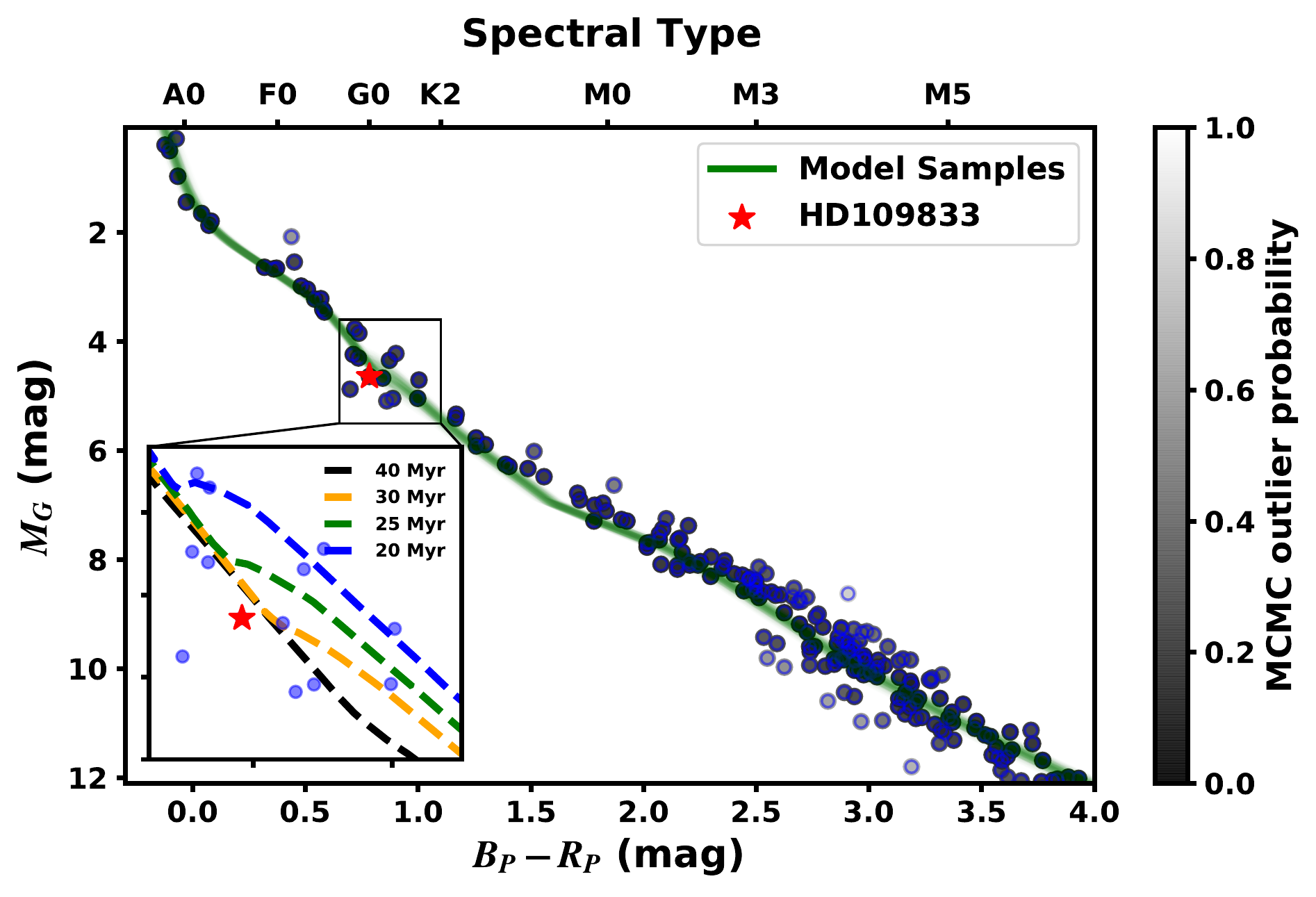}
    \caption{Comparison of the PARSEC isochrones to high-probability members of \assoc{} (circles). Stars are shaded by their outlier probability, as determined by the mixture model fit. Outliers are not necessarily non-members. The red star shows the planet host (\starname). The green lines are 100 random draws from the MCMC using the PARSEC isochrones and fit reddening. The inset shows the region around \starname{} and representative-age model predictions. }
    \label{fig:isochrone}
\end{figure} 

As we show in Figure~\ref{fig:isochrone}, the isochrone fits the sequence relatively well. One region of disagreement are the under-luminous G dwarfs around $B_P-R_P\simeq0.9$, which includes \starname. At 20-30\,Myr, this part of the CMD corresponds to stars' transition into He-3 burning, which leads to a rapid drop in the brightness of the star during the transition period. The resulting bend in the CMD is seen in the models as well as the similar-aged $\beta$~Pic \citep{mamajek_age_2014}. The fit is marginally consistent with the observations of \starname, but the three stars redward look like main-sequence interlopers. \starname's CMD position is an excellent match to the 30\,Myr (and older) isochrones, so the discrepancy may be due to a modest (2-3\,Myr) age spread, or other systematics in the models. Another possibility is that \starname{} is a young field interloper, which we discuss further in Section~\ref{sec:conclusion}. 

The best-fit age from our fit is $26.0\pm2.1$\,Myr. Repeating the analysis with the Dartmouth Stellar Evolution Program \citep[DSEP;][]{dotter_dartmouth_2008} with magnetic enhancement \citep{feiden_magnetic_2016} yields a consistent but slightly older age of $27.1\pm2.3$\,Myr. We use the age from the PARSEC fit because the DSEP models did not include stars above 1.7$M_\odot$ at this age. These high-mass stars provide an age constraint independent of the LDB age (which relies on mid-M dwarfs).  

Fitting the full population without the cuts above gives a younger age ($23\pm3$\,Myr). However, when using the full sample the outlier model fits the main-sequence interlopers, treating binaries and LCC interlopers as part of the main population (both of which bias the fit to younger ages). Similar small adjustments to the fitting method, such as non-solar composition, also change the resulting age at the $\simeq2$\,Myr level, generally preferring older ages. This suggests systematic errors are comparable to measurement errors. All ages are in excellent agreement with our $27\pm3$\,Myr age from the LDB (Section~\ref{sec:lithium}).

 % - - - - - - - - - - - - - - - - - - - - - - - - - - - - - - -
\section{Directly imaged planets in \assoc}\label{sec:direct_imaging} % Updated w/ DR3 probs 2022-08-08

Along with the newly-identified planet host \starname{} (discussed in Sections \ref{sec:stellar_properties} \& \ref{sec:planet_params}), three other candidate members of \assoc{} host four planetary-mass objects previously identified through direct imaging surveys. These systems represent a non-trivial fraction of all the directly imaged planetary-mass companions \citep{currie_directimaging_2022}, so a change in their age (and hence derived masses) could impact population-level statistics. 

To update the masses of the directly imaged planets, we analyze the reported luminosity or absolute magnitude in the original papers and compare them to the weak-non-equilibrium ATMO2020 models of \citet{phillips_atmosphere_2020} using linear interpolation. We first verify that when using the originally asserted ages, we recover consistent masses, so any difference is primarily a result of the revised (older) age. 

% - - - - - - - - - - - - - - - - 
\subsection{TYC 8998-760-1 (YSES 1)}
TYC 8998-760-1 is a young K-dwarf type star with two directly imaged, wide companions from the Young Suns Exoplanet Survey (YSES) direct imaging survey \citep{bohn_young_2020, bohn_two_2020}. The star has been classified as a $\simeq16$ Myr-old member of LCC  \citep[e.g.,][]{pecaut_star_2016}. However, using our updated \banyan{} parameters we find that it is a high-probability candidate member of \assoc{}. The star's position, proper motions, and velocity are near the center of \assoc{}, with a \banyan{} membership probability of $99.1\%$.

Assuming an age of $16.7\pm1.4$ Myr, \citet{bohn_young_2020, bohn_two_2020} measured masses of $14\pm3 M_{Jup}$ and $6\pm1 M_{Jup}$ for planets YSES-1 b and YSES-1 c, respectively. Using the age of MELANGE-4 ($27\pm3$ Myr, see Section \ref{sec:assoc_properties}), we estimate masses of $21.8\pm3 M_{Jup}$, and $7.2\pm0.7 M_{Jup}$. As expected, the new masses are much larger than the earlier values, although still ($<3\sigma$) consistent due to large uncertainties on the nearly vertical evolution of late-type pre-MS stars and brown dwarfs. 

% DR3 Prob -> 99.1%

% - - - - - - - - - - - - - - - - 
\subsection{TYC 8984-2245-1 (YSES 2)}

Like TYC 8998-760-1, TYC 8984-2245-1 is a young K-dwarf with a directly-imaged companion observed by the YSES survey \citep{bohn_discovery_2021}.
Several surveys of LCC have previously included TYC 8984-2245-1 as a member \citep[e.g.,][]{preibisch_nearest_2008, gagne_banyan_2018}. However, using the \gaia{} DR3 RV of $12.93$\kms, we find that it is a better match for \assoc{}, with a kinematic membership probability of $94.8\%$.

We re-estimate the mass of YSES-2 b using the H and K magnitudes and find a new mass of $8.4\pm1.5 M_{Jup}$. Again, this is a higher mass but still consistent with the prior estimate of $6.3^{+1.3}_{-0.9} M_{Jup}$ \citep{bohn_discovery_2021}.

% DR3 Prob -> 94.8

% - - - - - - - - - - - - - - - - 
\subsection{HD 95086}
HD 95086 is an A8 pre-MS star, with both a directly imaged planet and an imaged debris disk \citep{rameau_confirmation_2013,moor_resolved_2013}.  \citet{de_zeeuw_hipparcos_1999} and \citet{rizzuto_multidimensional_2011} both consider it to be a member of the LCC population, but \citet{booth_age_2021} argue that HD 95086 is instead a member of Carina, simultaneously proposing a younger (17\,Myr) age for Carina.

The RV of HD 95086 has been measured multiple times by different sources; all of them strongly favor membership in \assoc{}. \citet{madsen_astrometric_2002} estimate an astrometric RV of $10.1\pm1.2$\kms, which gives a \banyan{} membership probability of $98.7\%$ for \assoc{}, a $1.3\%$ probability of being a field star, and $< 1e-4\% $ for any of the LCC groups or Carina. 
\citet{moor_resolved_2013} measure $RV=17\pm2$\,\kms, with which the probability of membership in \assoc{} increases to $99.1\% $, with a probability of $0.66\%$ for Carina and $< 1e-4 \%$ for all of the LCC groups.
The \gaia{} DR3 velocity is a similar $RV=18.04\pm0.16$\,\kms, which gives a membership probability of $98.9\%$ for \assoc, $0.77\%$ for membership in Carina, and $0.30\%$ for field.

Using the previously assumed age of $17\pm2$\,Myr, \citet{de_rosa_spectroscopic_2016} derive a mass of $4.4\pm0.8 M_{Jup}$, placing HD 95086\,b among the least massive planets yet detected with direct imaging. Using the K-band luminosity and the new, older age, we estimate a higher mass of $7.2\pm0.7 M_{Jup}$. This shows more tension with the discovery value than for the other three planets, but the new value is still marginally consistent with the original ($2.6\sigma$).

% DR3 Prob -> 98.9

% NEW MASSES
% YSES-1 b     21.8 \pm 3   M_{Jup}  (Lbol)
% YSES-1 c      7.2 \pm 0.7 M_{Jup}  (Lbol)
% YSES-2 b      8.4 \pm 1.5 M_{Jup}  (H & K mag)
% HD 95086 b    7.2 \pm 0.7 M_{Jup}  (K mag)

% - - - - - - - - - - - - - - - - - - - - - - - - - - - - - - -
% - - - - - - - - - - - - - - - - - - - - - - - - - - - - - - -
\section{Parameters of TOI-1097}\label{sec:stellar_properties}

We summarize constraints on the candidate-planet host star in Table~\ref{tab:prop}, the details of which we provide in this section.

\begin{deluxetable}{lccc}
\centering
\tabletypesize{\scriptsize}
\tablewidth{0pt}
\tablecaption{Properties of the host star \starname. \label{tab:prop}}
\tablehead{\colhead{Parameter} & \colhead{Value} & \colhead{Source} }
\startdata
\multicolumn{3}{c}{Identifiers}\\
\hline
HD & 109833 & \\
HIP & 61723 & \textit{Hipparcos} \\
TOI & 1097 & \citet{guerrero_tess_2021}\\
Gaia & 5838450865699668736 &  \gaia\ DR3 \\
TIC &  360630575  &  \citet{stassun_tess_2018} \\
2MASS & J12390642-7434263  & 2MASS \\
\hline
\multicolumn{3}{c}{Astrometry}\\
\hline
$\alpha$  &  189.775832 & \emph{Gaia} DR3\\
$\delta$  &  -74.574021 & \emph{Gaia} DR3 \\
$\mu_\alpha$ (mas\,yr$^{-1}$)  & $-50.489 \pm 0.012$ & \emph{Gaia} DR3\\
$\mu_\delta$  (mas\,yr$^{-1}$) & $ -6.764 \pm 0.014$ & \emph{Gaia} DR3\\
$\pi$ (mas) & $12.5686 \pm 0.0118$ & \emph{Gaia} DR3\\
\hline
\multicolumn{3}{c}{Photometry}\\
\hline
G$_{Gaia}$ (mag)  & $9.145 \pm 0.003$ & \emph{Gaia} DR3\\
BP$_{Gaia}$ (mag) & $9.451 \pm 0.006$ & \emph{Gaia} DR3\\
RP$_{Gaia}$ (mag) & $8.668 \pm 0.004$ & \emph{Gaia} DR3\\
B$_T$ (mag)       & $10.082 \pm 0.027$ & Tycho-2 \\%
V$_T$ (mag)       & $9.380 \pm 0.020 $ & Tycho-2\\%
J (mag)  & $8.144 \pm 0.023$ &  2MASS\\
H (mag)  & $7.890 \pm 0.038$ & 2MASS\\
Ks (mag) & $7.820 \pm 0.026$  & 2MASS\\
W1 (mag) & $7.772 \pm 0.028$ & ALLWISE\\
W2 (mag) & $7.814 \pm 0.020 $ & ALLWISE\\
W3 (mag) & $7.787 \pm 0.018$ & ALLWISE\\ 
\hline
\multicolumn{3}{c}{Kinematics \& Position}\\
\hline
RV$_{\rm{Bary}}$ (km\, s$^{-1}$) & 10.73 $\pm$ 0.23 &  \emph{Gaia} DR3\\
X (pc)            &  $41.39 \pm 0.04$ & This work\\
Y (pc)            & $-66.00 \pm 0.06$ & This work\\
Z (pc)            & $-16.16 \pm 0.02$ & This work\\
U (km\, s$^{-1}$) & $-10.84 \pm 0.30$ & This work\\
V (km\, s$^{-1}$) & $-18.3  \pm 0.48$ & This work\\
W (km\, s$^{-1}$) & $-5.58  \pm 0.12$ & This work\\
\hline
\multicolumn{3}{c}{Physical Properties}\\
\hline
$P_{\rm{rot}}$ (days)  &  \red{$5.111 \pm 0.51$} & This work\\
\vsini (km\, s$^{-1}$) & $ 10.5 \pm 0.2 $ & This work\\
$i_*$ ($^\circ$)       & $ >84$           & This work\\
\fbol\,(erg\,cm$^{-2}$\,s$^{-1}$) & ($6 \pm 0.4)\times10^{-9}$ & This work \\
\teff\ (K)            & $5881 \pm 50$    & This work \\
\logg\ (dex)          & $4.45 \pm 0.10$  & This work \\
M$_\star$ (M$_\odot$) & $1.08 \pm 0.05$  & This work \\
R$_\star$ (R$_\odot$) & $1.00 \pm 0.04 $ & This work \\
L$_\star$ (L$_\odot$) & $1.18 \pm 0.08 $ & This work \\
$\rm{[M/H]}$ & $-0.07\pm0.08$ & This work \\
$\rho_\star$ ($\rho_\odot$) & $1.08\pm0.17$ & This work \\
Age (Myr) & $27\pm3$  & This work 
\enddata
\end{deluxetable}

\subsection{Fit to the spectral-energy distribution}

To determine \teff, $R_*$, and $L_*$ of \starname, we fit the spectral-energy-distribution (SED) following the methodology from \citet{mann_k2-33_2016}. To summarize, we compare the observed photometry to a grid of optical and NIR spectra of nearby unreddened stars. Most spectra are drawn from \citet{rayner_infrared_2009}, supplemented by Hubble's Next Generation Spectral Library \citep[NGSL;][]{heap_ngsl_2016}. To fill in gaps in the spectrum, we use BT-SETTL CIFIST atmospheric models \citep{baraffe_new_2015}, fitting to the template spectrum as outlined in \citet{gaidos_m_2014}. This also provides an estimate of \teff. We integrate the resulting full SED to determine the bolometric flux (\fbol), which combined with the \gaia{} DR3 parallax, gives us the total luminosity ($L_*$). We then use the  Stefan-Boltzmann relation to calculate $R_*$ from \teff{} and $L_*$. 

For our fit, we use photometry from Tycho-2 \citep{Hog_tycho_2000}, the Two Micron All Sky Survey \citep[2MASS;][]{skrutskie_two_2006, doi_2mass}, the Wide-field Infrared Survey Explorer \citep[WISE;][]{wright_wide-field_2010, doi_allwise}, and \gaia{} DR3 \citep{GaiaCollaboration2021}. We exclude $W3$ and $W4$ photometry in our fit because the star's young age allows the possibility of a cool debris disk. To account for variability in the star, we add 0.02 mags in quadrature to the errors of all optical photometry. In total, the fit includes six free parameters: the choice of template, $A_V$, three parameters that describe the atmospheric model selection ($\log~g$, \teff, and $[M/H]$), and a scale factor between the model and the photometry ($S$). 

The resulting fit gives \teff=$5950\pm90$\,K, \fbol=$(6.0\pm0.4)\times10^{-9}$ (erg\,cm$^{-2}$\,s$^{-1}$), $L_*=1.18\pm0.08L_\odot$, $R_* = 1.00\pm0.04R_\odot$, and a spectral type of G1V--G3V. The best-fit model predicts a $W3$ that is lower than the observed value by $10-20\%$, which suggests the presence of a debris disk. However, the excess is below significance for some templates, and no significant excess is seen in the less precise $W4$ point.

\begin{figure}[tb]
    \centering
    \includegraphics[width=0.49\textwidth]{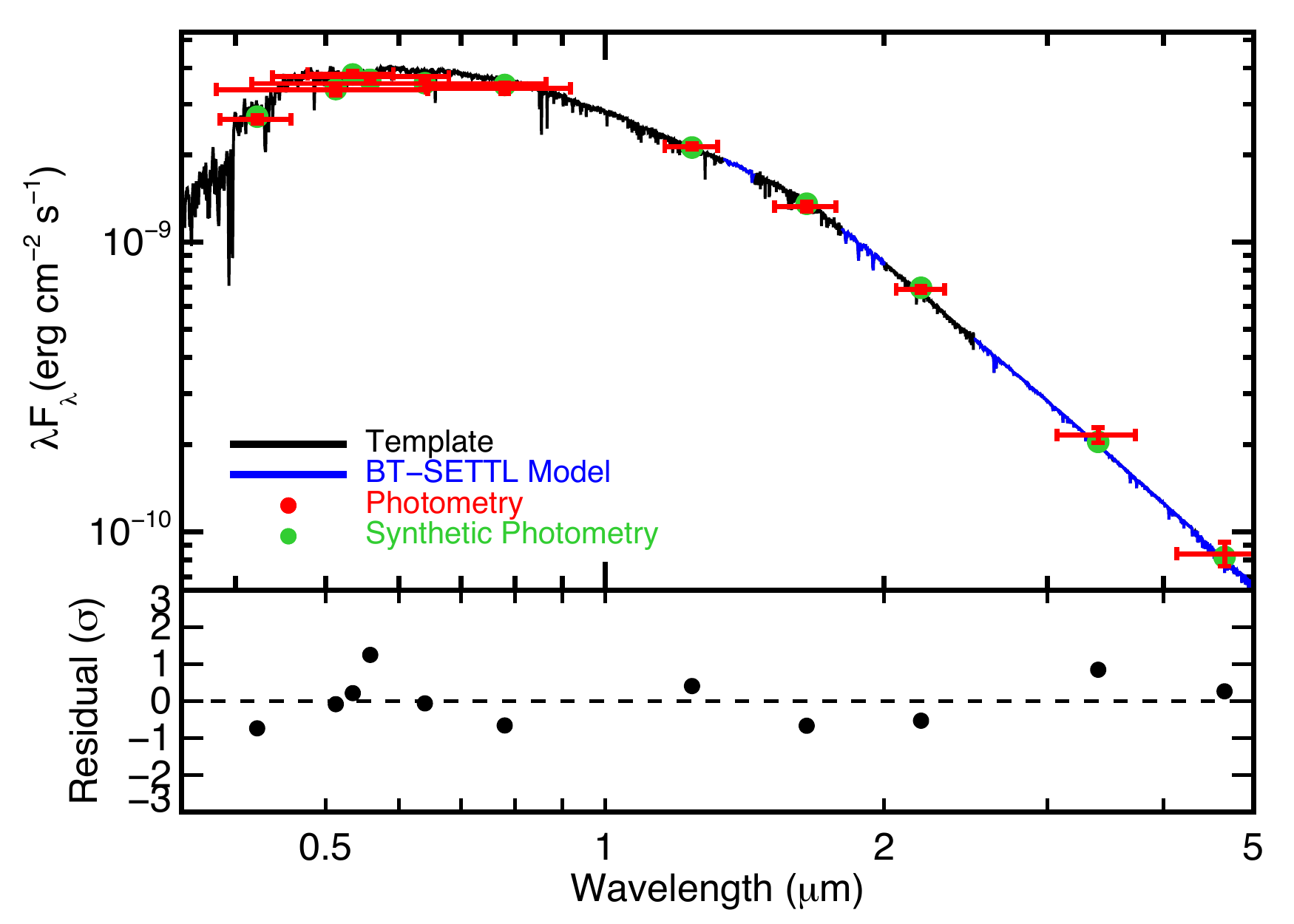}
    \caption{Best-fit template spectrum (G1V; black) and synthetic photometry (green) compared to the observed photometry of \starname{} (red). Errors on observed photometry are shown as vertical errors, while horizontal errors indicate the approximate width of the filter. BT-SETTL models (blue) were used to fill in regions of high telluric absorption or beyond the template range. The bottom panel shows the photometric residual in units of standard deviations. }
    \label{fig:sed}
\end{figure} 

\subsection{Fit to the high-resolution spectra}\label{sec:highres}

In addition to the SED fit described above, we derive the atmospheric parameters from both the CHIRON and HARPS spectra. These served as an independent test of \teff, while also providing constraints on metallicity, $\log~g$, and rotational broadening (\vsini). 

We derive spectral parameters ($T_{\mathrm{eff}}$, $\log{g}$, \vsini\ and [M/H]) from the CHIRON spectra of \starname{} using the Spectral Parameter Classification (SPC) tool \citep{buchhave_abundance_2012}. SPC cross correlates the observed spectrum against a grid of Kurucz atmospheric models \citep{kurucz_new_1993}. Four parameters are allowed to vary: \teff, \logg, bulk metallicity ([M/H]), and \vsini. We run each spectrum separately, then combine the results, yielding \teff = $5881 \pm 45$\,K, $\log~g= 4.445\pm0.041$, [M/H]$=-0.068\pm0.053$, and $v\sin i_\star = 10.5 \pm 0.2
{\rm km\,s}^{-1}$. The assigned errors reflect the scatter between spectra, and do not include systematic errors. Based on comparison with benchmark stars (e.g., asteroseismic targets), we adopt error floors of $50\,K$, $0.1$ dex in $\log~g$, and 0.08~dex in [M/H]. Even with the error floors, the resulting \teff{} estimate is more precise than those from our SED fit.

We separately derive spectral parameters from the HARPS spectra using the methodology described in \citet[][]{Sousa-14, Santos-13}. We first measure the equivalent widths (EW) of 224 FeI and 35 FeII lines using the ARES v2 code\footnote{\url{http://www.astro.up.pt/$\sim$sousasag/ares}} \citep{Sousa-15}. Then we use these EWs together with a grid of Kurucz model atmospheres \citep{Kurucz-93} and the radiative transfer code MOOG \citep{Sneden-73} to determine the parameters under the assumption of ionization and excitation equilibrium. The abundances of Mg and Si are also derived using the same tools and models as detailed in \citep[e.g.][]{Adibekyan-12, Adibekyan-15}. Although the EWs of the spectral lines are automatically measured with ARES, we performed a careful visual inspection of the EWs measurements as only three lines are available for the Mg abundance determination. This analysis gives us \teff$=5975\pm70$\,K, $\log~g= 4.58\pm0.11$, [Fe/H]$=0.08\pm0.05$, [Mg/H] = 0.02$\pm$0.06, and [Si/H] = 0.05$\pm$0.05.

The rotational projected velocity (\vsini = 9.9$\pm$0.8 \kms) is derived from the HARPS spectra by performing spectral synthesis with MOOG on 36 iron isolated lines and by fixing all the stellar parameters, macroturbulent velocity, and limb-darkening coefficient \citep{CostaSilva-20}. The limb-darkening coefficient (0.58) is determined using the stellar parameters as described in \citet{Espinoza-15} assuming a linear limb darkening law. The macroturbulent velocity ($3.6$\kms) is determined using the temperature and gravity-dependent empirical formula from \citet{Doyle-14}. 

Despite using different methods and data, the two sets of results above are all consistent within $2\sigma$, and both are consistent with our SED analysis. We use the CHIRON/SPC fit for \teff{} and \vsini, and the abundances from the HARPS/MOOG analysis. Using any set, or the average of the three, does not significantly change any results or conclusions of the paper.

% - - - - - - - - - - - - - - - - - - - - -
\subsection{Mass from stellar isochrones}
To determine the stellar mass ($M_*$) we compare the observed photometry (from \gaia\ $G$, $B_P$, $R_P$; 2MASS $J$, $H$, $K$; and Tycho $B_T$, $V_T$) to predictions from the DSEP magnetic models \citep{feiden_magnetic_2016}. The DSEP magnetic model covers ages from 1\,Myr$-$10\,Gyr and masses between 0.09\,\msun{} and 2.45\,\msun{}. To explore systematics between models, we also run a fit using the PARSEC (v1.2S) models \citep{bressan_parsec_2012}. We apply the comparison within an MCMC framework with \texttt{emcee}, simultaneously fitting for four parameters: age, $M_*$, $A_V$, and a factor describing the underestimation of the errors on the measured photometry ($f$). For both grids, we assume Solar metallicity.  

To alleviate the computational cost of bilinearly interpolating the model grid at every step, we pre-interpolate the grid using the \texttt{isochrones} package \citep{morton_isochrones_2015} to give it tighter spacing in age and mass (0.1\,Myr and 0.01\,\msun{}) than expected errors. Grid re-sampling also lets us enforce uniform sampling in age. During each sampling step, the procedure is as follows: first, we employ a hybrid interpolation method, which finds the nearest neighbor in age, then linearly interpolates in mass, to extract predicted photometry and stellar parameters (such as \teff{} and $R_*$). Second, the predicted photometry is corrected according to the given $A_V$ value, using a combination of \texttt{synphot} \citep{lim_synphot_2020} and the extinction model presented by \citet{cardelli_extinction_1989}. Lastly, we compare the corrected model photometry to the measured photometry in a Bayesian maximum-likelihood framework. 

We place Gaussian priors of $27 \pm 3$\,Myr on age and $5881 \pm 50$\,K on \teff{}, following our analysis of the spectrum. All other fit parameters evolve under uniform priors. For the DSEP magnetic model, we find {$\textnormal{age} = 34.5 \pm 1.7$\,Myr} and {$M_* = 1.13 \pm 0.02$\,\msun{}}, while PARSEC gives {$\textnormal{age} = 30.9 \pm 0.8$\,Myr} and {$M_* = 1.03 \pm 0.02\msun{}$}. The errors are statistical only, as evident by the $\simeq3\sigma$ disagreement on $M_*$ (0.10$\pm$0.03\msun) between the two methods. We adopt $M_*=1.08\pm0.05\msun{}$, which encompasses both values and more accurately reflects the systematic limits of the models \citep{tayer_guide_2022}.

\subsection{Stellar Inclination}

We use the combination of \vsini, $P_{\rm{rot}}$, and $R_*$ to estimate the stellar inclination ($i_*$). Since the transiting planets are nearly edge-on ($i>85^\circ $; Section~\ref{sec:planet_params}) this measures the alignment between the stellar spin and planetary orbit axes.
Simplistically, the equatorial velocity ($V$) in \vsini{} can be derived straight-forwardly using $V=2\pi R_*/P_{\rm{rot}}$. 
In practice, however, this calculation requires additional corrections due to the effects of sky-projection and measurement uncertainties, which could cause the appearance of \vsini$>V$. We followed the formalism from \citep{masuda_inference_2020}. Using either \vsini{} determined in Section~\ref{sec:highres} gave an inclination consistent with edge-on. The CHIRON \vsini{} gave a limit of $i_*>84^\circ $ at 95\% confidence, while the HARPS \vsini{} yielded $i_*>66^\circ $.

% - - - - - - - - - - - - - - - - - - - - - - - - -
% - - - - - - - - - - - - - - - - - - - - - - - - -
\section{Parameters of TOI-1097\,\lowercase{b} and TOI-1097\,\lowercase{c}}\label{sec:planet_params}

\subsection{Detection of the planetary signals}\label{sec:search}

The first planet signal, TOI-1097.01, was originally detected from the joint search of sectors 11 and 12 as part of the Quick-Look Pipeline (QLP) search \citep{huang_photometry_2020}. The candidate passed initial vetting and was alerted on 2021 Oct 29. 

To confirm the detection and search for additional planets, we use the Notch and LoCoR pipelines described in \citet{rizzuto_zodiacal_2017}\footnote{\url{https://github.com/arizzuto/Notch_and_LOCoR}}. To briefly summarize, the Notch filter fits a window of the light curve as a combination of an outlier-robust second-order polynomial and a trapezoidal notch. The window is shifted along the light curve until the variability is detrended (flattened) while preserving the planet signal. At each data point, Notch calculates the improvement from adding the trapezoidal notch based on the change in the Bayesian Information Criterion (BIC) compared to modeling just a polynomial. 

After running Notch, we perform a box-least-squares search on the BIC values and recovered both the initial 9.2\,day planet candidate from \tess{} and an additional signal at either 13.9 or 41 days (it was initially ambiguous). Additional short-cadence data from Sectors 38 and 39 made it clear that the shorter 13.9\,day period was the correct one. This candidate was later recovered by a \tess{} Science Processing Operations Center (SPOC) pipeline joint search of sectors 38 and 39 \citep{Jenkins_methods_2002, Jenkins_software_2010, Jenkins_handbook_2020} and designated TOI-1097.02 on 2022 Mar 24 by the TESS Science Office.

There were no other significant detections from our Notch search, other than those near aliases of the planets and/or rotation period. The BIC is sensitive to single-transit detections \citep{rizzuto_tess_2020}, but we did not identify any such signals that survived visual inspection.

\subsection{MCMC fit of light curves}

To determine the planet parameters, we compare a transit model to the \tess{} photometry using the \texttt{MISTTBORN} (MCMC Interface for Synthesis of Transits, Tomography, Binaries, and Others of a Relevant Nature) code\footnote{\url{https://github.com/captain-exoplanet/misttborn}}. \texttt{MISTTBORN} uses \texttt{BATMAN} \citep{kreidberg_batman_2015} to generate model light curves, \texttt{celerite} \citep{foreman-mackey_fast_2017} to model the stellar variability, and \texttt{emcee} \citep{foreman-mackey_emcee_2013} to explore the parameter space. More details on the code can be found in \citet{mann_k2_25_2016} and \citet{johnson_k2-260_2018}. 

% Limb-darkening = 
% %g1=0.4020±0.0003
% %g2=0.1379±0.0009
The standard implementation of \texttt{MISTTBORN} fits for four parameters for each transiting planet: time of periastron ($T_0$), orbital period of the planet ($P$), planet-to-star radius ratio ($R_p/R_\star$), and impact parameter ($b$). It also fits for three parameters specific to the host star: stellar density ($\rho_\star$) and two limb-darkening parameters ($q_1$, $q_2$) using the triangular sampling prescription from \citet{kipping_efficient_2013}. 

Stellar variability from the star was far stronger than the transit signal over all \tess{} data. We fit the variations using the Gaussian Process (GP) feature within \texttt{MISTTBORN}. We initially adopted the GP kernel based on a mixture of two simple harmonic oscillators but found that the parameters associated with the second oscillator were unconstrained and never fully converged. Instead, we adopt a single-term simple harmonic oscillator, \red{based on the model used by \citet{foreman-mackey_fast_2017}}, which includes three GP terms: the period ($\ln(P_{GP})$), amplitude ($\ln{\rm{Amp}}$), and the decay timescale for the variability (quality factor, $\ln{Q}$). 

\begin{figure*}[tb]
    \centering
    \includegraphics[width=0.95\textwidth]{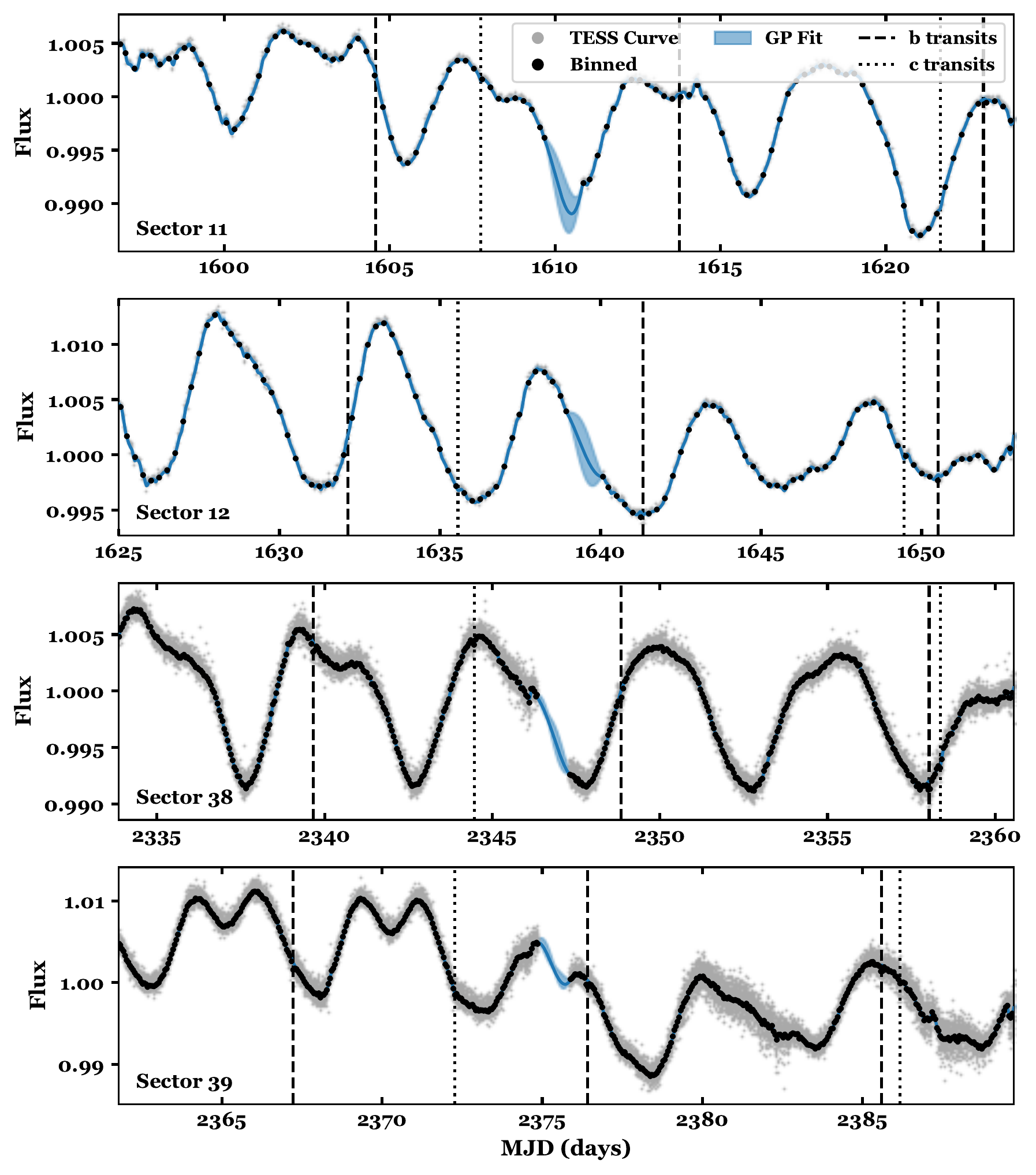}
    \caption{\tess{} light curves for Sectors (from top to bottom) 11, 12, 38, and 39. Data for each sector is shown as gray dots and binned values as black dots. The blue line indicates the model GP fit to the data with uncertainties (shaded region). The times of the transits are marked with dashed and dotted vertical lines for planets b and c, respectively.}
    \label{fig:gp_transit}
\end{figure*} 

% Figure: Transits
\begin{figure}[tb]
    \centering
    \includegraphics[width=0.99\linewidth]{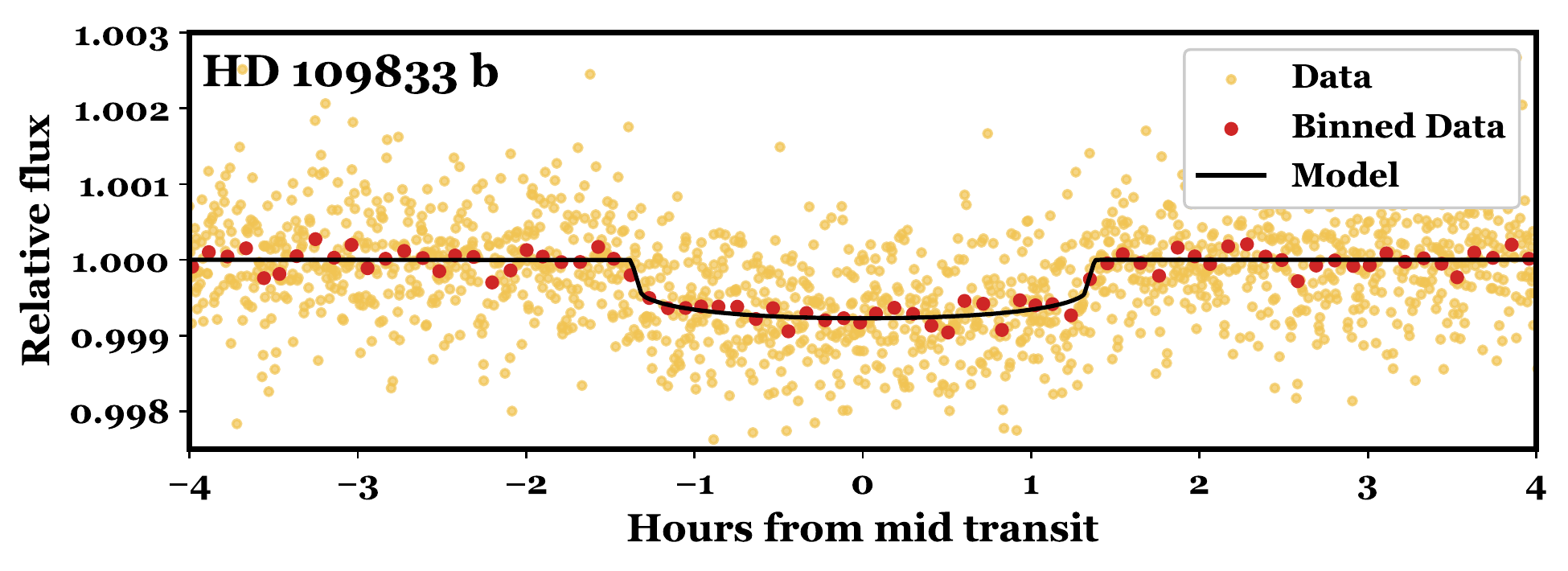}\\
    \includegraphics[width=0.99\linewidth]{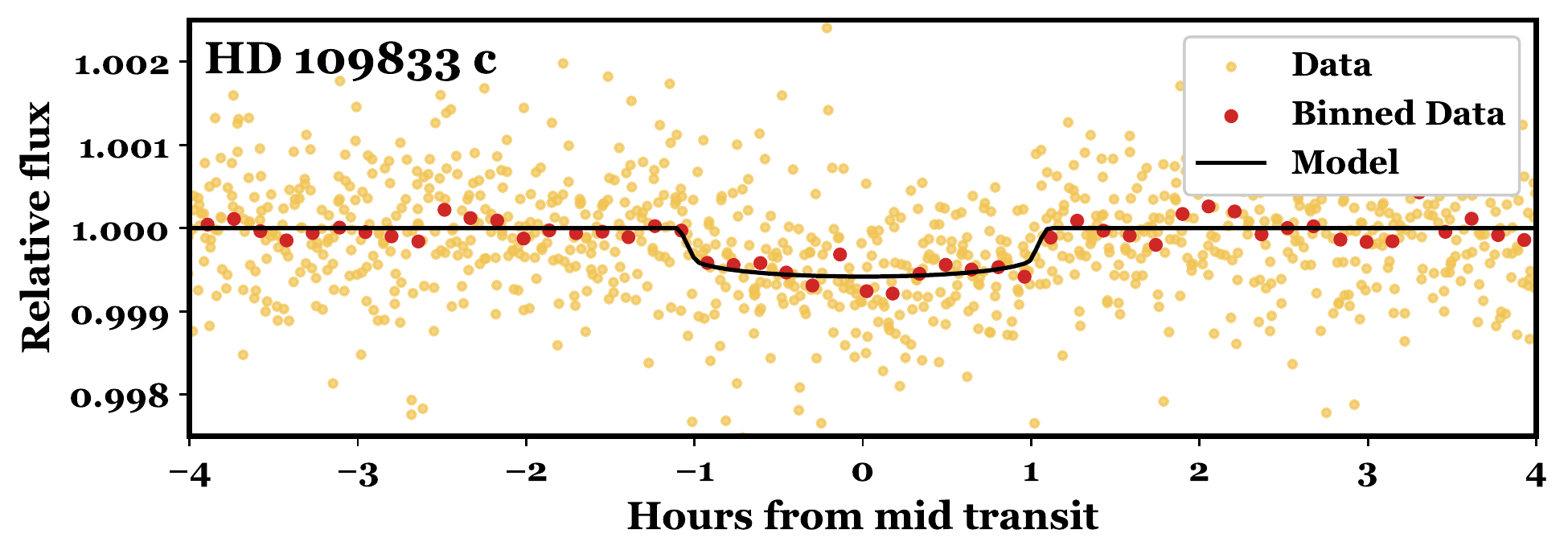}
    \caption{Phase-folded transits of \planetname{} and \planetnametwo. The data are shown in yellow, with the 20\,s cadence data from Sectors 38 and 39 binned to 2\,m. Red points show all data binned together, and the black line shows the \texttt{batman} model using the results from the $e=0$ \texttt{MISTTBORN} run.
    }
    \label{fig:transits}
\end{figure}

Although eccentricities of young planets are expected to be near zero due to gravitational interactions and drag from the circumstellar disk \citep{tanaka_three-dimensional_2004}, this young regime has few observational constraints. So, we run two fits, one with eccentricity locked at zero and a uniform prior on $\rho_\star$, and a second fitting two parameters describing eccentricity and argument of periastron ($\sqrt{e}*\cos(\omega)$, and $\sqrt{e}*\sin(\omega)$) with a uniform prior, and assuming a Gaussian prior for $\rho_\star$ drawn from Section~\ref{sec:stellar_properties}.

We apply Gaussian priors on the limb-darkening coefficients based on the values from the \texttt{LDTK} toolkit \citep{parviainen_ldtk_2015}, with errors accounting for the difference between these two estimates (which differ by $0.04-0.08$).

After an initial fit, we find a few walkers wandered off the transit signal, adjusting the GP signal to partially fit the transit. To prevent this, we place weak Gaussian priors on $T_0$, $P$, and $\ln(P_{GP})$ around the initially estimated values from a least-squares fit, and with widths of 0.1\,days, 0.1\,days, and 0.1\,dex (1 day), respectively. The width of these priors was much larger than the final uncertainties and had a negligible effect on the result (other than preventing the wandering walkers). 
All other parameters evolve under uniform priors with physically motivated limits.

For the first fit ($e$ locked at 0), we run the MCMC using 50 walkers for 100,000 steps including a burn-in of 20,000 steps. This setup is sufficient for convergence based on the autocorrelation time. For the second fit, which has a lower acceptance fraction, we use 200 walkers and 100,000 steps (the same burn-in). 

\begin{figure*}[tb]
    \begin{centering}
    \includegraphics[width=0.45\linewidth]{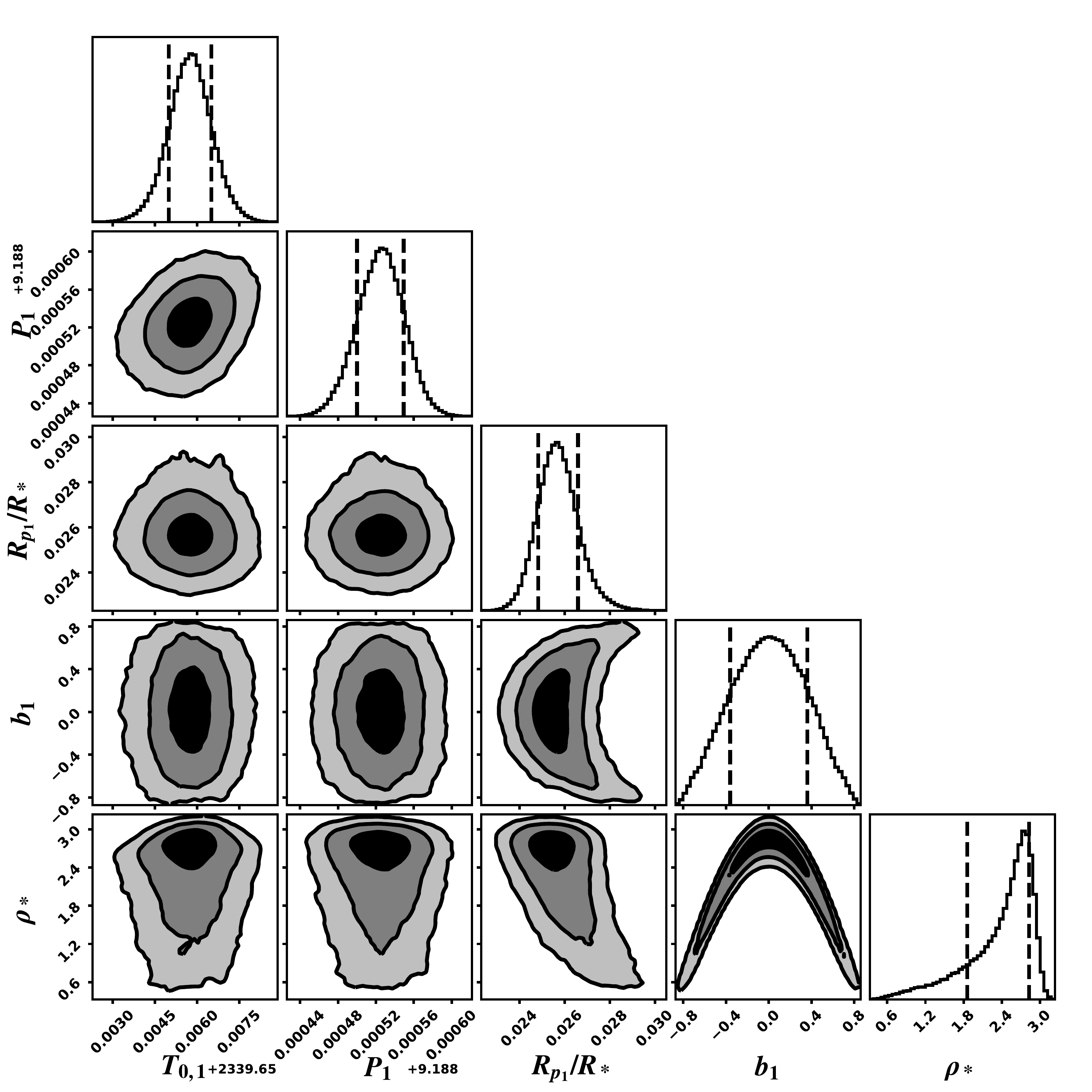}
    \includegraphics[width=0.46\linewidth]{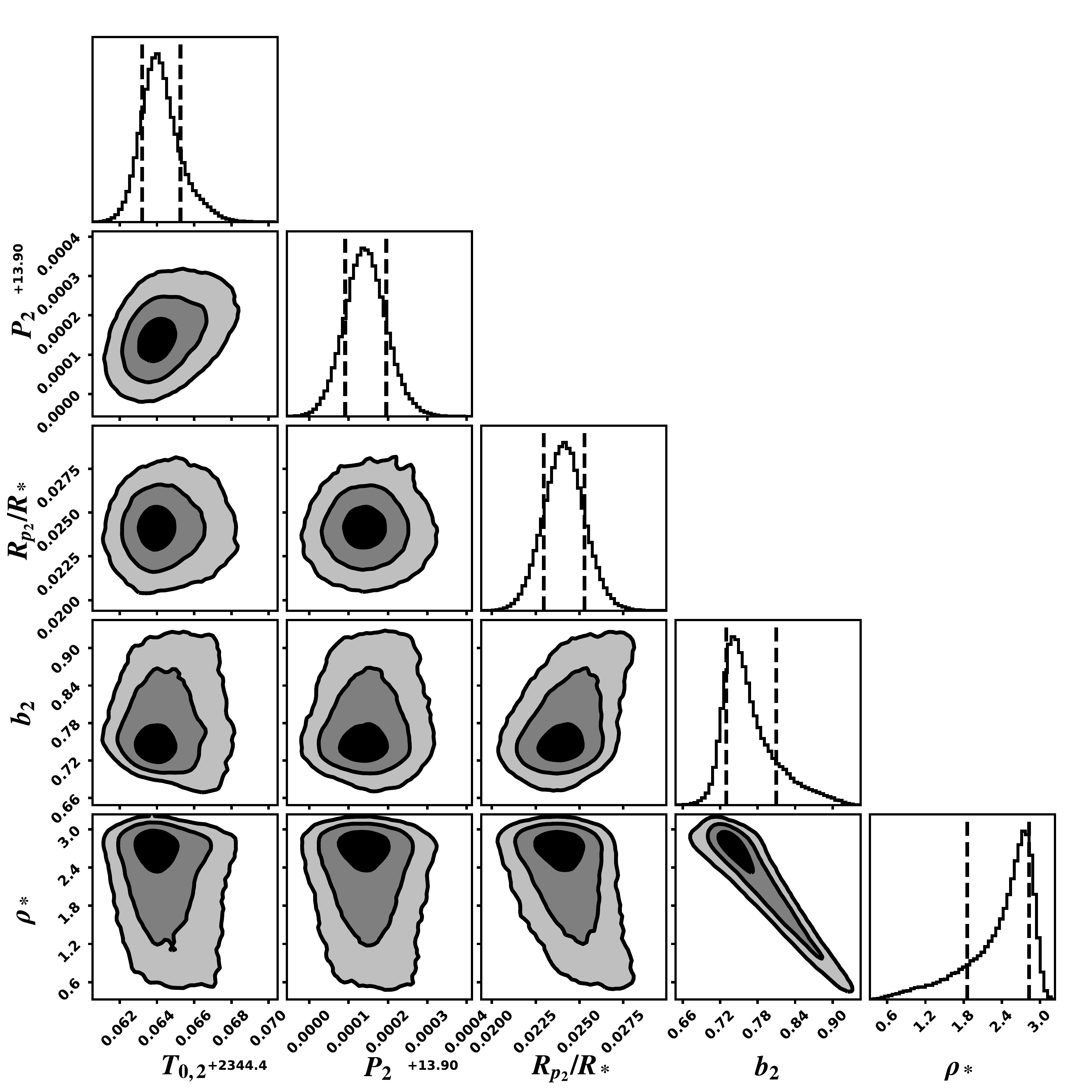}
    \end{centering}
    \caption{Corner plots of the \texttt{MISTTBORN} results for TOI-1097\,b (left) and TOI-1097\,c (right), with eccentricity locked to $e = 0$. Posterior distributions are shown as 2-dimensional contour plots, with levels corresponding to 1, 2, and 3$\sigma$, and as histograms with the $16th$ and $84th$ percentiles marked with dashed lines. Distributions are mostly Gaussian, with the exception of $\rho_*$, which has a long tail towards a less-massive stellar host. The distribution of $\rho_*$ is shown on both plots, but only a single value is explored in the fit. .}
    \label{fig:MISTTBORN}
\end{figure*}

\begin{figure}[tb]
    \begin{centering}
    \includegraphics[width=0.49\textwidth]{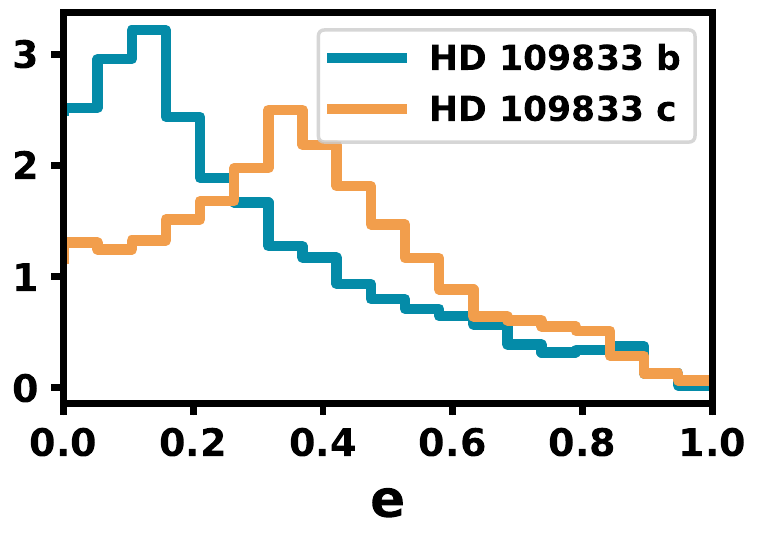}
    \end{centering}
    \caption{The posterior distribution of eccentricity resulting from \texttt{MISTTBORN} fit, using a uniform prior distribution on e, and Gaussian prior on $\rho_*$ from our stellar parameters. Both planets are consistent with zero eccentricity.}
    \label{fig:eccentricity}
\end{figure}

Both of the fits were broadly consistent, producing consistent $q_1, q_2$, and GP fit for the star, $T_0, P,$ and $R_P/R_*$ for both planets, and impact parameter for the outer planet. The impact parameter of the inner planet is higher when allowing non-zero eccentricities, but consistent within the uncertainties. \red{The GP fit found a stellar rotation period of $P_{GP}=5.64\pm1.08$ days, consistent with the rotation period found using a Lomb-Scargle periodigram analysis, $P_{rot}=5.11\pm0.51$ days (see Section 3.1.2).}
\red{The results are shown as corner plots in Figure \ref{fig:MISTTBORN}}. 
The eccentric fit cannot rule out a zero eccentricity for either planet but does suggest a potentially eccentric orbit, especially for the outer planet, which has a best-fit eccentricity $e_c = 0.3^{+0.21}_{-0.19}$. The posterior distribution of eccentricity for each planet is shown in Figure \ref{fig:eccentricity}.

\begin{deluxetable}{|l|c|c|}
\tablewidth{0pt}
\tablecaption{Parameters of \planetname{} and \planetnametwo \label{table:planetParams}
}
\tablehead{
    \colhead{} & \colhead{\bf{\starname}} & \colhead{} \\ 
    \colhead{Parameter} & \colhead{e=0} & \colhead{e float (preferred)}
}
\startdata
    \multicolumn{3}{|c|}{Measured Parameters }\\
    \hline
    $\rho_{\star}$ ($\rho_{\odot}$) & $2.544^{+0.287}_{-0.714}$ & $1.1027^{+0.1359}_{-0.1415}$\\ 
    $q_{1,1}$ & $0.303^{+0.101}_{-0.094}$  & $0.321^{+0.141}_{-0.115}$ \\ 
    $q_{2,1}$ & $0.364^{+0.077}_{-0.084}$ & $0.325^{+0.109}_{-0.142}$ \\
    $\ln(P_{GP})$ & $1.730^{+0.083}_{-0.081}$ & $1.729^{+0.092}_{-0.081}$ \\ 
    $\ln(Amp)$ & $-9.525^{+0.177}_{-0.164}$ & $-9.531^{+0.183}_{-0.170}$ \\ 
    $\ln(Q)$ & $0.618^{+0.035}_{-0.026}$ & $0.618^{+0.036}_{-0.028}$ \\
    \hline
    \hline
    % -  -  -  -  -  -  -  -  Planet b  -  -  -  -  -  -  -  -  -  -
    \multicolumn{3}{|c|}{\textbf{\planetname}} \\
    \multicolumn{1}{|c}{Parameter} & \multicolumn{1}{c}{e=0} & \multicolumn{1}{c|}{e float (preferred)} \\  % Inside multicolumn to control vertical lines
    \hline
    \multicolumn{3}{|c|}{Measured Parameters }\\
    \hline
    $T_0$ (BJD-2454833)  & $1604.57376^{+0.00074}_{-0.00076}$    & $1604.57374^{+0.00091}_{-0.00097}$ \\
    $P$ (days)           & $9.188525 \pm 2.5\times10^{-5}$       &  $9.188526 \pm 2.6\times10^{-5}$\\ 
    $R_P/R_{\star}$      & $0.02569^{+0.00091}_{-0.00085}$      & $0.0265^{+0.0014}_{-0.0012}$ \\ 
    $b$                  & $0.25^{+0.25}_{-0.18}$               & $0.61^{+0.15}_{-0.33}$\\
    $\sqrt{e}\sin\omega$ & --                                   & $0.02^{+0.23}_{-0.25}$ \\ 
    $\sqrt{e}\cos\omega$ & --                                   & $-0.03^{+0.51}_{-0.46}$\\ 
    \hline 
    \multicolumn{3}{|c|}{Derived Parameters }\\
    \hline
    $a/R_{\star}$        & $25.2^{+0.9}_{-3.0}$                 & $19.9^{+1.9}_{-2.1}$ \\ 
    $i$ ($^{\circ}$)     & $89.4^{+0.4}_{-0.7}$                 & $88.13^{+1.0}_{-0.55}$\\
    $e$                  & --                                   & $0.18^{+0.24}_{-0.12}$ \\ 
    $R_P$ ($R_\oplus$)   & $2.802^{+0.099}_{-0.093}$            & $2.888^{+0.152}_{-0.127}$ \\ 
    \hline
    \hline
    % -  -  -  -  -  -  -  -  Planet c  -  -  -  -  -  -  -  -  -  -
    \multicolumn{3}{|c|}{\textbf{\planetnametwo}} \\
    \multicolumn{1}{|c}{Parameter} & \multicolumn{1}{c}{e=0} & \multicolumn{1}{c|}{e float (preferred)} \\  
    \hline
    \multicolumn{3}{|c|}{Measured Parameters }\\
    \hline
    $T_0$ (BJD-2454833)     & $1607.75659^{+0.0012}_{-0.00092}$ & $1607.7567^{+0.0015}_{-0.0011}$ \\
    $P$ (days)              & $13.900142 \pm 5.3\times10^{-5}$  & $13.900148 \pm 5.7\times10^{-5}$\\ 
    $R_P/R_{\star}$         & $0.0241^{+0.0012}_{-0.0012}$      & $0.0237^{+0.0018}_{-0.0016}$ \\ 
    $b$                     & $0.757^{+0.055}_{-0.027}$         & $0.73^{+0.14}_{-0.32}$\\ 
    $\sqrt{e}\sin\omega$    & --                                & $0.2^{+0.24}_{-0.31}$ \\ 
    $\sqrt{e}\cos\omega$    & --                                & $-0.0^{+0.54}_{-0.56}$ \\ 
    \hline 
    \multicolumn{3}{|c|}{Derived Parameters }\\
    \hline
    $a/R_{\star}$       & $33.2^{+1.2}_{-3.5}$      & $29.4^{+4.5}_{-4.3}$\\ 
    $i$ ($^{\circ}$)    & $88.696^{+0.085}_{-0.3}$  & $88.24^{+0.73}_{-0.34}$ \\ 
    $e$ 		        & --                        & $0.3^{+0.21}_{-0.19}$ \\ 
    $R_P$ ($R_\oplus$)  & $2.63^{+0.128}_{-0.125}$  & $2.59^{+0.196}_{-0.175}$ \\
\enddata
\tablecomments{Results of the \texttt{MISTBORN} MCMC fitting of the planet transits.}
\end{deluxetable}

\subsection{False Positive Analysis}\label{sec:false_pos}

For our false positive analysis, we first calculate the magnitude limit ($\Delta m$) of a potential blended source (bound or background) that could reproduce the transit signal, using the source brightness constraints described by \citet{seager_unique_2003} and \citet{vanderburg_tess_2019}. This depends on the ingress or egress duration compared to the transit duration and reflects the true radius ratio, independent of whether there is contaminating flux: 

\begin{align}
    \Delta m \leq 2.5\log_{10}(\frac{T^2_{12}}{T^2_{13}\delta})\notag
\end{align}

Here, $\delta$ is the transit depth, $T_{12}$ is the ingress/egress duration and $T_{13}$ is the time between the first and third contact. We calculate $\Delta m$ for the posterior samples for our floating eccentricity transit fit and take the 99.7\% confidence limit. We find $\Delta m <3.5$ and $<5.3$ for \planetname{} and \planetnametwo, respectively. 

\begin{figure}[tb]
    \begin{centering}
    \includegraphics[width=0.49\textwidth]{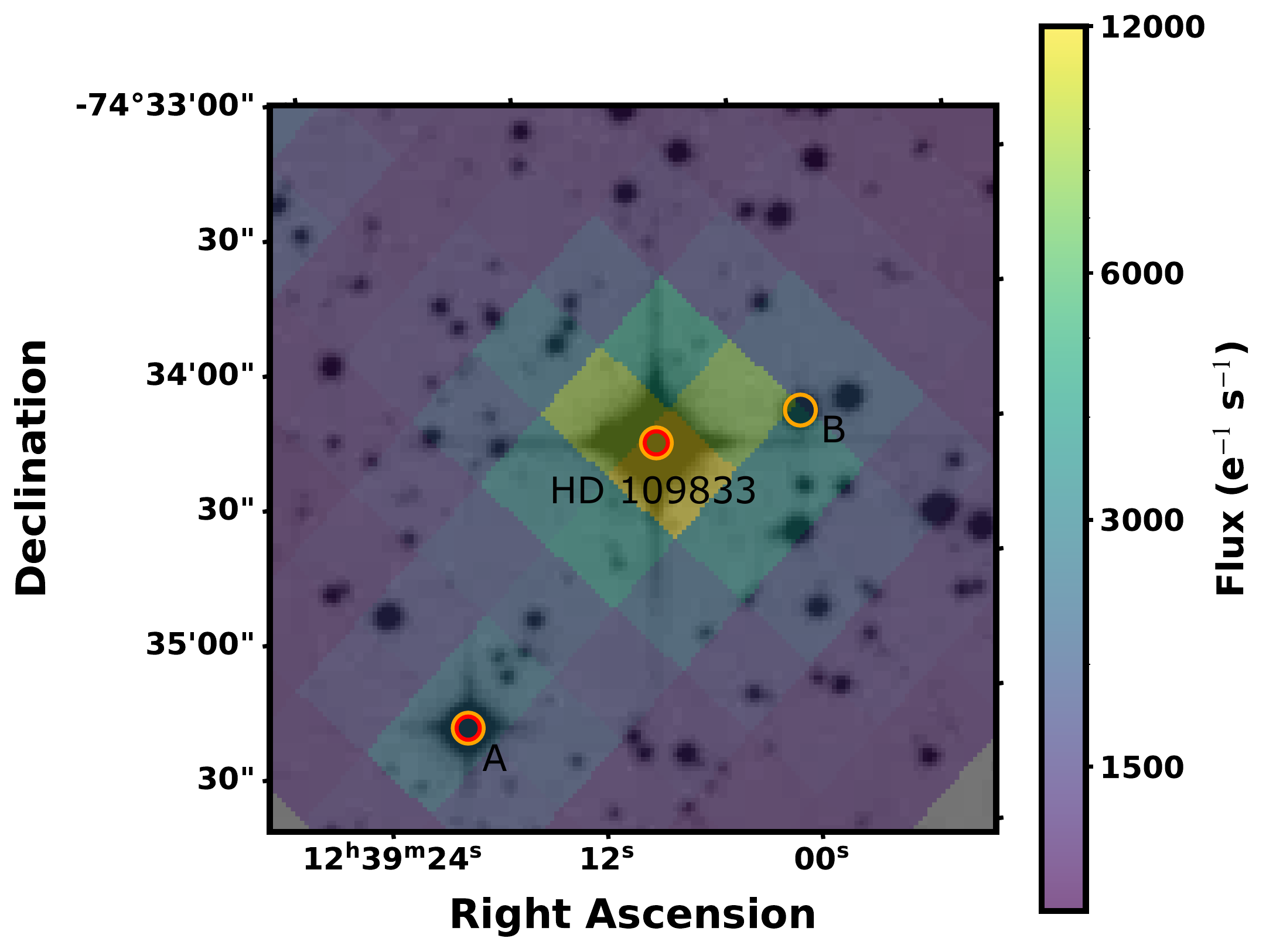}
    \end{centering}
    \caption{A \tess{} Sector 38 image colored by flux (see the colorbar) on top of a DSS image (greyscale). Red circles indicate the two stars that could reproduce the transit signal for planet b, while orange indicates those that could reproduce c. The bright central star is \starname.}
    \label{fig:aperture}
\end{figure}

Based on these magnitude limits, only two stars detected by \gaia{} (including \starname) could reproduce the transits of \planetname{}, and three stars could reproduce \planetnametwo{} (see Figure \ref{fig:aperture}). By selectively resizing the aperture, we rule out all stars other than \starname{} as the source of the planetary signals. \red{We also check this using the \texttt{tpfplotter} tool from \citet{aller_tpfplotter_2020}, and find four faint stars within the aperture, all with $\Delta T > 5$.}

%The star to the lower left of \starname{} is outside of the \tess{} aperture for most sectors, and can be ruled out as the source of the signal due to the consistent transit detection across all sectors. To test the effect of the star to the right of \starname{} , we extract a light curve using a smaller aperture centered on that star, and partially excluding \starname{}. If the signal were coming from this fainter star, we would expect to detect a deeper transit using a tighter aperture. However, we do not detect any transits from either planet, ruling it out as the signal source.

Separately, the SPOC data validation \citep{twicken_kepler_2018, Li_DataValidation_2019} centroid offsets for Sectors 38-39 exclude all TICv8.2 objects capable of producing the observed transit depths other than the target star. As with the transit duration and depth constraints above, this confirms that the only remaining false-positive scenarios involve objects unresolved with \starname. 

To handle unresolved sources, we use a combination of Multi Observational Limits on Unseen Stellar Companions \citep[MOLUSC;][]{wood_characterizing_2021} and a tool for vetting and validating TESS Objects of Interest \citep[Triceratops;][]{giacalone_vetting_2021}. \texttt{MOLUSC} provides limits on the range of possible companions allowed by the existing data, while \texttt{triceratops} combines the companion limits with information about the light curve to compute probabilities of given false-positive (FP) scenarios. 

\texttt{MOLUSC} generates simulated companions and compares them to RV measurements, contrast imaging, and \gaia{} RUWE and imaging constraints. We test two different scenarios, one of a transiting stellar companion, for which \red{the cosine of stellar inclination,} $cos(i)$, is locked to only transiting companions, and one of any stellar companion, for which $cos(i)$ was drawn from a uniform distribution. In the second scenario, we assume the transits are caused by planetary-mass objects, but there is an additional stellar companion in the system. For both scenarios we generate 5,000,000 companions, with orbital parameters drawn from realistic binary distributions (see \citet{wood_characterizing_2021} for details). Across both scenarios, $91.3\%$ of generated companions were ruled out.
In the transiting scenario, we found $3\sigma$ detection limits of $\sim 0.4-0.7 M_{Jup}$ at periods of $9-13$ days. This alone rules out an eclipsing stellar companion (orbiting \starname) as the cause of the transit signals. As the planets are not detected in any of the included datasets, the mass limits do not include the planets. The non-transiting companion scenario, while allowing the possibility of a stellar companion at moderate periods, rules out nearly all companions with $M_{comp} > 0.6 M_{\odot}$ or $P < 100$ days.

To calculate the probability that each signal is due to a planet, we run \texttt{triceratops} using all four sectors of \tess{} data and the output binaries generated from \texttt{MOLUSC}. Including the \texttt{MOLUSC} output effectively limited the possible FP scenarios to those consistent with the observational limits. Of the 18 different scenarios considered by \citet{giacalone_vetting_2021}, three are considered ``true positives", i.e. the transiting planet, unresolved binary with a transiting planet around the primary, and unresolved background star with a transiting planet around the primary. By default, \texttt{triceratops} considers signals that come from a bound companion as false positives, even if the signal is still a transiting planet. We consider cases including a planet and a bound companion as true positives, as the signal was still from a young planet even if the radii are likely significantly underestimated. 

For each planet we run \texttt{triceratops} 20 times and find the mean and standard deviation of the FPP. For \planetname{}, we find $FPP <0.0003 \red{(0.03\%)}$, with an additional $\simeq1\%$ probability that the planet orbits a bound companion. If \planetname{} is orbiting a secondary companion with the predicted mass of $\sim0.8 M_{sun}$, it would still have a radius consistent with a planet.

For \planetnametwo{}, \texttt{triceratops} yielded a higher FPP of $9.2\pm2.5\%$. Multi-planet systems are less likely to be false positives \citep[e.g.,][]{lissauer_almost_2012}. \citet{guerrero_tess_2021} recalculated this `multiplicity boost' for TOIs, i.e., the multiplicative factor that reflects the \textit{a priori} probability that a candidate in a multi-transiting system is a true positive. They estimate this to be $\simeq$50 for planets $R_p<6R_\oplus$. Even the more conservative factor of 20 for all TOIs is still sufficient to bring \planetnametwo{} below the 1\% required for validation. 

% - - - - - - - - - - - - - - - - - - - - - - - - -
% - - - - - - - - - - - - - - - - - - - - - - - - -
\section{Summary and Conclusions}\label{sec:conclusion}

In this paper, we present a new, $27\pm3 Myr$ old association (\assoc{}) on the outskirts of LCC. We initially identified the group from a population of pre-main-sequence M dwarfs co-moving with a candidate transiting-planet host from \tess{} (TOI \toi). We gather a wide range of ground-based follow-up and archival data with the goals of 1) improving the list of likely members, 2) measuring the age and basic parameters of the association, 3) confirming the group is distinct from known young populations (Carina and LCC), and 4) validating and characterizing the planetary system TOI \toi.

%% a brief summary part 1
We first perform a membership search for candidate members using an iterative process, 
%alternating between indicators of youth and kinematic alignment with the association, 
resulting in list of $306$ candidate members. %While the target list was selected to be relatively free of field or LCC interlopers, 
A small number ($\lesssim10$) of the candidates sit unusually high or low on the CMD (see Figure~\ref{fig:cmd}) or have low lithium levels for their spectral type (see Figure~\ref{fig:li_sequence}). Thus, the list is not totally clean, but is still sufficient to estimate the age and properties of the group. 

\assoc 's rotation, CMD, and lithium levels are consistent with an age of $25-30$\,Myr. Rotation provides only a qualitative check on the age (and membership) due to the large spread in rotation periods at this age \citep{rebull_rotation_2018}. A fit to the CMD with stellar models offers a consistent and arguably more precise age (26$\pm$2\,Myr), with an additional 2\,Myr error based on differences between model grids and input assumptions. The most reliable age constraint comes from the lithium depletion, which provides an age of 27$\pm3$\,Myr that is almost entirely independent of model selection (see Table~\ref{table:ldb_ages}).

The association is kinematically close to LCC but is distinct in position, velocity, dispersion, and age. 
%In $UVW$ velocity, \assoc{} is closest to LCC-C, but it is furthest from that sub-population in $XYZ$ position. Similarly, the sub-group closest in position (LCC-A) to \assoc{} is the most discrepant in $UVW$.
\red{The LCC subpopulation which is closest in $XYZ$ is most discrepant in velocity, and the group with the most similar velocity is furthest in $XYZ$.}
All of the known LCC sub-groups are also significantly younger. In Section~\ref{app:ages}, we analyzed all the LCC sub-groups using the same isochronal method we used for \assoc{} in Section~\ref{sec:isochrone}, and found the oldest LCC subgroup is $16.6\pm1.1$\,Myr, inconsistent with our isochronal age ($26\pm2$) and lithium-depletion age (27$\pm$3\,Myr) for \assoc. 

Of the other known moving groups in \citet{gagne_banyan_2018}, the two closest in kinematics are Carina ($\Delta V \simeq3.4$ \kms) and Columba ($\Delta V \simeq4.1$ \kms). The cores of these groups are more than 30\,pc from \assoc{} and are significantly older ($\simeq$45\,Myr) than \assoc{}. A more complicated possibility is that \assoc{} is a mix of members from Carina, Columba, and LCC. However, if this were the case we would expect Li-rich low mass stars to be preferentially closer to LCC and Li-poor ones close to Carina. We find instead that Li-rich PMS stars are spread throughout the association. We conclude that \assoc{} is distinct from any known young association.
%in position, kinematics, and age, making it one of many newly discovered subpopulations within larger associations \citep[e.g.,][]{pecaut_revised_2012, cantat-gaudin_expanding_2019, kerr_stars_2021, krolikowski_gaia_2021, bouma_del_lyra_2021}.

We report the discovery of two transiting planets around the Sun-like star \starname, which we identify as a \red{candidate} member of MELANGE-4. The first planet was first identified by \tess{}, while the second planet by our Notch analysis (and later by \tess). Both planets are super-Earth sized, with radii of $2.9 R_{E}$ and $2.6 R_{E}$, 
%and orbit close to the star, 
with periods of $9.2$ and $13.9$ days, close to a $3:2$ resonance. We validate the b planet as planetary in nature. The c planet was a weaker detection and had an unusually short duration; \texttt{triceratops} gave a FPP of 10\%. However, with the multiplicity boost, both planets  meet the requirements for statistical validation.  

Along with the newly discovered two-planet system around \starname{} (this paper), three high-probability candidate members have directly-imaged planetary-mass companions, TYC 8998-760-1, HD 95086, and TYC 8984-2245-1 \citep{rameau_confirmation_2013, bohn_two_2020, bohn_discovery_2021}. These systems were previously placed in LCC or the Carina moving group, \red{with assumed younger ages. Our older} age changes the masses and inferred properties of the planets. %This highlights the importance of mapping out such young stellar associations for direct imaging surveys.

\starname{} is {\it not} an unambiguous member of \assoc. This is especially surprising since the planet-host was the seed of our initial search that identified \assoc. 
%Most of our data favor membership, but \starname{} lands on the outskirts of the group in both $U$ and $V$ velocity. Additionally, \starname's CMD position is lower than models predict for this age.
Most of our data favor membership. \red{However \starname{} has a lower CMD position than predicted for this age, and lies on the outskirts of the group in $U$ and $V$ velocity.}
As discussed in Section~\ref{sec:isochrone}, this may be an issue with the models.%, as there are several other stars with similarly low CMD position at that color. 
While \starname{} is clearly young ($<200$\,Myr), the lithium levels are lower than expected for this age \citep{gutierrez_albarran_lithium_2020} and compared to similar stars in the same association (Figure~\ref{fig:li_sequence}). The rotation period matches expectations for members of \assoc, but is also consistent with ages up to 200\,Myr. The strongest evidence in favor of membership is the $99.8$\% \banyan{} membership probability. We conclude that \starname{} is likely part of \assoc{}, but we cannot reject the possibility that it is a young field star coincidentally moving with \assoc.

An older age for \starname{} does not significantly impact our inferred properties of the star or planet. A star of this \teff{} exhibits minimal change in CMD properties between hitting the main-sequence ($\simeq$30\,Myr) to the oldest ages consistent with Li and rotation (100-200\,Myr), and our \teff{} and $R_*$ estimates did not make any assumptions about the age. 

\begin{figure}[tb]
    \begin{centering}
    \includegraphics[width=0.49\textwidth]{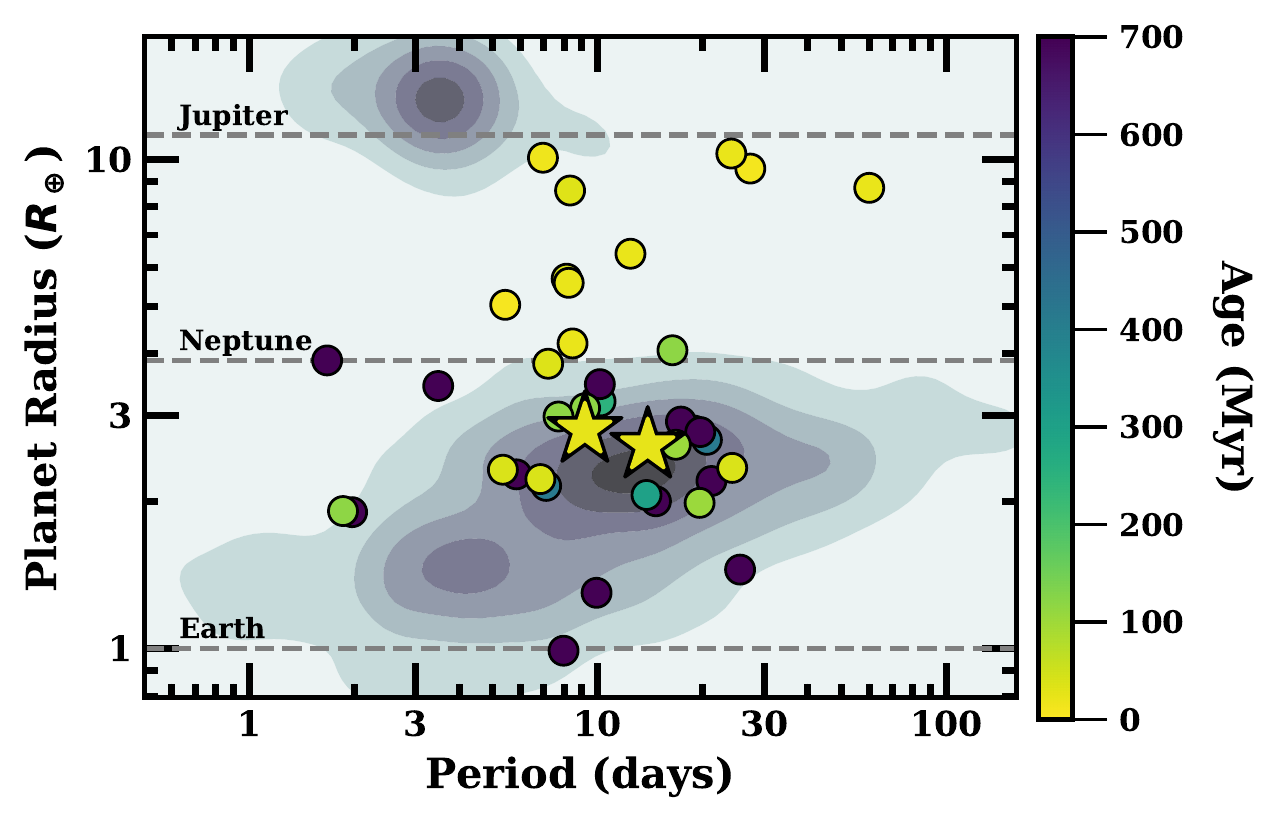}
    \end{centering}
    \caption{A comparison of the period-radius distribution of \starname{} planets, other transiting planets from young associations (colored dots), and from the field (background contours). The young planets are colored according to their ages. Planets orbiting field stars are shown as density contours. \planetname{} and \planetnametwo{} are shown as stars. Planet data is taken from Table 9 of \citet{newton_groupx_2022}, with the addition of the young planets discovered by \citet{zhou_abdor_2022}, \citet{bouma_kepler_2022}, and \citet{Barber_thyme_2022}}.
    \label{fig:radius}
\end{figure}

While the observed young-planet sample has grown dramatically in the last five years \citep[e.g.,][]{mann_k2-33_2016, david_neptune-sized_2016, newton_tess_2019, benatti_possibly_2019}, there are still few planets younger than $200 Myr$, and very few known multi-planet systems of that age, so this discovery radically expands our sample of young planetary systems. Interestingly, both planets have radii comparable to field-age stars, while most of the young ($< 100$\,Myr) transiting planets land in the sub-Saturn desert (4--10 $R_\oplus$; Figure~\ref{fig:radius}). This is less compelling if we adopt the older (100-200\,Myr) age, but most planets at that age still appear inflated compared to their older counterparts \citep[e.g.,][]{Newton_tess_2022}. 

\red{The small size of the planets may be caused by photoevaporation of their atmospheres by high energy radiation from the host star. However, comparison to the similar system V1298 Tau, a $23$ Myr, multi-planet system in Group 29 \cite{david_four_2019}, shows that this may be unlikely. \citet{poppenhaeger_x-ray_2021} find the x-ray luminosity, $L_X$, of V1298 Tau to be $L_X=10^{30.1} erg/s$. The x-ray luminosity of TOI 1097, calculated using the flux found by \citet{freund_stellar_2022}, is comparatively lower, at $L_X=10^{29.07}$. If photoevaporation is the driving factor of the planet's sizes we would expect \starname{}, being a similar age, and having lower $L_X$, to have larger planets than V1298 Tau.  However, V1298 Tau c and d, with periods of $8.25$ and $12.40$ days respectively, have sizes $R_{P,c} = 5.59 R_{\oplus}$, and $R_{P,d}=6.41 R_{\oplus}$, about twice the size of TOI 1097 b and c. }

The orbital periods of the planets also fall near a 3:2 mean motion resonance, making this system one of only a few known young systems near resonance \citep[e.g.,][]{feinstein_v1298_2022}. The mechanism responsible for resonant chains is still unknown, and establishing the timescale in which they form is critical for understanding this process.

% Future work/observing prospects
As a new and nearby association, with hundreds of candidate members, \assoc{} is an excellent subject for future observations and research. We do not expect that the membership presented here is either complete or contaminant-free, so additional studies on the membership and additional RV measurements are needed to better delineate the association members. Further observations of low-mass members to measure their Li abundance would improve the limits on the LDB age, which is currently limited by the small number of Li detections in the relevant mass range.
New planet searches focusing on candidate members may find more $\sim30$ Myr planets within the association, further increasing the sample of young planets, while future studies on the relationship between \assoc{} and the nearby young associations could improve understanding of the cloud collapse and star formation process.

The planet host, \starname, is also a promising subject for follow-up given its proximity to Earth and bright magnitude ($G=9.14$). Additional observations may help to solidify its membership in \assoc{}, or reject it as a member. Further characterization of the planets, including mass measurements and better constraints on eccentricity, may be possible with a search for transit timing variations. 

% - - - - - - - - - - - - - - - - - - - - - - - - -
% - - - - - - - - - - - - - - - - - - - - - - - - -
\acknowledgments

This work was made possible by grants to A.W.M. from the {\it TESS} Guest Investigator Program (80NSSC21K1054), NASA's Astrophysics Data Analysis Program (80NSSC19K0583), and the NSF CAREER grant (AST-2143763). M.L.W. and M.G.B were supported by the NC Space Grant Graduate Research program. SNQ acknowledges support from the TESS GI Program under award 80NSSC21K1056.

% TO's acknowledgment
Many thanks to Patricio, Carlos, Juan, Sergio, and Rodrigo at SOAR for helping through many nights of observations. 

% SOAR Acknowledgment
This research includes data from observations obtained at the Southern Astrophysical Research (SOAR) telescope, which is a joint project of the Minist\'{e}rio da Ci\^{e}ncia, Tecnologia e Inova\c{c}\~{o}es (MCTI/LNA) do Brasil, the US National Science Foundation’s NOIRLab, the University of North Carolina at Chapel Hill (UNC), and Michigan State University (MSU).

% TESS Acknowledgment
This paper includes data collected by the TESS mission, which are publicly available from the Mikulski Archive for Space Telescopes (MAST). Funding for the TESS mission is provided by NASA’s Science Mission Directorate.
We acknowledge the use of public TOI Release data from pipelines at the TESS Science Office and at the TESS Science Processing Operations Center.

% ExoFOP Acknowledgement
The Exoplanet Follow-up Observing Program (ExoFOP) website is designed to optimize resources and facilitate collaboration in follow-up studies of exoplanet candidates. ExoFOP-TESS serves as a repository for community-gathered follow-up data on TESS (Transiting Exoplanet Survey Satellite) planet candidates by allowing upload and display of data and derived astrophysical parameters.

This dataset or service is made available by the NASA Exoplanet Science Institute at IPAC, which is operated by the California Institute of Technology under contract with the National Aeronautics and Space Administration. 

% Gaia Acknowledgement
This work has made use of data from the European Space Agency (ESA) mission \emph{Gaia} \footnote{\url{https://www.cosmos.esa.int/gaia}}, processed by the \emph{Gaia} Data Processing and Analysis Consortium (DPAC)\footnote{\url{https://www.cosmos.esa.int/web/gaia/dpac/consortium}}. Funding for the DPAC has been provided by national institutions, in particular the institutions participating in the \emph{Gaia} Multilateral Agreement.  

% Vizier Acknowledgment
This research has made use of the VizieR catalogue access tool, CDS, Strasbourg, France. The original description of the VizieR service was published in A\&AS 143, 23. Resources supporting this work were provided by the NASA High-End Computing (HEC) Program through the NASA Advanced Supercomputing (NAS) Division at Ames Research Center for the production of the SPOC data products. 

% LCO Acknowledgement
This work makes use of observations from the LCOGT NRES network.

% Gemini Zorro Acknowledgment
Observations in the paper made use of the High-Resolution Imaging instrument(s) Zorro. Zorro was funded by the NASA Exoplanet Exploration Program and built at the NASA Ames Research Center by Steve B. Howell, Nic Scott, Elliott P. Horch, and Emmett Quigley. Zorro was mounted on the Gemini South telescope of the international Gemini Observatory, a program of NSF’s NOIRLab, which is managed by the Association of Universities for Research in Astronomy (AURA) under a cooperative agreement with the National Science Foundation. on behalf of the Gemini partnership: the National Science Foundation (United States), National Research Council (Canada), Agencia Nacional de Investigaci\'{o}n y Desarrollo (Chile), Ministerio de Ciencia, Tecnología e Innovación (Argentina), Minist\'{e}rio da Ci\^{e}ncia, Tecnologia e Inova\c{c}\~{o}es (Brazil), and Korea Astronomy and Space Science Institute (Republic of Korea).

% Other?
Based on observations made with the Italian {\it Telescopio Nazionale Galileo} (TNG) operated by the {\it Fundaci\'on Galileo Galilei} (FGG) of the {\it Istituto Nazionale di Astrofisica} (INAF) at the {\it Observatorio del Roque de los Muchachos} (La Palma, Canary Islands, Spain). Part of this research was carried out at the Jet Propulsion Laboratory, California Institute of Technology, under a contract with the National Aeronautics and Space Administration (NASA). The HARPS-N project has been funded by the Prodex Program of the Swiss Space Office (SSO), the Harvard University Origins of Life Initiative (HUOLI), the Scottish Universities Physics Alliance (SUPA), the University of Geneva, the Smithsonian Astrophysical Observatory (SAO), and the Italian National Astrophysical Institute (INAF), the University of St Andrews, Queens University Belfast, and the University of Edinburgh. DJA acknowledges support from the STFC via an Ernest Rutherford Fellowship (ST/R00384X/1). The work of HPO has been carried out within the framework of the NCCR PlanetS supported by the Swiss National Science Foundation under grants 51NF40\_182901 and 51NF40\_205606.

\vspace{5mm}
\facilities{TESS, SOAR 4m (Goodman HTS), LCOGT 1m (NRES), SMARTS 1.5m (CHIRON), TNG (HARPS-N), Gemini South (Zorro), }

\software{
% General
Astropy \citep{astropy_collaboration_astropy_2013, astropy_collaboration_astropy_2018},
Astroquery \citep{ginsburg_astroquery_2019},
% Plotting Tools
matplotlib \citep{hunter_matplotlib_2007}, 
\texttt{corner.py} \citep{foreman-mackey_corner.py:_2016},
% Association Membership
Comove \citep{tofflemire_tess_2021},
\banyan{} \citep{malo_bayesian_2012, gagne_banyan_2018},
% Spectra
BANZAI-NRES,
% Characterization
\texttt{misttborn.py} \citep{mann_k2_25_2016, johnson_k2-260_2018}, 
\textit{emcee} \citep{foreman-mackey_emcee_2013},
\textit{batman} \citep{kreidberg_batman_2015},
\texttt{celerite} \citep{foreman-mackey_fast_2017},
% False Positive Checking
\texttt{triceratops} \citep{giacalone_vetting_2021},
\texttt{MOLUSC} \citep{wood_characterizing_2021}
}

\clearpage

%% what goes in this table?
%% Name, EDR3 ID, RA, Dec, SpT(phot), Lithium, Friend?, Banyan prob, rotation period
%% do we want to denote the ones from:
%--https://ui.adsabs.harvard.edu/abs/2008hsf2.book..757T/abstract
%--https://ui.adsabs.harvard.edu/abs/2013MNRAS.435.1325M/abstract
%--https://arxiv.org/abs/2011.06621
%--https://ui.adsabs.harvard.edu/abs/2020AJ....159..166U/abstract
%--https://arxiv.org/abs/2102.05589
%--https://arxiv.org/abs/1807.02076 (this is the Crux one). 

\begin{longrotatetable}
    \begin{center}
    \begin{table}
    \caption{Members of \assoc}
    \label{table:members}
    \begin{tabular}{|ccccccccccc|}
    \hline
    DR3 ID & ra (\degree) & dec (\degree) &  $\pi$ (mas) & RUWE & $G$ & $G_{BP}-G_{RP}$ & $K_S$ & $FF^a$ & $P_{BANYAN}$ & TIC \\
    \hline
    Gaia DR3 5269346361575307264 & 121.3139 & -71.103 & 10.750 & 0.927 & 11.62 & 1.71 & 8.782 &  & 0.815 & TIC306779173 \\
    Gaia DR3 5314992445071183872 & 130.1828 & -57.550 & 13.887 & 1.354 & 14.22 & 2.74 & 10.338 &  & 0.554 & TIC45192378 \\
    Gaia DR3 5215182391566488448 & 135.4216 & -77.933 & 11.519 & 1.112 & 15.12 & 2.89 & 11.254 &  & 0.868 & TIC323574677 \\
    Gaia DR3 5304828971900544896 & 136.2097 & -56.294 & 12.413 & 1.159 & 12.64 & 2.15 & 9.334 &  & 0.706 & TIC384397468 \\
    Gaia DR3 5216186726719969792 & 139.3644 & -74.734 & 13.090 & 1.070 & 5.86 & -0.03 & 5.865 &  & 0.921 & TIC452468734 \\
    Gaia DR3 5219515292014933760 & 139.6958 & -70.615 & 10.086 & 1.078 & 15.67 & 3.13 & 11.636 &  & 0.540 & TIC303693668 \\
    Gaia DR3 5217812354662194048 & 140.9457 & -73.682 & 13.088 & 1.227 & 13.38 & 2.62 & 9.722 &  & 0.555 & TIC452522877 \\
    Gaia DR3 5250471114189790336 & 141.6836 & -63.023 & 11.940 & 2.249 & 14.26 & 2.94 & 10.325 &  & 0.546 & TIC360130454 \\
    Gaia DR3 5219351911459314048 & 142.6309 & -70.697 & 9.854 & 0.944 & 10.08 & 0.89 & 8.565 &  & 0.755 & TIC370330200 \\
    Gaia DR3 5217846851839896832 & 142.8546 & -73.747 & 12.832 & 1.589 & 9.19 & 0.80 & -- &  & 0.815 & TIC843283158 \\
    Gaia DR3 5217846817480160640 & 142.9057 & -73.751 & 12.804 & 1.966 & 13.90 & 3.13 & 9.823 &  & 0.728 & TIC452591875 \\
    Gaia DR3 5217554622264363008 & 143.5176 & -74.093 & 12.853 & 1.277 & 14.29 & 2.89 & 10.379 &  & 0.729 & TIC452604601 \\
    Gaia DR3 5250926999202194688 & 143.8354 & -62.367 & 12.017 & 1.206 & 14.45 & 2.86 & 10.552 &  & 0.730 & TIC361227905 \\
    Gaia DR3 5251098523021221376 & 144.8372 & -61.328 & 14.479 & 0.791 & 4.48 & -0.07 & 4.647 &  & 0.723 & TIC361834361 \\
    Gaia DR3 5244271552233795200 & 145.1523 & -67.755 & 13.104 & 1.108 & 16.46 & 3.81 & 11.824 &  & 0.618 & TIC370825476 \\
    Gaia DR3 5307852908070273792 & 145.5575 & -56.027 & 12.422 & 1.128 & 14.39 & 2.94 & 10.47 &  & 0.598 & TIC441744406 \\
    Gaia DR3 5218453026345066368 & 147.2531 & -71.634 & 12.754 & 2.566 & 13.90 & 2.55 & 9.5 &  & 0.582 & TIC371372421 \\
    Gaia DR3 5257836605156299776 & 147.5025 & -58.472 & 12.876 & 1.163 & 14.23 & 2.88 & 10.393 &  & 0.537 & TIC444536794 \\
    Gaia DR3 5218262707753445760 & 147.6715 & -71.783 & 11.126 & 1.111 & 14.79 & 2.94 & 10.878 &  & 0.887 & TIC371498920 \\
    Gaia DR3 5257392295070412032 & 148.5564 & -60.275 & 11.835 & 1.070 & 12.32 & 2.03 & 9.083 &  & 0.952 & TIC269691345 \\
    Gaia DR3 5259151277454731008 & 149.3857 & -58.390 & 10.059 & 14.572 & 15.49 & 3.31 & -- &  & 0.509 & TIC855167492 \\
    Gaia DR3 5258252942105518080 & 150.9937 & -59.401 & 10.204 & 1.285 & 14.23 & 2.93 & 10.285 &  & 0.552 & TIC462320070 \\
    Gaia DR3 5230321567172871040 & 151.3447 & -71.616 & 11.583 & 2.216 & 14.60 & 2.93 & 10.674 &  & 0.968 & TIC372515585 \\
    Gaia DR3 5230321361014440704 & 151.3556 & -71.624 & 11.671 & 5.079 & 12.45 & 2.56 & 8.802 &  & 0.952 & TIC372515598 \\
    Gaia DR3 5253295141117135104 & 152.1298 & -61.633 & 12.623 & 1.138 & 15.46 & 2.96 & 11.453 &  & 0.743 & TIC375752627 \\
    Gaia DR3 5246284685000163968 & 152.6949 & -65.380 & 9.227 & 1.214 & 12.94 & 2.02 & 9.756 &  & 0.502 & TIC376031475 \\
    Gaia DR3 5254979283697117440 & 154.0168 & -59.909 & 12.276 & 1.225 & 13.88 & 2.73 & 9.483 &  & 0.966 & TIC463556306 \\
    Gaia DR3 5255082603415689600 & 154.3064 & -59.640 & 12.378 & 1.181 & 15.42 & 3.23 & 11.297 &  & 0.968 & TIC463578988 \\
    Gaia DR3 5251591482193147776 & 154.7628 & -64.681 & 11.521 & 1.303 & 13.38 & 2.51 & 9.807 &  & 0.978 & TIC378035942 \\
    Gaia DR3 5251591477884660352 & 154.7696 & -64.676 & 11.567 & 1.119 & 6.48 & 0.08 & -- &  & 0.844 & TIC847861519 \\
    \hline
    \end{tabular}
    \end{table}
    \tablecomments{Table is truncated. The full version is available online.}
    \tablenotetext{a}{Friends are less than 30\,pc away and a tangential velocity within 3\kms\ of \starname.}
    \end{center}
\end{longrotatetable}

\clearpage

% Observation Table
\begin{table*}
    \begin{center}
    \caption{Observations of \assoc{} candidates.}
    \begin{tabular}{|ccccccc|}
    \hline
    Name & SpT & Telescope & ObsDate & $M_{K}$ & EW(Li) & $P_{BANYAN}$ \\
        &   &  & YYYYMMDD & mag & m\AA & \\
    \hline
    TIC 427036962 & M3 & Goodman/SOAR & 20210329, 20210919 & 6.52 & $413.0$ & 0.961 \\
    TIC 258101273 & M1 & Goodman/SOAR & 20210329 & 4.83 & $29.0$ & 0.979 \\
    TIC 259726904 & M3 & Goodman/SOAR & 20210329 & 5.01 & $< 10$ & 0.997 \\
    TIC 68515382  & M3 & Goodman/SOAR & 20210423 & 6.09 & $153.0$ & 0.993 \\
    TIC 378413560 & M3 & Goodman/SOAR & 20210423 & 4.98 & $< 10$ & 0.902 \\
    TIC 253067348 & M4 & Goodman/SOAR & 20210423 & 6.38 & $680.0$ & 0.992 \\
    TIC 303048907 & M4 & Goodman/SOAR & 20210423 & 5.90 & $< 10$ & 0.928 \\
    TIC 378126824 & M3 & Goodman/SOAR & 20210423 & 5.27 & $20.0$ & 0.852 \\
    TIC 406249571 & M3 & Goodman/SOAR & 20210423 & 5.73 & $583.0$ & 0.986 \\
    TIC 401561267 & M3 & Goodman/SOAR & 20210501 & 5.42 & $34.0$ & 0.962 \\
    TIC 453766186 & M3 & Goodman/SOAR & 20210501 & 5.63 & $< 10$ & 0.896 \\
    TIC 453808153 & M3 & Goodman/SOAR & 20210501 & 5.78 & $< 10$ & 0.934 \\
    TIC 401484858 & M1 & Goodman/SOAR & 20210507 & 6.11 & $16.0$ & 0.986 \\
    TIC 402030604 & M3 & Goodman/SOAR & 20210507 & 5.72 & $< 10$ & 0.897 \\
    TIC 335366271 & M3 & Goodman/SOAR & 20210507 & 5.93 & $33.0$ & 0.998 \\
    TIC 451425923 & M1 & Goodman/SOAR & 20210507 & 6.23 & $< 10$ & 0.821 \\
    TIC 402808278 & M3 & Goodman/SOAR & 20210507, 20210810 & $5.70$ & < 10 & 0.504 \\
    TIC 425871236 & M3 & Goodman/SOAR & 20210810 & 5.53 & $< 10$ & 0.974 \\
    TIC 299610396 & M3 & Goodman/SOAR & 20210810 & 5.78 & $< 10$ & 0.891 \\
    TIC 97882429  & M3 & Goodman/SOAR & 20210810 & 5.84 & $< 10$ & 0.663 \\
    Gaia5227091236372916864 & M4 & Goodman/SOAR & 20210810 & 6.09 & $414.0$ & 0.996 \\
    TIC256168939 & M3 & Goodman/SOAR & 20210810 & 5.89 & $204.0$ & 0.979 \\
    TIC189421351 & M0 & Goodman/SOAR & 20210919 & 4.07 & $313.0$ & 0.728 \\
    TIC461148251 & M3 & Goodman/SOAR & 20210919 & 5.17 & $< 10$ & 0.882 \\
    TIC443273186 & M3 & Goodman/SOAR & 20210920 & 6.90 & $< 10$ & 0.985 \\
    TIC361571108 & M5 & Goodman/SOAR & 20210920 & 6.87 & $456.0$ & 0.958 \\
    % - - - - - - - - - - - - - - - - - - - - - - - - - - - - - - - - - - - - - - 
    % 2MASS J13444279-6347495 & K4 & NRES/LCO & 20211231 & $-4.3\pm0.2$ & 9.44 & 356 & 0.847 \\
    % HD 95086    & A8 & NRES/LCO   & 20211216    & $20.1\pm0.4$ & 2.10 & $< 10$ & 0.990 \\
    % HD 124831   & G3 & NRES/LCO   & 20211227    & $12.67\pm0.36$ & 2.69 & 103.0 & 0.890 \\
    % HD 119269   & G3 & NRES/LCO   & 20211224    & $9.68\pm0.3$ & 2.83 & 110.9 & 0.981 \\
    % HD 104919   & G9 & NRES/LCO   & 20220512    & $15.37\pm0.36$ & 2.58 & 178.0 & 0.995 \\
    % TYC 8693-607-1 & -- & NRES/LCO & 20220515   & $7.55\pm 0.05$ & 3.57 & $< 10$ & 0.796 \\
    % CD-49 8410  & K0 & NRES/LCO   & 20211228    & $7.05\pm0.17$ & 2.92 & 424.7 & 0.529 \\
    % CD-48 8201  & K3 & NRES/LCO   & 20211228    & $8.33\pm0.34$ & 2.87 & 357.8 & 0.610 \\
    % GSC 08659-01804* & K7 & NRES/LCO & 20211227 & $42.20\pm0.13$ & 2.64 & 427.8 & 0.958 \\
    TYC 9034-968-1 & K2  & CHIRON/SMARTS    & 20210430 & 3.70 & 250.0 & 0.999 \\
    TYC 8992-346-1 &     & CHIRON/SMARTS    & 20210504 & 3.34 & 340.0 & 0.999 \\
    \hline
    \end{tabular}
    % \tablenotetext{*}{GSC 08659-01804 is an SB2 binary. EW(Li) measured by summing the EW of both lines.}
    \label{table:obs}
    \end{center}
\end{table*}

\appendix
\section{BANYAN Definitions of LCC Sub-Populations}\label{app:lcc}

Since \assoc{} is near to several other young moving groups, correctly defining this new population requires not only a description of it, but also an accurate description of the nearby groups. Since the publication of \citet{gagne_banyan_2018}, significant substructure has been found within LCC \cite[e.g.]{pecaut_star_2016, goldman_large_2018, kerr_stars_2021}. Some of the LCC sub-populations found by \citet{goldman_large_2018} and \citet{kerr_stars_2021} are near \assoc{} (spatially and kinematically), raising the risk of contamination. To handle this, we defined each of the populations as a separate moving group within \banyan{}, which should provide a more accurate relative probability of membership between groups. 

The LCC sub-populations were independently found by G18 and K21, who used different naming schemes for the groups. We use the names from G18, and note that the groups A0\footnote{also known as Musca \citep{mann_tess_2022}}, A, B, and C correspond to B, C, E and D in K21. K21's group LCC A has no counterpart in the Goldman paper, but is more commonly known as \epscha, and was already in \banyan{}.

\begin{table*}
\begin{center}
\caption{Parameters of LCC Sub-populations} 
\begin{tabular}{|c c c c c c c c c c c c|}
\hline 
% Get errors on these numbers
Goldman & Kerr & $N_{Goldman}$ & $N_{Kerr}$ & $N_{Total}$ & X & Y(pc) & Z & U & V & W & Age \\ 
Name & Name & & & & (pc) & (pc) & (pc) & (\kms) & (\kms) & (\kms) & (Myr)  \\
\hline 
A0 & B & 49 &  70 & 79 & 52 &-87 &-15 & -8.48 & -21.69 & -8.88 & 10.6$\pm$1.9\\
A & C & 149 & 197 & 211 & 53 & -91 & -3 & -9.80 & -19.71 & -7.79& 12.0$\pm$1.8 \\
B & E & 317 & 441 & 474 & 53 & -94 & 11 & -8.92 & -20.30 & -6.88& 14.9$\pm$1.6 \\
C & D & 487 &  69 & 494 & 61 & -96 & 22  & -8.55 & -20.36 & -6.14& 16.6$\pm$1.1 \\
\hline 
\end{tabular}
\end{center}
\label{table:lcc}
\end{table*}

First we define the membership of the LCC sub-populations by combining the membership lists from G18 and K21. We obtain the candidate members of groups A0, A, B, and C from G18 Table 2, and use the provided \gaia{} DR2 coordinates to crossmatch the sources with \gaia{} DR3, searching a 1 arcminute area around each star, and taking the closest source as a match. 
The K21 candidate members are obtained from Table 3 of K21, and the provided \gaia{} DR2 coordinates used to crossmatch the sources with \gaia{} eDR3, using the same radius as for the G18 match.
We then combine the membership lists for the two, using their \gaia{} DR3 source IDs. 

For sub-populations A0, A, and B the majority of the candidate members were recovered by both surveys, and the selections of each are similar, with a common core, and some variation on the outskirts. However, few of the sub-population C candidates were recovered by K21, so that the G18 membership list contains the vast majority of the members, and has a significantly larger extent than the K21 membership.

To calculate the center vector and covariance matrix of each sub-population we use the combined membership lists, cut to only those stars with \gaia DR2 RV measurements. The covariance matrices of all four populations are shown below, and the values of the center vector listed in Table \ref{table:lcc}.

Musca (LCC-A0)
\begin{align}
    \bar{\bar\Sigma}_{A0} &= \begin{bmatrix}
        6.774 & -3.706 & 12.025 & 1.906 & -2.431 & 1.222 \\
        -3.706 & 13.499 & -1.037 & 5.263 & -7.798 & -1.359 \\
        12.025 & -1.037 & 27.141 & 9.426 & -13.696 & 1.53  \\
        1.906 & 5.263 & 9.426 & 19.13 & -30.034 & -3.405 \\
        -2.431 & -7.798 & -13.696 & -30.034 & 47.339 & 5.521 \\
        1.222 & -1.359  & 1.53 & -3.405 & 5.521 & 1.094
    \end{bmatrix}.\notag
\end{align}

LCC-A
\begin{align}
    \bar{\bar\Sigma}_{A} &= \begin{bmatrix}
     5.364 & -0.873 & -0.807 & 0.165 & 0.614 & 0.16 \\
    -0.873 & 17.4   & -4.569 & 1.271 & -0.026 & -0.283 \\
    -0.807 & -4.569 & 9.034 & 0.815 & -1.742 & 0.378 \\
     0.165 &  1.271 & 0.815 & 5.792 & -9.457 & -0.28 \\
     0.614 & -0.026 & -1.742 & -9.457 & 16.059 & 0.449 \\
     0.16  & -0.283 & 0.378 & -0.28 & 0.449 & 0.149 
    \end{bmatrix}.\notag
\end{align}

LCC-B
\begin{align}
    \bar{\bar\Sigma}_{B} &= \begin{bmatrix}
     54.486 & 3.078 & -4.913 & 1.8 & 3.413 & -0.675 \\
      3.078 & 16.729 & -4.972 & 0.169 & 1.887 & -0.477 \\
     -4.913 & -4.972 & 28.202 & 0.205 & -1.079 & 2.185 \\
      1.8 & 0.169 & 0.205 & 2.052 & -3.02 & 0.348 \\
      3.413 & 1.887 & -1.079 & -3.02 & 6.021 & -0.673 \\
     -0.675 & -0.477 & 2.185 & 0.348 & -0.673 & 0.392
    \end{bmatrix}.\notag
\end{align}

LCC-C
\begin{align}
    \bar{\bar\Sigma}_{C} &= \begin{bmatrix}
    114.137 &  4.809  & 5.911 & 7.288 & 0.418 & 0.815 \\
     4.809 & 44.83  & -7.989 & 1.304 & 3.228 & -1.043 \\
     5.911 & -7.989 & 16.065 & 0.677 &-0.809 &  0.81  \\
     7.288 &  1.304 &  0.677 & 3.529 &-3.895 &  0.911 \\
     0.418 &  3.228 & -0.809 &-3.895 & 6.345 & -1.198 \\
     0.815 & -1.043 &  0.81  & 0.911 & -1.198 & 0.58 
    \end{bmatrix}.\notag
\end{align}

We add the groups to \banyan{} using the parameters listed above. To test the recovery of the initial samples we run \banyan{} on a sample of stars from \gaia{} EDR3 within 100pc of the central position of LCC \citet[using the original definition from ][]{gagne_banyan_2018}). We use all resulting candidates with a kinematic membership probability greater than $50\%$ as the output sample for each population.

We recover a majority of the input stars for all four populations, with A0 having the lowest recovery rate at $56\%$, and C having the highest, recovering $92\%$ of the initial sample. The A0 population has the smallest number of stars, and only a fraction of them have radial velocity measurements on which to base the \banyan{} definition, a likely contributor to the lower recovery fraction. For sub-populations A, B, and C, the recovery rates were $>75\%$. Many of the stars that are not recovered are placed into a different sub-population (so still part of LCC). In particular, there was significant cross-contamination between groups A and B, and between groups B and C. This was expected, as both groups show significant overlap in spatial and kinematic space, and it is likely the input lists were imperfect.

\section{Revised ages of the LCC sub-groups}\label{app:ages}

Our determination that \assoc{} is not part of the known LCC sub-groups was based on differences in both kinematics and age. The latter is complicated by the discrepant age in the literature; \citet{goldman_large_2018} assigned ages of 7--10\,Myr for all populations while \citet{kerr_stars_2021} find ages ranging from 13-23\,Myr for the same groups. Using the \citet{kerr_stars_2021} ages, the oldest group is marginally consistent with our age for \assoc{} (25--30\,Myr). These differences are likely a reflection of differences in methodology and models, as evident by the fact that both references agree on ordering of groups in terms of age. 
%\citet{mann_tess_2022} revisited the age of one subgroup (Musca, known as A0 in Goldman et al. 2018 and LCC B in Kerr et al. 2021) and found an age of 11$\pm$2, which is in between and marginally consistent with both (7\,Myr and 13\,Myr). 

To ensure a more robust comparison, we place the ages of each population on a consistent scale as was done for \assoc{} in Section~\ref{sec:isochrone}. We adopt a target selection for each of the four sub-groups using our updated \banyan{} model described in Section~\ref{app:lcc}. We then fit each group using a mixture identical to what was described in Section~\ref{sec:isochrone} and \citet{mann_tess_2022}. For simplicity, we restrict our analysis to the PARSEC models and solar metallicity. All fits are run with 20 walkers until they passed 50 times the autocorrelation time (checking every 5,000 steps), for a total of 10,000--30,000 steps.  

In all cases, our ages are between the values from \citet{goldman_large_2018} and \citet{kerr_stars_2021}. We summarize the results in Table~\ref{table:lcc}. Importantly, we find that the oldest group is 16.6$\pm$1.1\,Myr, inconsistent with our identically-derived isochronal age ($26.0\pm2.1$\,Myr) and our LDB age for \assoc{}.

Our fits are mildly sensitive to membership selection, assumed metallicity, or the model grid. Using the original membership list from \citet{goldman_large_2018} or \citet{kerr_stars_2021} changes our ages at the 1--2\,Myr level, small compared to the difference between the two literature ages (5--13\,Myr). Adjusting the assumed metallicity at the 0.1\,dex level or swapping to the DSEP-magnetic models also changes the derived ages by $\lesssim$2\,Myr, but impacted all groups in the same direction (including \assoc). 

\bibliography{full_library}{}

\begin{thebibliography}{}
\expandafter\ifx\csname natexlab\endcsname\relax\def\natexlab#1{#1}\fi
\providecommand{\url}[1]{\href{#1}{#1}}
\providecommand{\dodoi}[1]{doi:~\href{http://doi.org/#1}{\nolinkurl{#1}}}
\providecommand{\doeprint}[1]{\href{http://ascl.net/#1}{\nolinkurl{http://ascl.net/#1}}}
\providecommand{\doarXiv}[1]{\href{https://arxiv.org/abs/#1}{\nolinkurl{https://arxiv.org/abs/#1}}}

\bibitem[{{Adibekyan} {et~al.}(2015){Adibekyan}, {Figueira}, {Santos}, {Sousa},
  {Faria}, {Delgado-Mena}, {Oshagh}, {Tsantaki}, {Hakobyan}, {Gonz{\'a}lez
  Hern{\'a}ndez}, {Su{\'a}rez-Andr{\'e}s}, \& {Israelian}}]{Adibekyan-15}
{Adibekyan}, V., {Figueira}, P., {Santos}, N.~C., {et~al.} 2015, \aap, 583,
  A94, \dodoi{10.1051/0004-6361/201527120}

\bibitem[{{Adibekyan} {et~al.}(2012){Adibekyan}, {Sousa}, {Santos}, {Delgado
  Mena}, {Gonz{\'a}lez Hern{\'a}ndez}, {Israelian}, {Mayor}, \&
  {Khachatryan}}]{Adibekyan-12}
{Adibekyan}, V.~Z., {Sousa}, S.~G., {Santos}, N.~C., {et~al.} 2012, \aap, 545,
  A32, \dodoi{10.1051/0004-6361/201219401}

\bibitem[{{Aller} {et~al.}(2020){Aller}, {Lillo-Box}, {Jones}, {Miranda}, \&
  {Barcel{\'o} Forteza}}]{aller_tpfplotter_2020}
{Aller}, A., {Lillo-Box}, J., {Jones}, D., {Miranda}, L.~F., \& {Barcel{\'o}
  Forteza}, S. 2020, \aap, 635, A128, \dodoi{10.1051/0004-6361/201937118}

\bibitem[{{Andrews} {et~al.}(2022){Andrews}, {Curtis}, {Chanam{\'e}},
  {Ag{\"u}eros}, {Schuler}, {Kounkel}, \& {Covey}}]{andrews_young_2022}
{Andrews}, J.~J., {Curtis}, J.~L., {Chanam{\'e}}, J., {et~al.} 2022, \aj, 163,
  275, \dodoi{10.3847/1538-3881/ac6952}

\bibitem[{Asplund {et~al.}(2009)Asplund, Grevesse, Sauval, \&
  Scott}]{asplund_chemical_2009}
Asplund, M., Grevesse, N., Sauval, A.~J., \& Scott, P. 2009, Annual Review of
  Astronomy and Astrophysics, 47, 481,
  \dodoi{10.1146/annurev.astro.46.060407.145222}

\bibitem[{{Astropy Collaboration} {et~al.}(2013){Astropy Collaboration},
  Robitaille, Tollerud, Greenfield, Droettboom, Bray, Aldcroft, Davis,
  Ginsburg, Price-Whelan, Kerzendorf, Conley, Crighton, Barbary, Muna,
  Ferguson, Grollier, Parikh, Nair, Unther, Deil, Woillez, Conseil, Kramer,
  Turner, Singer, Fox, Weaver, Zabalza, Edwards, Azalee~Bostroem, Burke, Casey,
  Crawford, Dencheva, Ely, Jenness, Labrie, Lim, Pierfederici, Pontzen, Ptak,
  Refsdal, Servillat, \& Streicher}]{astropy_collaboration_astropy_2013}
{Astropy Collaboration}, Robitaille, T.~P., Tollerud, E.~J., {et~al.} 2013,
  Astronomy and Astrophysics, 558, A33, \dodoi{10.1051/0004-6361/201322068}

\bibitem[{{Astropy Collaboration} {et~al.}(2018){Astropy Collaboration},
  {Price-Whelan}, {Sip{\H{o}}cz}, {G{\"u}nther}, {Lim}, {Crawford}, {Conseil},
  {Shupe}, {Craig}, {Dencheva}, {Ginsburg}, {VanderPlas}, {Bradley},
  {P{\'e}rez-Su{\'a}rez}, {de Val-Borro}, {Aldcroft}, {Cruz}, {Robitaille},
  {Tollerud}, {Ardelean}, {Babej}, {Bach}, {Bachetti}, {Bakanov}, {Bamford},
  {Barentsen}, {Barmby}, {Baumbach}, {Berry}, {Biscani}, {Boquien}, {Bostroem},
  {Bouma}, {Brammer}, {Bray}, {Breytenbach}, {Buddelmeijer}, {Burke},
  {Calderone}, {Cano Rodr{\'\i}guez}, {Cara}, {Cardoso}, {Cheedella}, {Copin},
  {Corrales}, {Crichton}, {D'Avella}, {Deil}, {Depagne}, {Dietrich}, {Donath},
  {Droettboom}, {Earl}, {Erben}, {Fabbro}, {Ferreira}, {Finethy}, {Fox},
  {Garrison}, {Gibbons}, {Goldstein}, {Gommers}, {Greco}, {Greenfield},
  {Groener}, {Grollier}, {Hagen}, {Hirst}, {Homeier}, {Horton}, {Hosseinzadeh},
  {Hu}, {Hunkeler}, {Ivezi{\'c}}, {Jain}, {Jenness}, {Kanarek}, {Kendrew},
  {Kern}, {Kerzendorf}, {Khvalko}, {King}, {Kirkby}, {Kulkarni}, {Kumar},
  {Lee}, {Lenz}, {Littlefair}, {Ma}, {Macleod}, {Mastropietro}, {McCully},
  {Montagnac}, {Morris}, {Mueller}, {Mumford}, {Muna}, {Murphy}, {Nelson},
  {Nguyen}, {Ninan}, {N{\"o}the}, {Ogaz}, {Oh}, {Parejko}, {Parley}, {Pascual},
  {Patil}, {Patil}, {Plunkett}, {Prochaska}, {Rastogi}, {Reddy Janga},
  {Sabater}, {Sakurikar}, {Seifert}, {Sherbert}, {Sherwood-Taylor}, {Shih},
  {Sick}, {Silbiger}, {Singanamalla}, {Singer}, {Sladen}, {Sooley},
  {Sornarajah}, {Streicher}, {Teuben}, {Thomas}, {Tremblay}, {Turner},
  {Terr{\'o}n}, {van Kerkwijk}, {de la Vega}, {Watkins}, {Weaver}, {Whitmore},
  {Woillez}, {Zabalza}, \& {Astropy
  Contributors}}]{astropy_collaboration_astropy_2018}
{Astropy Collaboration}, {Price-Whelan}, A.~M., {Sip{\H{o}}cz}, B.~M., {et~al.}
  2018, \aj, 156, 123, \dodoi{10.3847/1538-3881/aabc4f}

\bibitem[{Baraffe \& Chabrier(2010)}]{baraffe_effect_2010}
Baraffe, I., \& Chabrier, G. 2010, Astronomy \& Astrophysics, 521, A44,
  \dodoi{10.1051/0004-6361/201014979}

\bibitem[{Baraffe {et~al.}(2015)Baraffe, Homeier, Allard, \&
  Chabrier}]{baraffe_new_2015}
Baraffe, I., Homeier, D., Allard, F., \& Chabrier, G. 2015, Astronomy \&
  Astrophysics, 577, A42, \dodoi{10.1051/0004-6361/201425481}

\bibitem[{{Barber} {et~al.}(2022){Barber}, {Mann}, {Bush}, {Tofflemire},
  {Kraus}, {Krolikowski}, {Vanderburg}, {Fields}, {Newton}, {Owens}, \&
  {Thao}}]{Barber_thyme_2022}
{Barber}, M.~G., {Mann}, A.~W., {Bush}, J.~L., {et~al.} 2022, arXiv e-prints,
  arXiv:2206.08383

\bibitem[{Bell {et~al.}(2015)Bell, Mamajek, \&
  Naylor}]{bell_self-consistent_2015}
Bell, C. P.~M., Mamajek, E.~E., \& Naylor, T. 2015, Monthly Notices of the
  Royal Astronomical Society, 454, 593, \dodoi{10.1093/mnras/stv1981}

\bibitem[{Belokurov {et~al.}(2020)Belokurov, Penoyre, Oh, Iorio, Hodgkin,
  Evans, Everall, Koposov, Tout, Izzard, Clarke, \&
  Brown}]{belokurov_unresolved_2020}
Belokurov, V., Penoyre, Z., Oh, S., {et~al.} 2020, Monthly Notices of the Royal
  Astronomical Society, 496, 1922, \dodoi{10.1093/mnras/staa1522}

\bibitem[{Benatti {et~al.}(2019)Benatti, Nardiello, Malavolta, Desidera,
  Borsato, Nascimbeni, Damasso, D'Orazi, Mesa, Messina, Esposito, Bignamini,
  Claudi, Covino, Lovis, \& Sabotta}]{benatti_possibly_2019}
Benatti, S., Nardiello, D., Malavolta, L., {et~al.} 2019, arXiv:1904.01591
  [astro-ph].
\newblock \url{http://arxiv.org/abs/1904.01591}

\bibitem[{Binks \& Jeffries(2014)}]{binks_lithium_2014}
Binks, A.~S., \& Jeffries, R.~D. 2014, Monthly Notices of the Royal
  Astronomical Society: Letters, 438, L11, \dodoi{10.1093/mnrasl/slt141}

\bibitem[{Binks {et~al.}(2021)Binks, Jeffries, Jackson, Franciosini, Sacco,
  Bayo, Magrini, Randich, Arancibia-Silva, Bergemann, Bragaglia, Gilmore,
  Gonneau, Hourihane, Jofr{\'e}, Korn, Morbidelli, Prisinzano, Worley, \&
  Zaggia}]{binks_gaia-eso_2021}
Binks, A.~S., Jeffries, R.~D., Jackson, R.~J., {et~al.} 2021, Monthly Notices
  of the Royal Astronomical Society, 505, 1280, \dodoi{10.1093/mnras/stab1351}

\bibitem[{Bochanski {et~al.}(2007)Bochanski, West, Hawley, \&
  Covey}]{bochanski_low-mass_2007}
Bochanski, J.~J., West, A.~A., Hawley, S.~L., \& Covey, K.~R. 2007, The
  Astronomical Journal, 133, 531, \dodoi{10.1086/510240}

\bibitem[{Bohn {et~al.}(2020{\natexlab{a}})Bohn, Kenworthy, Ginski, Rieder,
  Mamajek, Meshkat, Pecaut, Reggiani, Boer, Keller, Snik, \&
  Southworth}]{bohn_two_2020}
Bohn, A.~J., Kenworthy, M.~A., Ginski, C., {et~al.} 2020{\natexlab{a}}, The
  Astrophysical Journal, 898, L16, \dodoi{10.3847/2041-8213/aba27e}

\bibitem[{Bohn {et~al.}(2020{\natexlab{b}})Bohn, Kenworthy, Ginski, Manara,
  Pecaut, de~Boer, Keller, Mamajek, Meshkat, Reggiani, Todorov, \&
  Snik}]{bohn_young_2020}
---. 2020{\natexlab{b}}, Monthly Notices of the Royal Astronomical Society,
  492, 431, \dodoi{10.1093/mnras/stz3462}

\bibitem[{Bohn {et~al.}(2021)Bohn, Ginski, Kenworthy, Mamajek, Pecaut,
  Mugrauer, Vogt, Adam, Meshkat, Reggiani, \& Snik}]{bohn_discovery_2021}
Bohn, A.~J., Ginski, C., Kenworthy, M.~A., {et~al.} 2021, Astronomy \&
  Astrophysics, 648, A73, \dodoi{10.1051/0004-6361/202140508}

\bibitem[{Booth {et~al.}(2021)Booth, del Burgo, \& Hambaryan}]{booth_age_2021}
Booth, M., del Burgo, C., \& Hambaryan, V.~V. 2021, Monthly Notices of the
  Royal Astronomical Society, 500, 5552, \dodoi{10.1093/mnras/staa3631}

\bibitem[{{Bouma} {et~al.}(2022){Bouma}, {Kerr}, {Curtis}, {Isaacson},
  {Hillenbrand}, {Howard}, {Kraus}, {Bieryla}, {Latham}, {Petigura}, \&
  {Huber}}]{bouma_kepler_2022}
{Bouma}, L.~G., {Kerr}, R., {Curtis}, J.~L., {et~al.} 2022, arXiv e-prints,
  arXiv:2205.01112.
\newblock \doarXiv{2205.01112}

\bibitem[{Bressan {et~al.}(2012)Bressan, Marigo, Girardi, Salasnich, Dal~Cero,
  Rubele, \& Nanni}]{bressan_parsec_2012}
Bressan, A., Marigo, P., Girardi, L., {et~al.} 2012, Monthly Notices of the
  Royal Astronomical Society, 427, 127,
  \dodoi{10.1111/j.1365-2966.2012.21948.x}

\bibitem[{Buchhave {et~al.}(2012)Buchhave, Latham, Johansen, Bizzarro, Torres,
  Rowe, Batalha, Borucki, Brugamyer, Caldwell, Bryson, Ciardi, Cochran, Endl,
  Esquerdo, Ford, Geary, Gilliland, Hansen, Isaacson, Laird, Lucas, Marcy,
  Morse, Robertson, Shporer, Stefanik, Still, \&
  Quinn}]{buchhave_abundance_2012}
Buchhave, L.~A., Latham, D.~W., Johansen, A., {et~al.} 2012, Nature, 486, 375,
  \dodoi{10.1038/nature11121}

\bibitem[{Burke {et~al.}(2004)Burke, Pinsonneault, \&
  Sills}]{burke_theoretical_2004}
Burke, C.~J., Pinsonneault, M.~H., \& Sills, A. 2004, The Astrophysical
  Journal, 604, 272, \dodoi{10.1086/381242}

\bibitem[{{Cao} {et~al.}(2022){Cao}, {Pinsonneault}, {Hillenbrand}, \&
  {Kuhn}}]{cao_age_2022}
{Cao}, L., {Pinsonneault}, M.~H., {Hillenbrand}, L.~A., \& {Kuhn}, M.~A. 2022,
  \apj, 924, 84, \dodoi{10.3847/1538-4357/ac307f}

\bibitem[{{Cardelli} {et~al.}(1989){Cardelli}, {Clayton}, \&
  {Mathis}}]{cardelli_extinction_1989}
{Cardelli}, J.~A., {Clayton}, G.~C., \& {Mathis}, J.~S. 1989, \apj, 345, 245,
  \dodoi{10.1086/167900}

\bibitem[{{Castelli} \& {Kurucz}(2004)}]{castelli_new_2004}
{Castelli}, F., \& {Kurucz}, R.~L. 2004, ArXiv Astrophysics e-prints

\bibitem[{{Clemens} {et~al.}(2004){Clemens}, {Crain}, \&
  {Anderson}}]{clemens_goodman_2004}
{Clemens}, J.~C., {Crain}, J.~A., \& {Anderson}, R. 2004, in Society of
  Photo-Optical Instrumentation Engineers (SPIE) Conference Series, Vol. 5492,
  Ground-based Instrumentation for Astronomy, ed. A.~F.~M. {Moorwood} \&
  M.~{Iye}, 331--340, \dodoi{10.1117/12.550069}

\bibitem[{{Costa Silva} {et~al.}(2020){Costa Silva}, {Delgado Mena}, \&
  {Tsantaki}}]{CostaSilva-20}
{Costa Silva}, A.~R., {Delgado Mena}, E., \& {Tsantaki}, M. 2020, \aap, 634,
  A136, \dodoi{10.1051/0004-6361/201936523}

\bibitem[{{Currie} {et~al.}(2022){Currie}, {Biller}, {Lagrange}, {Marois},
  {Guyon}, {Nielsen}, {Bonnefoy}, \& {De Rosa}}]{currie_directimaging_2022}
{Currie}, T., {Biller}, B., {Lagrange}, A.-M., {et~al.} 2022, arXiv e-prints,
  arXiv:2205.05696.
\newblock \doarXiv{2205.05696}

\bibitem[{{da Silva} {et~al.}(2009){da Silva}, {Torres}, {de La Reza}, {Quast},
  {Melo}, \& {Sterzik}}]{da_silva_search_2009}
{da Silva}, L., {Torres}, C.~A.~O., {de La Reza}, R., {et~al.} 2009, \aap, 508,
  833, \dodoi{10.1051/0004-6361/200911736}

\bibitem[{David {et~al.}(2019)David, Petigura, Luger, Foreman-Mackey,
  Livingston, Mamajek, \& Hillenbrand}]{david_four_2019}
David, T.~J., Petigura, E.~A., Luger, R., {et~al.} 2019, The Astrophysical
  Journal Letters, 885, L12, \dodoi{10.3847/2041-8213/ab4c99}

\bibitem[{David {et~al.}(2016)David, Hillenbrand, Petigura, Carpenter,
  Crossfield, Hinkley, Ciardi, Howard, Isaacson, Cody, Schlieder, Beichman, \&
  Barenfeld}]{david_neptune-sized_2016}
David, T.~J., Hillenbrand, L.~A., Petigura, E.~A., {et~al.} 2016, Nature, 534,
  658, \dodoi{10.1038/nature18293}

\bibitem[{de~Rosa {et~al.}(2016)de~Rosa, Rameau, Patience, Graham, Doyon,
  Lafreni{\`e}re, Macintosh, Pueyo, Rajan, Wang, Ward-Duong, Hung, Maire,
  Nielsen, Ammons, Bulger, Cardwell, Chilcote, Galvez, Gerard, Goodsell,
  Hartung, Hibon, Ingraham, Johnson-Groh, Kalas, Konopacky, Marchis, Marois,
  Metchev, Morzinski, Oppenheimer, Perrin, Rantakyr{\"o}, Savransky, \&
  Thomas}]{de_rosa_spectroscopic_2016}
de~Rosa, R.~J., Rameau, J., Patience, J., {et~al.} 2016, The Astrophysical
  Journal, 824, 121, \dodoi{10.3847/0004-637X/824/2/121}

\bibitem[{de~Zeeuw {et~al.}(1999)de~Zeeuw, Hoogerwerf, Bruijne, Brown, \&
  Blaauw}]{de_zeeuw_hipparcos_1999}
de~Zeeuw, P.~T., Hoogerwerf, R., Bruijne, J. H. J.~d., Brown, A. G.~A., \&
  Blaauw, A. 1999, The Astronomical Journal, 117, 354, \dodoi{10.1086/300682}

\bibitem[{{Donati} {et~al.}(1997){Donati}, {Semel}, {Carter}, {Rees}, \&
  {Collier Cameron}}]{donati_spectropolarimetric_1997}
{Donati}, J.-F., {Semel}, M., {Carter}, B.~D., {Rees}, D.~E., \& {Collier
  Cameron}, A. 1997, \mnras, 291, 658, \dodoi{10.1093/mnras/291.4.658}

\bibitem[{{Donati} {et~al.}(2016){Donati}, {Moutou}, {Malo}, {Baruteau}, {Yu},
  {H{\'e}brard}, {Hussain}, {Alencar}, {M{\'e}nard}, {Bouvier}, {Petit},
  {Takami}, {Doyon}, \& {Collier Cameron}}]{donati_hot_2016}
{Donati}, J.~F., {Moutou}, C., {Malo}, L., {et~al.} 2016, \nat, 534, 662,
  \dodoi{10.1038/nature18305}

\bibitem[{Dotter {et~al.}(2008)Dotter, Chaboyer, Jevremovi{\'c}, Kostov, Baron,
  \& Ferguson}]{dotter_dartmouth_2008}
Dotter, A., Chaboyer, B., Jevremovi{\'c}, D., {et~al.} 2008, The Astrophysical
  Journal Supplement Series, 178, 89, \dodoi{10.1086/589654}

\bibitem[{{Doyle} {et~al.}(2014){Doyle}, {Davies}, {Smalley}, {Chaplin}, \&
  {Elsworth}}]{Doyle-14}
{Doyle}, A.~P., {Davies}, G.~R., {Smalley}, B., {Chaplin}, W.~J., \&
  {Elsworth}, Y. 2014, \mnras, 444, 3592, \dodoi{10.1093/mnras/stu1692}

\bibitem[{{Espinoza} \& {Jord{\'a}n}(2015)}]{Espinoza-15}
{Espinoza}, N., \& {Jord{\'a}n}, A. 2015, \mnras, 450, 1879,
  \dodoi{10.1093/mnras/stv744}

\bibitem[{Feiden(2016)}]{feiden_magnetic_2016}
Feiden, G.~A. 2016, Astronomy \& Astrophysics, 593, A99,
  \dodoi{10.1051/0004-6361/201527613}

\bibitem[{Feinstein {et~al.}(2022)Feinstein, David, Montet, Foreman-Mackey,
  Livingston, \& Mann}]{feinstein_v1298_2022}
Feinstein, A.~D., David, T.~J., Montet, B.~T., {et~al.} 2022, The Astrophysical
  Journal, 925, L2, \dodoi{10.3847/2041-8213/ac4745}

\bibitem[{Foreman-Mackey(2016)}]{foreman-mackey_corner.py:_2016}
Foreman-Mackey, D. 2016, corner.py: {Scatterplot} matrices in {Python},
  \dodoi{10.21105/joss.00024}

\bibitem[{Foreman-Mackey {et~al.}(2017)Foreman-Mackey, Agol, Ambikasaran, \&
  Angus}]{foreman-mackey_fast_2017}
Foreman-Mackey, D., Agol, E., Ambikasaran, S., \& Angus, R. 2017, The
  Astronomical Journal, 154, 220, \dodoi{10.3847/1538-3881/aa9332}

\bibitem[{Foreman-Mackey {et~al.}(2013)Foreman-Mackey, Hogg, Lang, \&
  Goodman}]{foreman-mackey_emcee_2013}
Foreman-Mackey, D., Hogg, D.~W., Lang, D., \& Goodman, J. 2013, Publications of
  the Astronomical Society of the Pacific, 125, 306, \dodoi{10.1086/670067}

\bibitem[{Freund {et~al.}(2022)Freund, Czesla, Robrade, Schneider, \&
  Schmitt}]{freund_stellar_2022}
Freund, S., Czesla, S., Robrade, J., Schneider, P.~C., \& Schmitt, J. H. M.~M.
  2022, Astronomy \& Astrophysics, 664, A105,
  \dodoi{10.1051/0004-6361/202142573}

\bibitem[{Gagn{\'e} {et~al.}(2020)Gagn{\'e}, David, Mamajek, Mann, Faherty, \&
  B{\'e}dard}]{gagne_mu_2020}
Gagn{\'e}, J., David, T.~J., Mamajek, E.~E., {et~al.} 2020, The Astrophysical
  Journal, 903, 96, \dodoi{10.3847/1538-4357/abb77e}

\bibitem[{Gagn{\'e} {et~al.}(2021)Gagn{\'e}, Faherty, Moranta, \&
  Popinchalk}]{gagne_number_2021}
Gagn{\'e}, J., Faherty, J.~K., Moranta, L., \& Popinchalk, M. 2021, The
  Astrophysical Journal Letters, 915, L29, \dodoi{10.3847/2041-8213/ac0e9a}

\bibitem[{Gagn{\'e} {et~al.}(2018)Gagn{\'e}, Mamajek, Malo, Riedel, Rodriguez,
  Lafreni{\`e}re, Faherty, Roy-Loubier, Pueyo, Robin, \&
  Doyon}]{gagne_banyan_2018}
Gagn{\'e}, J., Mamajek, E.~E., Malo, L., {et~al.} 2018, The Astrophysical
  Journal, 856, 23, \dodoi{10.3847/1538-4357/aaae09}

\bibitem[{{Gaia Collaboration} {et~al.}(2016){Gaia Collaboration}, Prusti,
  Bruijne, Brown, Vallenari, Babusiaux, Bailer-Jones, Bastian, Biermann, Evans,
  Eyer, Jansen, Jordi, Klioner, Lammers, Lindegren, Luri, Mignard, Milligan,
  Panem, Poinsignon, Pourbaix, Randich, Sarri, Sartoretti, Siddiqui, Soubiran,
  Valette, Leeuwen, Walton, Aerts, Arenou, Cropper, Drimmel, H{\o}g, Katz,
  Lattanzi, O'Mullane, Grebel, Holland, Huc, Passot, Bramante, Cacciari,
  Casta{\~n}eda, Chaoul, Cheek, Angeli, Fabricius, Guerra, Hern{\'a}ndez,
  Jean-Antoine-Piccolo, Masana, Messineo, Mowlavi, Nienartowicz,
  Ord{\'o}{\~n}ez-Blanco, Panuzzo, Portell, Richards, Riello, Seabroke, Tanga,
  Th{\'e}venin, Torra, Els, Gracia-Abril, Comoretto, Garcia-Reinaldos, Lock,
  Mercier, Altmann, Andrae, Astraatmadja, Bellas-Velidis, Benson, Berthier,
  Blomme, Busso, Carry, Cellino, Clementini, Cowell, Creevey, Cuypers,
  Davidson, Ridder, Torres, Delchambre, Dell'Oro, Ducourant, Fr{\'e}mat,
  Garc{\'\i}a-Torres, Gosset, Halbwachs, Hambly, Harrison, Hauser, Hestroffer,
  Hodgkin, Huckle, Hutton, Jasniewicz, Jordan, Kontizas, Korn, Lanzafame,
  Manteiga, Moitinho, Muinonen, Osinde, Pancino, Pauwels, Petit, Recio-Blanco,
  Robin, Sarro, Siopis, Smith, Smith, Sozzetti, Thuillot, Reeven, Viala, Abbas,
  Aramburu, Accart, Aguado, Allan, Allasia, Altavilla, {\'A}lvarez, Alves,
  Anderson, Andrei, Varela, Antiche, Antoja, Ant{\'o}n, Arcay, Atzei, Ayache,
  Bach, Baker, Balaguer-N{\'u}{\~n}ez, Barache, Barata, Barbier, Barblan,
  Baroni, Navascu{\'e}s, Barros, Barstow, Becciani, Bellazzini, Bellei,
  Garc{\'\i}a, Belokurov, Bendjoya, Berihuete, Bianchi, Bienaym{\'e},
  Billebaud, Blagorodnova, Blanco-Cuaresma, Boch, Bombrun, Borrachero,
  Bouquillon, Bourda, Bouy, Bragaglia, Breddels, Brouillet, Br{\"u}semeister,
  Bucciarelli, Budnik, Burgess, Burgon, Burlacu, Busonero, Buzzi, Caffau,
  Cambras, Campbell, Cancelliere, Cantat-Gaudin, Carlucci, Carrasco,
  Castellani, Charlot, Charnas, Charvet, Chassat, Chiavassa, Clotet, Cocozza,
  Collins, Collins, Costigan, Crifo, Cross, Crosta, Crowley, Dafonte, Damerdji,
  Dapergolas, David, David, Cat, Felice, Laverny, Luise, March, Martino, Souza,
  Debosscher, Pozo, Delbo, Delgado, Delgado, Marco, Matteo, Diakite, Distefano,
  Dolding, Anjos, Drazinos, Dur{\'a}n, Dzigan, Ecale, Edvardsson, Enke,
  Erdmann, Escolar, Espina, Evans, Bontemps, Fabre, Fabrizio, Faigler,
  Falc{\~a}o, Casas, Faye, Federici, Fedorets, Fern{\'a}ndez-Hern{\'a}ndez,
  Fernique, Fienga, Figueras, Filippi, Findeisen, Fonti, Fouesneau, Fraile,
  Fraser, Fuchs, Furnell, Gai, Galleti, Galluccio, Garabato,
  Garc{\'\i}a-Sedano, Gar{\'e}, Garofalo, Garralda, Gavras, Gerssen, Geyer,
  Gilmore, Girona, Giuffrida, Gomes, Gonz{\'a}lez-Marcos,
  Gonz{\'a}lez-N{\'u}{\~n}ez, Gonz{\'a}lez-Vidal, Granvik, Guerrier, Guillout,
  Guiraud, G{\'u}rpide, Guti{\'e}rrez-S{\'a}nchez, Guy, Haigron,
  Hatzidimitriou, Haywood, Heiter, Helmi, Hobbs, Hofmann, Holl, Holland, Hunt,
  Hypki, Icardi, Irwin, Fombelle, Jofr{\'e}, Jonker, Jorissen, Julbe,
  Karampelas, Kochoska, Kohley, Kolenberg, Kontizas, Koposov, Kordopatis,
  Koubsky, Kowalczyk, Krone-Martins, Kudryashova, Kull, Bachchan,
  Lacoste-Seris, Lanza, Lavigne, Poncin-Lafitte, Lebreton, Lebzelter, Leccia,
  Leclerc, Lecoeur-Taibi, Lemaitre, Lenhardt, Leroux, Liao, Licata,
  Lindstr{\o}m, Lister, Livanou, Lobel, L{\"o}ffler, L{\'o}pez, Lopez-Lozano,
  Lorenz, Loureiro, MacDonald, Fernandes, Managau, Mann, Mantelet, Marchal,
  Marchant, Marconi, Marie, Marinoni, Marrese, Marschalk{\'o}, Marshall,
  Mart{\'\i}n-Fleitas, Martino, Mary, Matijevi{\v c}, Mazeh, McMillan, Messina,
  Mestre, Michalik, Millar, Miranda, Molina, Molinaro, Molinaro, Moln{\'a}r,
  Moniez, Montegriffo, Monteiro, Mor, Mora, Morbidelli, Morel, Morgenthaler,
  Morley, Morris, Mulone, Muraveva, Musella, Narbonne, Nelemans, Nicastro,
  Noval, Ord{\'e}novic, Ordieres-Mer{\'e}, Osborne, Pagani, Pagano, Pailler,
  Palacin, Palaversa, Parsons, Paulsen, Pecoraro, Pedrosa, Pentik{\"a}inen,
  Pereira, Pichon, Piersimoni, Pineau, Plachy, Plum, Poujoulet, Pr{\v s}a,
  Pulone, Ragaini, Rago, Rambaux, Ramos-Lerate, Ranalli, Rauw, Read, Regibo,
  Renk, Reyl{\'e}, Ribeiro, Rimoldini, Ripepi, Riva, Rixon, Roelens,
  Romero-G{\'o}mez, Rowell, Royer, Rudolph, Ruiz-Dern, Sadowski, Sell{\'e}s,
  Sahlmann, Salgado, Salguero, Sarasso, Savietto, Schnorhk, Schultheis,
  Sciacca, Segol, Segovia, Segransan, Serpell, Shih, Smareglia, Smart, Smith,
  Solano, Solitro, Sordo, Nieto, Souchay, Spagna, Spoto, Stampa, Steele,
  Steidelm{\"u}ller, Stephenson, Stoev, Suess, S{\"u}veges, Surdej, Szabados,
  Szegedi-Elek, Tapiador, Taris, Tauran, Taylor, Teixeira, Terrett, Tingley,
  Trager, Turon, Ulla, Utrilla, Valentini, Elteren, Hemelryck, Leeuwen, Varadi,
  Vecchiato, Veljanoski, Via, Vicente, Vogt, Voss, Votruba, Voutsinas,
  Walmsley, Weiler, Weingrill, Werner, Wevers, Whitehead, Wyrzykowski, Yoldas,
  {\v Z}erjal, Zucker, Zurbach, Zwitter, Alecu, Allen, Prieto, Amorim,
  Anglada-Escud{\'e}, Arsenijevic, Azaz, Balm, Beck, Bernstein, Bigot, Bijaoui,
  Blasco, Bonfigli, Bono, Boudreault, Bressan, Brown, Brunet, Bunclark,
  Buonanno, Butkevich, Carret, Carrion, Chemin, Ch{\'e}reau, Corcione,
  Darmigny, Boer, Teodoro, Zeeuw, Luche, Domingues, Dubath, Fodor,
  Fr{\'e}zouls, Fries, Fustes, Fyfe, Gallardo, Gallegos, Gardiol, Gebran,
  Gomboc, G{\'o}mez, Grux, Gueguen, Heyrovsky, Hoar, Iannicola, Parache,
  Janotto, Joliet, Jonckheere, Keil, Kim, Klagyivik, Klar, Knude, Kochukhov,
  Kolka, Kos, Kutka, Lainey, LeBouquin, Liu, Loreggia, Makarov, Marseille,
  Martayan, Martinez-Rubi, Massart, Meynadier, Mignot, Munari, Nguyen,
  Nordlander, Ocvirk, O'Flaherty, Sanz, Ortiz, Osorio, Oszkiewicz, Ouzounis,
  Palmer, Park, Pasquato, Peltzer, Peralta, P{\'e}turaud, Pieniluoma, Pigozzi,
  Poels, Prat, Prod'homme, Raison, Rebordao, Risquez, Rocca-Volmerange, Rosen,
  Ruiz-Fuertes, Russo, Sembay, Vizcaino, Short, Siebert, Silva, Sinachopoulos,
  Slezak, Soffel, Sosnowska, Strai{\v z}ys, Linden, Terrell, Theil, Tiede,
  Troisi, Tsalmantza, Tur, Vaccari, {F. Vachier}, Valles, Hamme, Veltz,
  Virtanen, Wallut, Wichmann, Wilkinson, Ziaeepour, \&
  Zschocke}]{gaia_collaboration_gaia_2016}
{Gaia Collaboration}, Prusti, T., Bruijne, J. H. J.~d., {et~al.} 2016,
  Astronomy \& Astrophysics, 595, A1, \dodoi{10.1051/0004-6361/201629272}

\bibitem[{{Gaia Collaboration} {et~al.}(2021){Gaia Collaboration}, {Brown},
  {Vallenari}, {Prusti}, {de Bruijne}, {Babusiaux}, {Biermann}, {Creevey},
  {Evans}, {Eyer}, \& et~al.}]{GaiaCollaboration2021}
{Gaia Collaboration}, {Brown}, A.~G.~A., {Vallenari}, A., {et~al.} 2021, \aap,
  649, A1, \dodoi{10.1051/0004-6361/202039657}

\bibitem[{Gaia~Collaboration {et~al.}(2022)Gaia~Collaboration, Brown, \&
  Prusti}]{gaia_collaboration_gaia_2022}
Gaia~Collaboration, Vallenari, A., Brown, A. G.~A., \& Prusti, T. 2022,
  Astronomy \& Astrophysics, \dodoi{10.1051/0004-6361/202243940}

\bibitem[{Gaidos \& Mann(2014)}]{gaidos_m_2014}
Gaidos, E., \& Mann, A.~W. 2014, The Astrophysical Journal, 791, 54,
  \dodoi{10.1088/0004-637X/791/1/54}

\bibitem[{Giacalone {et~al.}(2021)Giacalone, Dressing, Jensen, Collins, Ricker,
  Vanderspek, Seager, Winn, Jenkins, Barclay, Barkaoui, Cadieux, Charbonneau,
  Collins, Conti, Doyon, Evans, Ghachoui, Gillon, Guerrero, Hart, Jehin,
  Kielkopf, McLean, Murgas, Palle, Parviainen, Pozuelos, Relles, Shporer,
  Socia, Stockdale, Tan, Torres, Twicken, Waalkes, \&
  Waite}]{giacalone_vetting_2021}
Giacalone, S., Dressing, C.~D., Jensen, E. L.~N., {et~al.} 2021, The
  Astronomical Journal, 161, 24, \dodoi{10.3847/1538-3881/abc6af}

\bibitem[{Ginsburg {et~al.}(2019)Ginsburg, Sip{\textbackslash}Hocz, Brasseur,
  Cowperthwaite, Craig, Deil, Guillochon, Guzman, Liedtke, Lim, Lockhart,
  Mommert, Morris, Norman, Parikh, Persson, Robitaille, Segovia, Singer,
  Tollerud, Val-Borro, Valtchanov, \& and}]{ginsburg_astroquery_2019}
Ginsburg, A., Sip{\textbackslash}Hocz, B.~M., Brasseur, C.~E., {et~al.} 2019,
  The Astronomical Journal, 157, 98, \dodoi{10.3847/1538-3881/aafc33}

\bibitem[{Goldman {et~al.}(2018)Goldman, R{\"o}ser, Schilbach, Mo{\'o}r, \&
  Henning}]{goldman_large_2018}
Goldman, B., R{\"o}ser, S., Schilbach, E., Mo{\'o}r, A.~C., \& Henning, T.
  2018, The Astrophysical Journal, 868, 32, \dodoi{10.3847/1538-4357/aae64c}

\bibitem[{{Gray} \& {Corbally}(1994)}]{gray_calibration_1994}
{Gray}, R.~O., \& {Corbally}, C.~J. 1994, \aj, 107, 742, \dodoi{10.1086/116893}

\bibitem[{Grevesse \& Sauval(1998)}]{grevesse_standard_1998}
Grevesse, N., \& Sauval, A. 1998, Space Science Reviews, 85, 161,
  \dodoi{10.1023/A:1005161325181}

\bibitem[{{Guerrero} {et~al.}(2021){Guerrero}, {Seager}, {Huang}, {Vanderburg},
  {Garcia Soto}, {Mireles}, {Hesse}, {Fong}, {Glidden}, {Shporer}, {Latham},
  {Collins}, {Quinn}, {Burt}, {Dragomir}, {Crossfield}, {Vanderspek},
  {Fausnaugh}, {Burke}, {Ricker}, {Daylan}, {Essack}, {G{\"u}nther}, {Osborn},
  {Pepper}, {Rowden}, {Sha}, {Villanueva}, {Yahalomi}, {Yu}, {Ballard},
  {Batalha}, {Berardo}, {Chontos}, {Dittmann}, {Esquerdo}, {Mikal-Evans},
  {Jayaraman}, {Krishnamurthy}, {Louie}, {Mehrle}, {Niraula}, {Rackham},
  {Rodriguez}, {Rowden}, {Sousa-Silva}, {Watanabe}, {Wong}, {Zhan},
  {Zivanovic}, {Christiansen}, {Ciardi}, {Swain}, {Lund}, {Mullally},
  {Fleming}, {Rodriguez}, {Boyd}, {Quintana}, {Barclay}, {Col{\'o}n},
  {Rinehart}, {Schlieder}, {Clampin}, {Jenkins}, {Twicken}, {Caldwell},
  {Coughlin}, {Henze}, {Lissauer}, {Morris}, {Rose}, {Smith}, {Tenenbaum},
  {Ting}, {Wohler}, {Bakos}, {Bean}, {Berta-Thompson}, {Bieryla}, {Bouma},
  {Buchhave}, {Butler}, {Charbonneau}, {Doty}, {Ge}, {Holman}, {Howard},
  {Kaltenegger}, {Kane}, {Kjeldsen}, {Kreidberg}, {Lin}, {Minsky}, {Narita},
  {Paegert}, {P{\'a}l}, {Palle}, {Sasselov}, {Spencer}, {Sozzetti}, {Stassun},
  {Torres}, {Udry}, \& {Winn}}]{guerrero_tess_2021}
{Guerrero}, N.~M., {Seager}, S., {Huang}, C.~X., {et~al.} 2021, \apjs, 254, 39,
  \dodoi{10.3847/1538-4365/abefe1}

\bibitem[{{Gully-Santiago} {et~al.}(2017){Gully-Santiago}, {Herczeg},
  {Czekala}, {Somers}, {Grankin}, {Covey}, {Donati}, {Alencar}, {Hussain},
  {Shappee}, {Mace}, {Lee}, {Holoien}, {Jose}, \&
  {Liu}}]{Gully-Santiago_spots_2017}
{Gully-Santiago}, M.~A., {Herczeg}, G.~J., {Czekala}, I., {et~al.} 2017, \apj,
  836, 200

\bibitem[{{Guti{\'e}rrez Albarr{\'a}n} {et~al.}(2020){Guti{\'e}rrez
  Albarr{\'a}n}, {Montes}, {G{\'o}mez Garrido}, {Tabernero}, {Gonz{\'a}lez
  Hern{\'a}ndez}, {Marfil}, {Frasca}, {Lanzafame}, {Klutsch}, {Franciosini},
  {Randich}, {Smiljanic}, {Korn}, {Gilmore}, {Alfaro}, {Baratella}, {Bayo},
  {Bensby}, {Bonito}, {Carraro}, {Delgado Mena}, {Feltzing}, {Gonneau},
  {Heiter}, {Hourihane}, {Jim{\'e}nez Esteban}, {Jofre}, {Masseron}, {Monaco},
  {Morbidelli}, {Prisinzano}, {Roccatagliata}, {Sousa}, {Van der Swaelmen},
  {Worley}, \& {Zaggia}}]{gutierrez_albarran_lithium_2020}
{Guti{\'e}rrez Albarr{\'a}n}, M.~L., {Montes}, D., {G{\'o}mez Garrido}, M.,
  {et~al.} 2020, \aap, 643, A71, \dodoi{10.1051/0004-6361/202037620}

\bibitem[{Haisch {et~al.}(2001)Haisch, Lada, \& Lada}]{haisch_disk_2001}
Haisch, Jr., K.~E., Lada, E.~A., \& Lada, C.~J. 2001, The Astrophysical
  Journal, 553, L153, \dodoi{10.1086/320685}

\bibitem[{{Hattori} {et~al.}(2021){Hattori}, {Foreman-Mackey}, {Hogg},
  {Montet}, {Angus}, {Pritchard}, {Curtis}, \&
  {Sch{\"o}lkopf}}]{hattori_unpopular_2021}
{Hattori}, S., {Foreman-Mackey}, D., {Hogg}, D.~W., {et~al.} 2021, arXiv
  e-prints, arXiv:2106.15063.
\newblock \doarXiv{2106.15063}

\bibitem[{{Heap} \& {Lindler}(2016)}]{heap_ngsl_2016}
{Heap}, S.~R., \& {Lindler}, D. 2016, in Astronomical Society of the Pacific
  Conference Series, Vol. 503, The Science of Calibration, ed. S.~{Deustua},
  S.~{Allam}, D.~{Tucker}, \& J.~A. {Smith}, 211

\bibitem[{{Hinkley} {et~al.}(2015){Hinkley}, {Kraus}, {Ireland}, {Cheetham},
  {Carpenter}, {Tuthill}, {Lacour}, {Evans}, \&
  {Haubois}}]{hinkley_companions_2015}
{Hinkley}, S., {Kraus}, A.~L., {Ireland}, M.~J., {et~al.} 2015, \apjl, 806, L9,
  \dodoi{10.1088/2041-8205/806/1/L9}

\bibitem[{{H{\o}g} {et~al.}(2000){H{\o}g}, {Fabricius}, {Makarov}, {Urban},
  {Corbin}, {Wycoff}, {Bastian}, {Schwekendiek}, \& {Wicenec}}]{Hog_tycho_2000}
{H{\o}g}, E., {Fabricius}, C., {Makarov}, V.~V., {et~al.} 2000, \aap, 355, L27

\bibitem[{Hogg {et~al.}(2010)Hogg, Bovy, \& Lang}]{hogg_data_2010}
Hogg, D.~W., Bovy, J., \& Lang, D. 2010, Data analysis recipes: {Fitting} a
  model to data, Tech. Rep. arXiv:1008.4686, arXiv,
  \dodoi{10.48550/arXiv.1008.4686}

\bibitem[{Howell {et~al.}(2011)Howell, Everett, Sherry, Horch, \&
  Ciardi}]{howell_speckle_2011}
Howell, S.~B., Everett, M.~E., Sherry, W., Horch, E., \& Ciardi, D.~R. 2011,
  The Astronomical Journal, 142, 19, \dodoi{10.1088/0004-6256/142/1/19}

\bibitem[{Huang {et~al.}(2020)Huang, Vanderburg, P{\'a}l, Sha, Yu, Fong,
  Fausnaugh, Shporer, Guerrero, Vanderspek, \& Ricker}]{huang_photometry_2020}
Huang, C.~X., Vanderburg, A., P{\'a}l, A., {et~al.} 2020, Research Notes of the
  American Astronomical Society, 4, 204, \dodoi{10.3847/2515-5172/abca2e}

\bibitem[{Hunter(2007)}]{hunter_matplotlib_2007}
Hunter, J.~D. 2007, Computing in Science \& Engineering, 9, 90,
  \dodoi{10.1109/MCSE.2007.55}

\bibitem[{Husser {et~al.}(2013)Husser, von Berg, Dreizler, Homeier, Reiners,
  Barman, \& Hauschildt}]{husser_new_2013}
Husser, T.-O., von Berg, S.~W., Dreizler, S., {et~al.} 2013, Astronomy \&
  Astrophysics, 553, A6, \dodoi{10.1051/0004-6361/201219058}

\bibitem[{{Jenkins}(2002)}]{Jenkins_methods_2002}
{Jenkins}, J.~M. 2002, \apj, 575, 493, \dodoi{10.1086/341136}

\bibitem[{{Jenkins} {et~al.}(2020){Jenkins}, {Tenenbaum}, {Seader}, {Burke},
  {McCauliff}, {Smith}, {Twicken}, \& {Chandrasekaran}}]{Jenkins_handbook_2020}
{Jenkins}, J.~M., {Tenenbaum}, P., {Seader}, S., {et~al.} 2020, {Kepler Data
  Processing Handbook: Transiting Planet Search}, Kepler Science Document
  KSCI-19081-003, id. 9. Edited by Jon M. Jenkins.

\bibitem[{{Jenkins} {et~al.}(2010){Jenkins}, {Chandrasekaran}, {McCauliff},
  {Caldwell}, {Tenenbaum}, {Li}, {Klaus}, {Cote}, \&
  {Middour}}]{Jenkins_software_2010}
{Jenkins}, J.~M., {Chandrasekaran}, H., {McCauliff}, S.~D., {et~al.} 2010, in
  Society of Photo-Optical Instrumentation Engineers (SPIE) Conference Series,
  Vol. 7740, Software and Cyberinfrastructure for Astronomy, ed. N.~M.
  {Radziwill} \& A.~{Bridger}, 77400D, \dodoi{10.1117/12.856764}

\bibitem[{{Jenkins} {et~al.}(2016){Jenkins}, {Twicken}, {McCauliff},
  {Campbell}, {Sanderfer}, {Lung}, {Mansouri-Samani}, {Girouard}, {Tenenbaum},
  {Klaus}, {Smith}, {Caldwell}, {Chacon}, {Henze}, {Heiges}, {Latham},
  {Morgan}, {Swade}, {Rinehart}, \& {Vanderspek}}]{jenkins_tess_2016}
{Jenkins}, J.~M., {Twicken}, J.~D., {McCauliff}, S., {et~al.} 2016, in Society
  of Photo-Optical Instrumentation Engineers (SPIE) Conference Series, Vol.
  9913, Software and Cyberinfrastructure for Astronomy IV, ed. G.~{Chiozzi} \&
  J.~C. {Guzman}, 99133E, \dodoi{10.1117/12.2233418}

\bibitem[{Johnson {et~al.}(2018)Johnson, Dai, Justesen, Gandolfi, Hatzes,
  Nowak, Endl, Cochran, Hidalgo, Watanabe, Parviainen, Hirano, Villanueva,
  Prieto-Arranz, Narita, Palle, Guenther, Barrag{\'a}n, Trifonov, Niraula,
  MacQueen, Cabrera, Csizmadia, Eigm{\"u}ller, Grziwa, Korth, P{\"a}tzold,
  Smith, Albrecht, Alonso, Deeg, Erikson, Esposito, Fridlund, Fukui, Kusakabe,
  Kuzuhara, Livingston, Monta{\~n}es~Rodriguez, Nespral, Persson, Purismo,
  Raimundo, Rauer, Ribas, Tamura, Van~Eylen, \& Winn}]{johnson_k2-260_2018}
Johnson, M.~C., Dai, F., Justesen, A.~B., {et~al.} 2018, Monthly Notices of the
  Royal Astronomical Society, 481, 596, \dodoi{10.1093/mnras/sty2238}

\bibitem[{{Kapteyn}(1914)}]{kapteyn_individual_1914}
{Kapteyn}, J.~C. 1914, \apj, 40, 43, \dodoi{10.1086/142098}

\bibitem[{Kerr {et~al.}(2021)Kerr, Rizzuto, Kraus, \& Offner}]{kerr_stars_2021}
Kerr, R. M.~P., Rizzuto, A.~C., Kraus, A.~L., \& Offner, S. S.~R. 2021, The
  Astrophysical Journal, 917, 23, \dodoi{10.3847/1538-4357/ac0251}

\bibitem[{{Kipping}(2013)}]{kipping_efficient_2013}
{Kipping}, D.~M. 2013, \mnras, 435, 2152, \dodoi{10.1093/mnras/stt1435}

\bibitem[{{Klein} {et~al.}(2022){Klein}, {Zicher}, {Kavanagh}, {Nielsen},
  {Aigrain}, {Vidotto}, {Barrag{\'a}n}, {Strugarek}, {Nicholson}, {Donati}, \&
  {Bouvier}}]{klein_one_2022}
{Klein}, B., {Zicher}, N., {Kavanagh}, R.~D., {et~al.} 2022, \mnras, 512, 5067,
  \dodoi{10.1093/mnras/stac761}

\bibitem[{Kounkel {et~al.}(2019)Kounkel, Covey, Moe, Kratter, Su{\'a}rez,
  Stassun, Rom{\'a}n-Z{\'u}{\~n}iga, Hernandez, Kim, Ram{\'\i}rez, Roman-Lopes,
  Stringfellow, Jaehnig, Borissova, Tofflemire, Krolikowski, Rizzuto, Kraus,
  Badenes, Longa-Pe{\~n}a, Chew, Barba, Nidever, Brown, Lee, Pan, Bizyaev,
  Oravetz, \& Oravetz}]{kounkel_close_2019}
Kounkel, M., Covey, K., Moe, M., {et~al.} 2019, The Astronomical Journal, 157,
  196, \dodoi{10.3847/1538-3881/ab13b1}

\bibitem[{Kraus {et~al.}(2015)Kraus, Cody, Covey, Rizzuto, Mann, \&
  Ireland}]{kraus_mass-radius_2015}
Kraus, A.~L., Cody, A.~M., Covey, K.~R., {et~al.} 2015, The Astrophysical
  Journal, 807, 3, \dodoi{10.1088/0004-637X/807/1/3}

\bibitem[{Kraus {et~al.}(2017)Kraus, Herczeg, Rizzuto, Mann, Slesnick,
  Carpenter, Hillenbrand, \& Mamajek}]{kraus_greater_2017}
Kraus, A.~L., Herczeg, G.~J., Rizzuto, A.~C., {et~al.} 2017, The Astrophysical
  Journal, 838, 150, \dodoi{10.3847/1538-4357/aa62a0}

\bibitem[{Kraus {et~al.}(2014)Kraus, Shkolnik, Allers, \&
  Liu}]{kraus_stellar_2014}
Kraus, A.~L., Shkolnik, E.~L., Allers, K.~N., \& Liu, M.~C. 2014, The
  Astronomical Journal, 147, 146, \dodoi{10.1088/0004-6256/147/6/146}

\bibitem[{Kreidberg(2015)}]{kreidberg_batman_2015}
Kreidberg, L. 2015, Publications of the Astronomical Society of the Pacific,
  127, 1161, \dodoi{10.1086/683602}

\bibitem[{Krolikowski {et~al.}(2021)Krolikowski, Kraus, \&
  Rizzuto}]{krolikowski_gaia_2021}
Krolikowski, D.~M., Kraus, A.~L., \& Rizzuto, A.~C. 2021, arXiv:2105.13370
  [astro-ph].
\newblock \url{http://arxiv.org/abs/2105.13370}

\bibitem[{Kurucz(1993)}]{kurucz_new_1993}
Kurucz, R.~L. 1993, International Astronomical Union Colloquium, 138, 87,
  \dodoi{10.1017/S0252921100020327}

\bibitem[{{Kurucz}(1993)}]{Kurucz-93}
{Kurucz}, R.~L. 1993, {SYNTHE spectrum synthesis programs and line data}

\bibitem[{{Li} {et~al.}(2019){Li}, {Tenenbaum}, {Twicken}, {Burke}, {Jenkins},
  {Quintana}, {Rowe}, \& {Seader}}]{Li_DataValidation_2019}
{Li}, J., {Tenenbaum}, P., {Twicken}, J.~D., {et~al.} 2019, \pasp, 131, 024506,
  \dodoi{10.1088/1538-3873/aaf44d}

\bibitem[{Lim(2020)}]{lim_synphot_2020}
Lim, P.~L. 2020, \dodoi{10.5281/zenodo.3971036}

\bibitem[{Lindegren {et~al.}(2021)Lindegren, Klioner, Hern{\'a}ndez, Bombrun,
  Ramos-Lerate, Steidelm{\"u}ller, Bastian, Biermann, Torres, Gerlach, Geyer,
  Hilger, Hobbs, Lammers, McMillan, Stephenson, Casta{\~n}eda, Davidson,
  Fabricius, Gracia-Abril, Portell, Rowell, Teyssier, Torra, Bartolom{\'e},
  Clotet, Garralda, Gonz{\'a}lez-Vidal, Torra, Abbas, Altmann, Varela,
  Balaguer-N{\'u}{\~n}ez, Balog, Barache, Becciani, Bernet, Bertone, Bianchi,
  Bouquillon, Brown, Bucciarelli, Busonero, Butkevich, Buzzi, Cancelliere,
  Carlucci, Charlot, Cioni, Crosta, Crowley, Peloso, Pozo, Drimmel, Esquej,
  Fienga, Fraile, Gai, Garcia-Reinaldos, Guerra, Hambly, Hauser, Jan{\ss}en,
  Jordan, Kostrzewa-Rutkowska, Lattanzi, Liao, Licata, Lister, L{\"o}ffler,
  Marchant, Masip, Mignard, Mints, Molina, Mora, Morbidelli, Murphy, Pagani,
  Panuzzo, Esteller, Poggio, Fiorentin, Riva, Sell{\'e}s, Gimenez, Sarasso,
  Sciacca, Siddiqui, Smart, Souami, Spagna, Steele, Taris, Utrilla, Reeven, \&
  Vecchiato}]{lindegren_gaia_2021}
Lindegren, L., Klioner, S.~A., Hern{\'a}ndez, J., {et~al.} 2021, Astronomy \&
  Astrophysics, 649, A2, \dodoi{10.1051/0004-6361/202039709}

\bibitem[{{Lissauer} {et~al.}(2012){Lissauer}, {Marcy}, {Rowe}, {Bryson},
  {Adams}, {Buchhave}, {Ciardi}, {Cochran}, {Fabrycky}, {Ford}, {Fressin},
  {Geary}, {Gilliland}, {Holman}, {Howell}, {Jenkins}, {Kinemuchi}, {Koch},
  {Morehead}, {Ragozzine}, {Seader}, {Tanenbaum}, {Torres}, \&
  {Twicken}}]{lissauer_almost_2012}
{Lissauer}, J.~J., {Marcy}, G.~W., {Rowe}, J.~F., {et~al.} 2012, \apj, 750,
  112, \dodoi{10.1088/0004-637X/750/2/112}

\bibitem[{Lomb(1976)}]{lomb_least_1976}
Lomb, N.~R. 1976, \apss, 39, 447, \dodoi{10.1007/BF00648343}

\bibitem[{{Madsen} {et~al.}(2002){Madsen}, {Dravins}, \&
  {Lindegren}}]{madsen_astrometric_2002}
{Madsen}, S., {Dravins}, D., \& {Lindegren}, L. 2002, \aap, 381, 446,
  \dodoi{10.1051/0004-6361:20011458}

\bibitem[{Malo {et~al.}(2012)Malo, Doyon, Lafrenière, Artigau, Gagné, Baron,
  \& Riedel}]{malo_bayesian_2012}
Malo, L., Doyon, R., Lafrenière, D., {et~al.} 2012, The Astrophysical Journal,
  762, 88, \dodoi{10.1088/0004-637X/762/2/88}

\bibitem[{Mamajek \& Bell(2014)}]{mamajek_age_2014}
Mamajek, E.~E., \& Bell, C. P.~M. 2014, Monthly Notices of the Royal
  Astronomical Society, 445, 2169, \dodoi{10.1093/mnras/stu1894}

\bibitem[{Mamajek {et~al.}(2002)Mamajek, Meyer, \& Liebert}]{mamajek_post_2002}
Mamajek, E.~E., Meyer, M.~R., \& Liebert, J. 2002, The Astronomical Journal,
  124, 1670, \dodoi{10.1086/341952}

\bibitem[{Mann {et~al.}(2016{\natexlab{a}})Mann, Newton, Rizzuto, Irwin,
  Feiden, Gaidos, Mace, Kraus, James, Ansdell, Charbonneau, Covey, Ireland,
  Jaffe, Johnson, Kidder, \& Vanderburg}]{mann_k2-33_2016}
Mann, A.~W., Newton, E.~R., Rizzuto, A.~C., {et~al.} 2016{\natexlab{a}}, The
  Astronomical Journal, 152, 61, \dodoi{10.3847/0004-6256/152/3/61}

\bibitem[{Mann {et~al.}(2016{\natexlab{b}})Mann, Gaidos, Mace, Johnson, Bowler,
  LaCourse, Jacobs, Vanderburg, Kraus, Kaplan, \& Jaffe}]{mann_k2_25_2016}
Mann, A.~W., Gaidos, E., Mace, G.~N., {et~al.} 2016{\natexlab{b}}, The
  Astrophysical Journal, 818, 46, \dodoi{10.3847/0004-637X/818/1/46}

\bibitem[{Mann {et~al.}(2019)Mann, Dupuy, Kraus, Gaidos, Ansdell, Ireland,
  Rizzuto, Hung, Dittmann, Factor, Feiden, Martinez, Ru{\'\i}z-Rodr{\'\i}guez,
  \& Thao}]{mann_how_2019}
Mann, A.~W., Dupuy, T., Kraus, A.~L., {et~al.} 2019, The Astrophysical Journal,
  871, 63, \dodoi{10.3847/1538-4357/aaf3bc}

\bibitem[{Mann {et~al.}(2022)Mann, Wood, Schmidt, Barber, Owen, Tofflemire,
  Newton, Mamajek, Bush, Mace, Kraus, Thao, Vanderburg, Llama, Johns-Krull,
  Prato, Stahl, Tang, Fields, Collins, Collins, Gan, Jensen, Kamler, Schwarz,
  Furlan, Gnilka, Howell, Lester, Owens, Suarez, Mekarnia, Guillot, Abe,
  Triaud, Johnson, Milburn, Rizzuto, Quinn, Kerr, Ricker, Vanderspek, Latham,
  Seager, Winn, Jenkins, Guerrero, Shporer, Schlieder, McLean, \&
  Wohler}]{mann_tess_2022}
Mann, A.~W., Wood, M.~L., Schmidt, S.~P., {et~al.} 2022, The Astronomical
  Journal, 163, 156, \dodoi{10.3847/1538-3881/ac511d}

\bibitem[{Masuda \& Winn(2020)}]{masuda_inference_2020}
Masuda, K., \& Winn, J.~N. 2020, The Astronomical Journal, 159, 81,
  \dodoi{10.3847/1538-3881/ab65be}

\bibitem[{{McInnes} \& {Healy}(2017)}]{mcinnes_hierarchical_2017}
{McInnes}, L., \& {Healy}, J. 2017, arXiv e-prints, arXiv:1705.07321.
\newblock \doarXiv{1705.07321}

\bibitem[{Meingast {et~al.}(2019)Meingast, Alves, \&
  F{\"u}rnkranz}]{meingast_extended_2019}
Meingast, S., Alves, J., \& F{\"u}rnkranz, V. 2019, Astronomy \& Astrophysics,
  622, L13, \dodoi{10.1051/0004-6361/201834950}

\bibitem[{{Messina} {et~al.}(2022){Messina}, {Nardiello}, {Desidera},
  {Baratella}, {Benatti}, {Biazzo}, \& {D'Orazi}}]{Messina_gyro_2022}
{Messina}, S., {Nardiello}, D., {Desidera}, S., {et~al.} 2022, \aap, 657, L3,
  \dodoi{10.1051/0004-6361/202142276}

\bibitem[{{Mo{\'o}r} {et~al.}(2013){Mo{\'o}r}, {{\'A}brah{\'a}m},
  {K{\'o}sp{\'a}l}, {Szab{\'o}}, {Apai}, {Balog}, {Csengeri}, {Grady},
  {Henning}, {Juh{\'a}sz}, {Kiss}, {Pascucci}, {Szul{\'a}gyi}, \&
  {Vavrek}}]{moor_resolved_2013}
{Mo{\'o}r}, A., {{\'A}brah{\'a}m}, P., {K{\'o}sp{\'a}l}, {\'A}., {et~al.} 2013,
  \apjl, 775, L51, \dodoi{10.1088/2041-8205/775/2/L51}

\bibitem[{{Morton}(2015)}]{morton_isochrones_2015}
{Morton}, T.~D. 2015, {isochrones: Stellar model grid package}, Astrophysics
  Source Code Library, record ascl:1503.010.
\newblock \doeprint{1503.010}

\bibitem[{Murphy {et~al.}(2013)Murphy, Lawson, \&
  Bessell}]{murphy_re-examining_2013}
Murphy, S.~J., Lawson, W.~A., \& Bessell, M.~S. 2013, Monthly Notices of the
  Royal Astronomical Society, 435, 1325, \dodoi{10.1093/mnras/stt1375}

\bibitem[{Newton {et~al.}(2019)Newton, Mann, Tofflemire, Pearce, Rizzuto,
  Vanderburg, Martinez, Wang, Ruffio, Kraus, Johnson, Thao, Wood, Rampalli,
  Nielsen, Collins, Dragomir, Hellier, Anderson, Barclay, Brown, Feiden, Hart,
  Isopi, Kielkopf, Mallia, Nelson, Rodriguez, Stockdale, Waite, Wright,
  Lissauer, Ricker, Vanderspek, Latham, Seager, Winn, Jenkins, Bouma, Burke,
  Davies, Fausnaugh, Li, Morris, Mukai, Villase{\~n}or, Villeneuva, Rosa,
  Macintosh, Mengel, Okumura, \& Wittenmyer}]{newton_tess_2019}
Newton, E.~R., Mann, A.~W., Tofflemire, B.~M., {et~al.} 2019, The Astrophysical
  Journal, 880, L17, \dodoi{10.3847/2041-8213/ab2988}

\bibitem[{{Newton} {et~al.}(2022{\natexlab{a}}){Newton}, {Rampalli}, {Kraus},
  {Mann}, {Curtis}, {Vanderburg}, {Krolikowski}, {Huber}, {Petter}, {Bieryla},
  {Tofflemire}, {Thao}, {Wood}, {Kerr}, {Safanov}, {Strakhov}, {Ciardi},
  {Giacalone}, {Dressing}, {Gill}, {Savel}, {Collins}, {Brown}, {Murgas},
  {Isogai}, {Narita}, {Palle}, {Quinn}, {Eastman}, {F{\H{u}}r{\'e}sz}, {Shiao},
  {Daylan}, {Caldwell}, {Ricker}, {Vanderspek}, {Seager}, {Winn}, {Jenkins}, \&
  {Latham}}]{Newton_tess_2022}
{Newton}, E.~R., {Rampalli}, R., {Kraus}, A.~L., {et~al.} 2022{\natexlab{a}},
  arXiv e-prints, arXiv:2206.06254.
\newblock \doarXiv{2206.06254}

\bibitem[{{Newton} {et~al.}(2022{\natexlab{b}}){Newton}, {Rampalli}, {Kraus},
  {Mann}, {Curtis}, {Vanderburg}, {Krolikowski}, {Huber}, {Petter}, {Bieryla},
  {Tofflemire}, {Thao}, {Wood}, {Kerr}, {Safanov}, {Strakhov}, {Ciardi},
  {Giacalone}, {Dressing}, {Gill}, {Savel}, {Collins}, {Brown}, {Murgas},
  {Isogai}, {Narita}, {Palle}, {Quinn}, {Eastman}, {F{\H{u}}r{\'e}sz}, {Shiao},
  {Daylan}, {Caldwell}, {Ricker}, {Vanderspek}, {Seager}, {Winn}, {Jenkins}, \&
  {Latham}}]{newton_groupx_2022}
---. 2022{\natexlab{b}}, arXiv e-prints, arXiv:2206.06254.
\newblock \doarXiv{2206.06254}

\bibitem[{Oh {et~al.}(2017)Oh, Price-Whelan, Hogg, Morton, \&
  Spergel}]{oh_comoving_2017}
Oh, S., Price-Whelan, A.~M., Hogg, D.~W., Morton, T.~D., \& Spergel, D.~N.
  2017, The Astronomical Journal, 153, 257, \dodoi{10.3847/1538-3881/aa6ffd}

\bibitem[{Parviainen \& Aigrain(2015)}]{parviainen_ldtk_2015}
Parviainen, H., \& Aigrain, S. 2015, Monthly Notices of the Royal Astronomical
  Society, 453, 3821, \dodoi{10.1093/mnras/stv1857}

\bibitem[{Pecaut \& Mamajek(2013)}]{pecaut_intrinsic_2013}
Pecaut, M.~J., \& Mamajek, E.~E. 2013, The Astrophysical Journal Supplement
  Series, 208, 9, \dodoi{10.1088/0067-0049/208/1/9}

\bibitem[{Pecaut \& Mamajek(2016)}]{pecaut_star_2016}
---. 2016, Monthly Notices of the Royal Astronomical Society, 461, 794,
  \dodoi{10.1093/mnras/stw1300}

\bibitem[{Pecaut {et~al.}(2012)Pecaut, Mamajek, \& Bubar}]{pecaut_revised_2012}
Pecaut, M.~J., Mamajek, E.~E., \& Bubar, E.~J. 2012, The Astrophysical Journal,
  746, 154, \dodoi{10.1088/0004-637X/746/2/154}

\bibitem[{Phillips {et~al.}(2020)Phillips, Tremblin, Baraffe, Chabrier, Allard,
  Spiegelman, Goyal, Drummond, \& H{\'e}brard}]{phillips_atmosphere_2020}
Phillips, M.~W., Tremblin, P., Baraffe, I., {et~al.} 2020, \aap, 637, A38,
  \dodoi{10.1051/0004-6361/201937381}

\bibitem[{Poppenhaeger {et~al.}(2021)Poppenhaeger, Ketzer, \&
  Mallonn}]{poppenhaeger_x-ray_2021}
Poppenhaeger, K., Ketzer, L., \& Mallonn, M. 2021, Monthly Notices of the Royal
  Astronomical Society, 500, 4560, \dodoi{10.1093/mnras/staa1462}

\bibitem[{{Preibisch} \& {Mamajek}(2008)}]{preibisch_nearest_2008}
{Preibisch}, T., \& {Mamajek}, E. 2008, in Handbook of Star Forming Regions,
  Volume II, ed. B.~{Reipurth}, Vol.~5, 235

\bibitem[{{Press} \& {Rybicki}(1989)}]{press_fast_1989}
{Press}, W.~H., \& {Rybicki}, G.~B. 1989, \apj, 338, 277,
  \dodoi{10.1086/167197}

\bibitem[{Rameau {et~al.}(2013)Rameau, Chauvin, Lagrange, Meshkat, Boccaletti,
  Quanz, Currie, Mawet, Girard, Bonnefoy, \&
  Kenworthy}]{rameau_confirmation_2013}
Rameau, J., Chauvin, G., Lagrange, A.-M., {et~al.} 2013, The Astrophysical
  Journal, 779, L26, \dodoi{10.1088/2041-8205/779/2/L26}

\bibitem[{Rampalli {et~al.}(2021)Rampalli, Ag{\"u}eros, Curtis, Douglas,
  N{\'u}{\~n}ez, Cargile, Covey, Gosnell, Kraus, Law, \&
  Mann}]{rampalli_three_2021}
Rampalli, R., Ag{\"u}eros, M.~A., Curtis, J.~L., {et~al.} 2021, The
  Astrophysical Journal, 921, 167, \dodoi{10.3847/1538-4357/ac0c1e}

\bibitem[{Rayner {et~al.}(2009)Rayner, Cushing, \&
  Vacca}]{rayner_infrared_2009}
Rayner, J.~T., Cushing, M.~C., \& Vacca, W.~D. 2009, The Astrophysical Journal
  Supplement Series, 185, 289, \dodoi{10.1088/0067-0049/185/2/289}

\bibitem[{{Rebull} {et~al.}(2018){Rebull}, {Stauffer}, {Cody}, {Hillenbrand},
  {David}, \& {Pinsonneault}}]{rebull_rotation_2018}
{Rebull}, L.~M., {Stauffer}, J.~R., {Cody}, A.~M., {et~al.} 2018, \aj, 155,
  196, \dodoi{10.3847/1538-3881/aab605}

\bibitem[{{Ricker} {et~al.}(2015){Ricker}, {Winn}, {Vanderspek}, {Latham},
  {Bakos}, {Bean}, {Berta-Thompson}, {Brown}, {Buchhave}, {Butler}, {Butler},
  {Chaplin}, {Charbonneau}, {Christensen-Dalsgaard}, {Clampin}, {Deming},
  {Doty}, {De Lee}, {Dressing}, {Dunham}, {Endl}, {Fressin}, {Ge}, {Henning},
  {Holman}, {Howard}, {Ida}, {Jenkins}, {Jernigan}, {Johnson}, {Kaltenegger},
  {Kawai}, {Kjeldsen}, {Laughlin}, {Levine}, {Lin}, {Lissauer}, {MacQueen},
  {Marcy}, {McCullough}, {Morton}, {Narita}, {Paegert}, {Palle}, {Pepe},
  {Pepper}, {Quirrenbach}, {Rinehart}, {Sasselov}, {Sato}, {Seager},
  {Sozzetti}, {Stassun}, {Sullivan}, {Szentgyorgyi}, {Torres}, {Udry}, \&
  {Villasenor}}]{ricker_transiting_2015}
{Ricker}, G.~R., {Winn}, J.~N., {Vanderspek}, R., {et~al.} 2015, Journal of
  Astronomical Telescopes, Instruments, and Systems, 1, 014003,
  \dodoi{10.1117/1.JATIS.1.1.014003}

\bibitem[{Riello {et~al.}(2021)Riello, Angeli, Evans, Montegriffo, Carrasco,
  Busso, Palaversa, Burgess, Diener, Davidson, Rowell, Fabricius, Jordi,
  Bellazzini, Pancino, Harrison, Cacciari, Leeuwen, Hambly, Hodgkin, Osborne,
  Altavilla, Barstow, Brown, Castellani, Cowell, Luise, Gilmore, Giuffrida,
  Hidalgo, Holland, Marinoni, Pagani, Piersimoni, Pulone, Ragaini, Rainer,
  Richards, Sanna, Walton, Weiler, \& Yoldas}]{riello_gaia_2021}
Riello, M., Angeli, F.~D., Evans, D.~W., {et~al.} 2021, Astronomy \&
  Astrophysics, 649, A3, \dodoi{10.1051/0004-6361/202039587}

\bibitem[{Rizzuto {et~al.}(2011)Rizzuto, Ireland, \&
  Robertson}]{rizzuto_multidimensional_2011}
Rizzuto, A.~C., Ireland, M.~J., \& Robertson, J.~G. 2011, Monthly Notices of
  the Royal Astronomical Society, 416, 3108,
  \dodoi{10.1111/j.1365-2966.2011.19256.x}

\bibitem[{Rizzuto {et~al.}(2017)Rizzuto, Mann, Vanderburg, Kraus, \&
  Covey}]{rizzuto_zodiacal_2017}
Rizzuto, A.~C., Mann, A.~W., Vanderburg, A., Kraus, A.~L., \& Covey, K.~R.
  2017, The Astronomical Journal, 154, 224, \dodoi{10.3847/1538-3881/aa9070}

\bibitem[{{Rizzuto} {et~al.}(2020){Rizzuto}, {Newton}, {Mann}, {Tofflemire},
  {Vanderburg}, {Kraus}, {Wood}, {Quinn}, {Zhou}, {Thao}, {Law}, {Ziegler}, \&
  {Brice{\~n}o}}]{rizzuto_tess_2020}
{Rizzuto}, A.~C., {Newton}, E.~R., {Mann}, A.~W., {et~al.} 2020, \aj, 160, 33,
  \dodoi{10.3847/1538-3881/ab94b7}

\bibitem[{Rockcliffe {et~al.}(2021)Rockcliffe, Newton, Youngblood, Bourrier,
  Mann, Berta-Thompson, Ag{\"u}eros, N{\'{u} }{\~{n}}ez, \&
  Charbonneau}]{rockcliffe_lyman_2021}
Rockcliffe, K.~E., Newton, E.~R., Youngblood, A., {et~al.} 2021, The
  Astronomical Journal, 162, 116, \dodoi{10.3847/1538-3881/ac126f}

\bibitem[{{Santos} {et~al.}(2013){Santos}, {Sousa}, {Mortier}, {Neves},
  {Adibekyan}, {Tsantaki}, {Delgado Mena}, {Bonfils}, {Israelian}, {Mayor}, \&
  {Udry}}]{Santos-13}
{Santos}, N.~C., {Sousa}, S.~G., {Mortier}, A., {et~al.} 2013, \aap, 556, A150,
  \dodoi{10.1051/0004-6361/201321286}

\bibitem[{{Savanov} {et~al.}(2018){Savanov}, {Dmitrienko}, {Karmakar}, \&
  {Pandey}}]{savanov_activity_2018}
{Savanov}, I.~S., {Dmitrienko}, E.~S., {Karmakar}, S., \& {Pandey}, J.~C. 2018,
  Astronomy Reports, 62, 532, \dodoi{10.1134/S1063772918080073}

\bibitem[{{Scargle}(1982)}]{scargle_studies_1982}
{Scargle}, J.~D. 1982, \apj, 263, 835, \dodoi{10.1086/160554}

\bibitem[{{Scott} {et~al.}(2021){Scott}, {Howell}, {Gnilka}, {Stephens},
  {Salinas}, {Matson}, {Furlan}, {Horch}, {Everett}, {Ciardi}, {Mills}, \&
  {Quigley}}]{scott_twin_2021}
{Scott}, N.~J., {Howell}, S.~B., {Gnilka}, C.~L., {et~al.} 2021, Frontiers in
  Astronomy and Space Sciences, 8, 138, \dodoi{10.3389/fspas.2021.716560}

\bibitem[{{Seager} \& {Mall{\'e}n-Ornelas}(2003)}]{seager_unique_2003}
{Seager}, S., \& {Mall{\'e}n-Ornelas}, G. 2003, \apj, 585, 1038,
  \dodoi{10.1086/346105}

\bibitem[{Sergison {et~al.}(2013)Sergison, Mayne, Naylor, Jeffries, \&
  Bell}]{sergison_no_2013}
Sergison, D.~J., Mayne, N.~J., Naylor, T., Jeffries, R.~D., \& Bell, C. P.~M.
  2013, Monthly Notices of the Royal Astronomical Society, 434, 966,
  \dodoi{10.1093/mnras/stt973}

\bibitem[{Shkolnik {et~al.}(2017)Shkolnik, Allers, Kraus, Liu, \&
  Flagg}]{shkolnik_all-sky_2017}
Shkolnik, E.~L., Allers, K.~N., Kraus, A.~L., Liu, M.~C., \& Flagg, L. 2017,
  The Astronomical Journal, 154, 69, \dodoi{10.3847/1538-3881/aa77fa}

\bibitem[{{Siverd} {et~al.}(2018){Siverd}, {Brown}, {Barnes}, {Bowman}, {De
  Vera}, {Foale}, {Harbeck}, {Henderson}, {Hygelund}, {Kirby}, {McCully},
  {Nation}, {Smith}, {Taylor}, \& {Tufts}}]{siverd_nres_2018}
{Siverd}, R.~J., {Brown}, T.~M., {Barnes}, S., {et~al.} 2018, in Society of
  Photo-Optical Instrumentation Engineers (SPIE) Conference Series, Vol. 10702,
  Ground-based and Airborne Instrumentation for Astronomy VII, ed. C.~J.
  {Evans}, L.~{Simard}, \& H.~{Takami}, 107026C, \dodoi{10.1117/12.2312800}

\bibitem[{Skrutskie {et~al.}(2006)Skrutskie, Cutri, Stiening, Weinberg,
  Schneider, Carpenter, Beichman, Capps, Chester, Elias, Huchra, Liebert,
  Lonsdale, Monet, Price, Seitzer, Jarrett, Kirkpatrick, Gizis, Howard, Evans,
  Fowler, Fullmer, Hurt, Light, Kopan, Marsh, McCallon, Tam, Dyk, \&
  Wheelock}]{skrutskie_two_2006}
Skrutskie, M.~F., Cutri, R.~M., Stiening, R., {et~al.} 2006, The Astronomical
  Journal, 131, 1163, \dodoi{10.1086/498708}

\bibitem[{{Skrutskie, M. F.; Cutri, R. M.; Stiening, R.; Weinberg, M. D.;
  Schneider, S.; Carpenter, J. M.; Beichman, C.; Capps, R.; Chester, T.; Elias,
  J.; Huchra, J.; Liebert, J.; Lonsdale, C.; Monet, D. G.; Price, S.; Seitzer,
  P.; Jarrett, T.; Kirkpatrick, J. D.; Gizis, J. E.; Howard, E.; Evans, T.;
  Fowler, J.; Fullmer, L.; Hurt, R.; Light, R.; Kopan, E. L.; Marsh, K. A.;
  McCallon, H. L.; Tam, R.; Van Dyk, S.; Wheelock, S.}(2003)}]{doi_2mass}
{Skrutskie, M. F.; Cutri, R. M.; Stiening, R.; Weinberg, M. D.; Schneider, S.;
  Carpenter, J. M.; Beichman, C.; Capps, R.; Chester, T.; Elias, J.; Huchra,
  J.; Liebert, J.; Lonsdale, C.; Monet, D. G.; Price, S.; Seitzer, P.; Jarrett,
  T.; Kirkpatrick, J. D.; Gizis, J. E.; Howard, E.; Evans, T.; Fowler, J.;
  Fullmer, L.; Hurt, R.; Light, R.; Kopan, E. L.; Marsh, K. A.; McCallon, H.
  L.; Tam, R.; Van Dyk, S.; Wheelock, S.} 2003, 2MASS All-Sky Point Source
  Catalog,  IPAC, \dodoi{10.26131/IRSA2}

\bibitem[{{Smith} {et~al.}(2012){Smith}, {Stumpe}, {Van Cleve}, {Jenkins},
  {Barclay}, {Fanelli}, {Girouard}, {Kolodziejczak}, {McCauliff}, {Morris}, \&
  {Twicken}}]{smith_kepler_2012}
{Smith}, J.~C., {Stumpe}, M.~C., {Van Cleve}, J.~E., {et~al.} 2012, \pasp, 124,
  1000, \dodoi{10.1086/667697}

\bibitem[{{Sneden}(1973)}]{Sneden-73}
{Sneden}, C.~A. 1973, PhD thesis, THE UNIVERSITY OF TEXAS AT AUSTIN.

\bibitem[{Soderblom {et~al.}(2014)Soderblom, Hillenbrand, Jeffries, Mamajek, \&
  Naylor}]{soderblom_ages_2014}
Soderblom, D.~R., Hillenbrand, L.~A., Jeffries, R.~D., Mamajek, E.~E., \&
  Naylor, T. 2014, arXiv:1311.7024 [astro-ph],
  \dodoi{10.2458/azu_uapress_9780816531240-ch010}

\bibitem[{Somers {et~al.}(2020)Somers, Cao, \&
  Pinsonneault}]{somers_spots_2020}
Somers, G., Cao, L., \& Pinsonneault, M.~H. 2020, The Astrophysical Journal,
  891, 29, \dodoi{10.3847/1538-4357/ab722e}

\bibitem[{Somers {et~al.}(2019)Somers, Pinsonneault, \&
  Cao}]{somers_garrett_2019}
Somers, G., Pinsonneault, M.~H., \& Cao, L. 2019, {The SPOTS Models: A Grid of
  Theoretical Stellar Evolution Tracks and Isochrones For Testing The Effects
  of Starspots on Structure and Colors}, 1.0,  Zenodo,
  \dodoi{10.5281/zenodo.3593339}

\bibitem[{{Sousa}(2014)}]{Sousa-14}
{Sousa}, S.~G. 2014, [arXiv:1407.5817].
\newblock \doarXiv{1407.5817}

\bibitem[{{Sousa} {et~al.}(2015){Sousa}, {Santos}, {Adibekyan}, {Delgado-Mena},
  \& {Israelian}}]{Sousa-15}
{Sousa}, S.~G., {Santos}, N.~C., {Adibekyan}, V., {Delgado-Mena}, E., \&
  {Israelian}, G. 2015, \aap, 577, A67, \dodoi{10.1051/0004-6361/201425463}

\bibitem[{{Stassun} {et~al.}(2014){Stassun}, {Feiden}, \&
  {Torres}}]{stassun_empirical_2014}
{Stassun}, K.~G., {Feiden}, G.~A., \& {Torres}, G. 2014, \nar, 60, 1,
  \dodoi{10.1016/j.newar.2014.06.001}

\bibitem[{Stassun {et~al.}(2018)Stassun, Oelkers, Pepper, Paegert, De~Lee,
  Torres, Latham, Charpinet, Dressing, Huber, Kane, L{\'e}pine, Mann, Muirhead,
  Rojas-Ayala, Silvotti, Fleming, Levine, \& Plavchan}]{stassun_tess_2018}
Stassun, K.~G., Oelkers, R.~J., Pepper, J., {et~al.} 2018, \aj, 156, 102,
  \dodoi{10.3847/1538-3881/aad050}

\bibitem[{{Stumpe} {et~al.}(2014){Stumpe}, {Smith}, {Catanzarite}, {Van Cleve},
  {Jenkins}, {Twicken}, \& {Girouard}}]{stump_multiscale_2014}
{Stumpe}, M.~C., {Smith}, J.~C., {Catanzarite}, J.~H., {et~al.} 2014, \pasp,
  126, 100, \dodoi{10.1086/674989}

\bibitem[{{Stumpe} {et~al.}(2012){Stumpe}, {Smith}, {Van Cleve}, {Twicken},
  {Barclay}, {Fanelli}, {Girouard}, {Jenkins}, {Kolodziejczak}, {McCauliff}, \&
  {Morris}}]{stump_kepler_2012}
{Stumpe}, M.~C., {Smith}, J.~C., {Van Cleve}, J.~E., {et~al.} 2012, \pasp, 124,
  985, \dodoi{10.1086/667698}

\bibitem[{Tanaka \& Ward(2004)}]{tanaka_three-dimensional_2004}
Tanaka, H., \& Ward, W.~R. 2004, The Astrophysical Journal, 602, 388,
  \dodoi{10.1086/380992}

\bibitem[{{Tayar} {et~al.}(2022){Tayar}, {Claytor}, {Huber}, \& {van
  Saders}}]{tayer_guide_2022}
{Tayar}, J., {Claytor}, Z.~R., {Huber}, D., \& {van Saders}, J. 2022, \apj,
  927, 31, \dodoi{10.3847/1538-4357/ac4bbc}

\bibitem[{Tofflemire {et~al.}(2021)Tofflemire, Rizzuto, Newton, Kraus, Mann,
  Vanderburg, Nelson, Hawkins, Wood, Zhou, Quinn, Howell, Collins, Schwarz,
  Stassun, Bouma, Essack, Osborn, Boyd, F{\textbackslash}Hur{\'e}sz, Glidden,
  Twicken, Wohler, McLean, Ricker, Vanderspek, Latham, Seager, Winn, \&
  Jenkins}]{tofflemire_tess_2021}
Tofflemire, B.~M., Rizzuto, A.~C., Newton, E.~R., {et~al.} 2021, The
  Astronomical Journal, 161, 171, \dodoi{10.3847/1538-3881/abdf53}

\bibitem[{Tokovinin(2018)}]{tokovinin_ten_2018}
Tokovinin, A. 2018, \pasp, 130, 035002, \dodoi{10.1088/1538-3873/aaa7d9}

\bibitem[{Tokovinin {et~al.}(2013)Tokovinin, Fischer, Bonati, Giguere, Moore,
  Schwab, Spronck, \& Szymkowiak}]{tokovinin_chironfiber_2013}
Tokovinin, A., Fischer, D.~A., Bonati, M., {et~al.} 2013, Publications of the
  Astronomical Society of the Pacific, 125, 1336, \dodoi{10.1086/674012}

\bibitem[{{Torres} {et~al.}(2006){Torres}, {Quast}, {da Silva}, {de La},
  {Melo}, \& {Sterzik}}]{torres_vizier_2006}
{Torres}, C.~A.~O., {Quast}, G.~R., {da Silva}, L., {et~al.} 2006, VizieR
  Online Data Catalog, J/A+A/460/695

\bibitem[{Torres {et~al.}(2008)Torres, Quast, Melo, \&
  Sterzik}]{torres_young_2008}
Torres, C. A.~O., Quast, G.~R., Melo, C. H.~F., \& Sterzik, M.~F. 2008,
  Handbook of Star Forming Regions, Volume II, 757.
\newblock \url{http://adsabs.harvard.edu/abs/2008hsf2.book..757T}

\bibitem[{{Twicken} {et~al.}(2018){Twicken}, {Catanzarite}, {Clarke},
  {Girouard}, {Jenkins}, {Klaus}, {Li}, {McCauliff}, {Seader}, {Tenenbaum},
  {Wohler}, {Bryson}, {Burke}, {Caldwell}, {Haas}, {Henze}, \&
  {Sanderfer}}]{twicken_kepler_2018}
{Twicken}, J.~D., {Catanzarite}, J.~H., {Clarke}, B.~D., {et~al.} 2018, \pasp,
  130, 064502, \dodoi{10.1088/1538-3873/aab694}

\bibitem[{Vanderburg {et~al.}(2019)Vanderburg, Huang, Rodriguez, Becker,
  Ricker, Vanderspek, Latham, Seager, Winn, Jenkins, Addison, Bieryla,
  Briceño, Bowler, Brown, Burke, Burt, Caldwell, Clark, Crossfield, Dittmann,
  Dynes, Fulton, Guerrero, Harbeck, Horner, Kane, Kielkopf, Kraus, Kreidberg,
  Law, Mann, Mengel, Morton, Okumura, Pearce, Plavchan, Quinn, Rabus, Rose,
  Rowden, Shporer, Siverd, Smith, Stassun, Tinney, Wittenmyer, Wright, Zhang,
  Zhou, \& Ziegler}]{vanderburg_tess_2019}
Vanderburg, A., Huang, C.~X., Rodriguez, J.~E., {et~al.} 2019, The
  Astrophysical Journal, 881, L19, \dodoi{10.3847/2041-8213/ab322d}

\bibitem[{Wang {et~al.}(2016)Wang, Hogg, Foreman-Mackey, \&
  Sch{\"o}lkopf}]{wang_causal_2016}
Wang, D., Hogg, D.~W., Foreman-Mackey, D., \& Sch{\"o}lkopf, B. 2016, \pasp,
  128, 094503, \dodoi{10.1088/1538-3873/128/967/094503}

\bibitem[{Wood {et~al.}(in prep)Wood, Mann, Bush, Fields, \&
  Thao}]{wood_carina_prep}
Wood, M.~L., Mann, A.~W., Bush, J., Fields, M., \& Thao, P.~C. in prep

\bibitem[{Wood {et~al.}(2021)Wood, Mann, \& Kraus}]{wood_characterizing_2021}
Wood, M.~L., Mann, A.~W., \& Kraus, A.~L. 2021, The Astronomical Journal, 162,
  128, \dodoi{10.3847/1538-3881/ac0ae9}

\bibitem[{Wright {et~al.}(2010)Wright, Eisenhardt, Mainzer, Ressler, Cutri,
  Jarrett, Kirkpatrick, Padgett, McMillan, Skrutskie, Stanford, Cohen, Walker,
  Mather, Leisawitz, Gautier, McLean, Benford, Lonsdale, Blain, Mendez, Irace,
  Duval, Liu, Royer, Heinrichsen, Howard, Shannon, Kendall, Walsh, Larsen,
  Cardon, Schick, Schwalm, Abid, Fabinsky, Naes, \&
  Tsai}]{wright_wide-field_2010}
Wright, E.~L., Eisenhardt, P. R.~M., Mainzer, A.~K., {et~al.} 2010, The
  Astronomical Journal, 140, 1868, \dodoi{10.1088/0004-6256/140/6/1868}

\bibitem[{Wright \& Mamajek(2018)}]{wright_kinematics_2018}
Wright, N.~J., \& Mamajek, E.~E. 2018, Monthly Notices of the Royal
  Astronomical Society, 476, 381, \dodoi{10.1093/mnras/sty207}

\bibitem[{{Wright, Edward L.; Eisenhardt, Peter R. M.; Mainzer, Amy K.;
  Ressler, Michael E.; Cutri, Roc M.; Jarrett, Thomas; Kirkpatrick, J. Davy;
  Padgett, Deborah; McMillan, Robert S.; Skrutskie, Michael; Stanford, S. A.;
  Cohen, Martin; Walker, Russell G.; Mather, John C.; Leisawitz, David;
  Gautier, Thomas N., III; McLean, Ian; Benford, Dominic; Lonsdale, Carol J.;
  Blain, Andrew; Mendez, Bryan; Irace, William R.; Duval, Valerie; Liu,
  Fengchuan; Royer, Don; Heinrichsen, Ingolf; Howard,
  Joan}(2019)}]{doi_allwise}
{Wright, Edward L.; Eisenhardt, Peter R. M.; Mainzer, Amy K.; Ressler, Michael
  E.; Cutri, Roc M.; Jarrett, Thomas; Kirkpatrick, J. Davy; Padgett, Deborah;
  McMillan, Robert S.; Skrutskie, Michael; Stanford, S. A.; Cohen, Martin;
  Walker, Russell G.; Mather, John C.; Leisawitz, David; Gautier, Thomas N.,
  III; McLean, Ian; Benford, Dominic; Lonsdale, Carol J.; Blain, Andrew;
  Mendez, Bryan; Irace, William R.; Duval, Valerie; Liu, Fengchuan; Royer, Don;
  Heinrichsen, Ingolf; Howard, Joan}. 2019, AllWISE Source Catalog,  IPAC,
  \dodoi{10.26131/IRSA1}

\bibitem[{Zapatero~Osorio {et~al.}(2002)Zapatero~Osorio, B{\'e}jar, Pavlenko,
  Rebolo, Allende~Prieto, Mart{\'\i}n, \&
  Garc{\'\i}a~L{\'o}pez}]{zapatero_osorio_lithium_2002}
Zapatero~Osorio, M.~R., B{\'e}jar, V. J.~S., Pavlenko, Y., {et~al.} 2002,
  Astronomy and Astrophysics, 384, 937, \dodoi{10.1051/0004-6361:20020046}

\bibitem[{Zhang {et~al.}(2022)Zhang, Knutson, Wang, Dai, Santos, Fossati,
  Henry, Ehrenreich, Alibert, Hoyer, Wilson, \&
  Bonfanti}]{zhang_detection_2022}
Zhang, M., Knutson, H.~A., Wang, L., {et~al.} 2022, The Astronomical Journal,
  163, 68, \dodoi{10.3847/1538-3881/ac3f3b}

\bibitem[{{Zhou} {et~al.}(2018){Zhou}, {Rodriguez}, {Vanderburg}, {Quinn},
  {Irwin}, {Huang}, {Latham}, {Bieryla}, {Esquerdo}, {Berlind}, \&
  {Calkins}}]{zhou_warm_2018}
{Zhou}, G., {Rodriguez}, J.~E., {Vanderburg}, A., {et~al.} 2018, \aj, 156, 93,
  \dodoi{10.3847/1538-3881/aad085}

\bibitem[{{Zhou} {et~al.}(2022){Zhou}, {Wirth}, {Huang}, {Venner}, {Franson},
  {Quinn}, {Bouma}, {Kraus}, {Mann}, {Newton}, {Dragomir}, {Heitzmann},
  {Lowson}, {Douglas}, {Battley}, {Gillen}, {Triaud}, {Latham}, {Howell},
  {Hartman}, {Tofflemire}, {Wittenmyer}, {Bowler}, {Horner}, {Kane},
  {Kielkopf}, {Plavchan}, {Wright}, {Addison}, {Mengel}, {Okumura}, {Ricker},
  {Vanderspek}, {Seager}, {Jenkins}, {Winn}, {Daylan}, {Fausnaugh}, \&
  {Kunimoto}}]{zhou_abdor_2022}
{Zhou}, G., {Wirth}, C.~P., {Huang}, C.~X., {et~al.} 2022, \aj, 163, 289,
  \dodoi{10.3847/1538-3881/ac69e3}

\bibitem[{Ziegler {et~al.}(2019)Ziegler, Tokovinin, Briceño, Mang, Law, \&
  Mann}]{ziegler_soar_2019}
Ziegler, C., Tokovinin, A., Briceño, C., {et~al.} 2019, The Astronomical
  Journal, 159, 19, \dodoi{10.3847/1538-3881/ab55e9}

\bibitem[{Ziegler {et~al.}(2018)Ziegler, Law, Baranec, Morton, Riddle, De~Lee,
  Huber, Mahadevan, \& Pepper}]{ziegler_measuring_2018}
Ziegler, C., Law, N.~M., Baranec, C., {et~al.} 2018, The Astronomical Journal,
  156, 259, \dodoi{10.3847/1538-3881/aad80a}

\bibitem[{Zucker {et~al.}(2022)Zucker, Goodman, Alves, Bialy, Foley, Speagle,
  Groβschedl, Finkbeiner, Burkert, Khimey, \& Swiggum}]{zucker_star_2022}
Zucker, C., Goodman, A.~A., Alves, J., {et~al.} 2022, Nature, 1,
  \dodoi{10.1038/s41586-021-04286-5}

\end{thebibliography}
\bibliographystyle{aasjournal}

\end{document}